\documentclass[11pt]{article}
\usepackage{epsfig}
\usepackage{amssymb,amsmath}
\usepackage{epigraph}
\usepackage[lined,boxed,commentsnumbered]{algorithm2e}

\usepackage{graphicx}				
\usepackage{amssymb}

\def\Cov{\text{Cov}}
\def\Var{\text{Var}}

\def\arg{\text{arg}}

\newcommand{\beq}{\begin{equation}}
\newcommand{\eeq}{\end{equation}}
\newcommand{\bea}{\begin{eqnarray}}
\newcommand{\eea}{\end{eqnarray}}

\begin{document}



\begin{center}

{\LARGE
Market Self-Learning of Signals, Impact and Optimal Trading:
\vskip0.5cm 
Invisible Hand Inference with Free Energy
\vskip0.7cm
}
{\Large
(or, How We Learned to Stop Worrying and Love Bounded Rationality)
}

\vskip1.0cm
{\Large Igor Halperin  and Ilya Feldshteyn} \\
\vskip0.5cm
NYU Tandon School of Engineering \\
\vskip0.5cm
{\small e-mail: igor.halperin@nyu.edu,  if@q-rd.com}
\vskip0.5cm
\today \\

\vskip1.0cm
{\Large Abstract:\\}
\end{center}
\parbox[t]{\textwidth}{
We present a simple model of a non-equilibrium self-organizing market where asset prices are partially driven by investment decisions of a bounded-rational agent. The agent acts in a stochastic market environment driven by various exogenous "alpha" signals, agent's own actions (via market impact), and noise. 
Unlike traditional agent-based models, our agent aggregates {\it all} traders in the market, rather than being a {\it representative} agent. 
Therefore, it can be identified with a bounded-rational component of the market {\it itself}, providing a particular implementation of an Invisible Hand market mechanism. In such setting, market dynamics are modeled as a fictitious {\it self-play} of such bounded-rational market-agent in its adversarial stochastic environment. As rewards obtained by such self-playing market agent are not observed from market data, we formulate and solve a simple model of such market dynamics based on a neuroscience-inspired Bounded Rational Information Theoretic Inverse Reinforcement Learning (BRIT-IRL). 
This results in effective asset price dynamics with a non-linear mean reversion - which in our model is generated {\it dynamically}, rather than being postulated.
We argue that our model can be used in a similar way to the Black-Litterman model. In particular, it represents, in a simple modeling framework, market views of common predictive signals, market impacts and implied optimal dynamic portfolio allocations, and can be used to assess values of private signals. 
Moreover, it allows one to quantify a "market-implied" optimal investment strategy, along with a measure of market rationality. 
Our approach is numerically light, and can be implemented using standard off-the-shelf software such as TensorFlow.
 }
 \newcounter{helpfootnote}
\setcounter{helpfootnote}{\thefootnote} 
\renewcommand{\thefootnote}{\fnsymbol{footnote}}
\setcounter{footnote}{0}
\footnotetext{
We would like to thank Ernest Baver,  Eric Berger, Jean-Philippe Bouchaud, Peter Carr, Sergei Esipov, Andrey Itkin, Vivek Kapoor,  Dan Nudelman and Nikolai Zaitsev for stimulating remarks and discussions.
All errors are ours.
}     

 \renewcommand{\thefootnote}{\arabic{footnote}}
\setcounter{footnote}{\thehelpfootnote} 

\newpage
 
\section{Introduction}

This paper presents a simple 'structural' model of price dynamics in a financial market.
Though based on concepts not commonly used in Finance ( Reinforcement Learning, Information Theory, Physics etc. see below), the model we suggest is mathematically rather 
simple {\it at the end} (see Eq.(\ref{GMR})), after getting through a 'story' behind its structure. It is designed as both a {\it practical tool} for market practitioners, and a  
{\it theoretical model} of a financial market that can be explored further using  simulations and/or analytical methods. 
 For definitiveness, we focus in this paper on stock markets, though the same approach can be applied to other markets in the same way.

In a way, the main idea of a model presented below can be formulated as a {\it dynamic} and {\it data-driven} extension of an approach to modeling excess returns that was suggested in the seminal Black-Litterman (BL) model \cite{BL}. As will be shown below, a structural asset return model arising in our solution to this problem has some interesting properties such as the presence of {\it mean reversion} in stock prices, which in our framework appears as a result of joint actions of all traders in the market that {\it dynamically} implement Markowitz-type mean-variance portfolio strategies.  

In essence, the BL model flips the Markowitz optimal portfolio theory
\cite{Markowitz} on its head, and considers an {\it inverse} optimization problem. 
Namely, it starts with an observation that a {\it market  portfolio} (as typically represented by the S\&P500 index) is, by definition, the optimal "market-implied" portfolio. 
Therefore, if we consider such a given market portfolio as an {\it optimal} portfolio, then 
we can {\it invert} the portfolio optimization problem, and ask what is the optimal asset allocation {\it policy} that corresponds to this optimal market portfolio.
Within the framework of a single-period Markowitz mean-variance optimization \cite{Markowitz}, this translates into market-implied values of expected returns and covariances of returns. 

Respectively, this framework was suggested by Black and Litterman as a way to assess values of private "alpha" signals in generating excess returns. The BL model was explicitly re-interpreted as an {\it inverse} portfolio optimization problem by Bertsimas  {\it et. al.} 
\cite{Bertsimas_2012}, along with proposing some extensions such as robust inverse optimization. Note that the inverse optimization in \cite{Bertsimas_2012} is still performed in a single-period (one time step) setting, the same as in the original BL model \cite{BL} and in the Markowitz mean-variance portfolio model \cite{Markowitz}.
   
A model suggested in this paper extends such inverse optimization view of the market portfolio to a {\it dynamic}, multi-period setting. 
While this requires some new mathematical tools, the {\it outputs} of the model can be used in essentially the same way as the outputs of the BL model: to assess the value of private "alpha" signals, and design trading strategies according to own assessments of joint effects of signals and market impacts from trades on expected excess returns. 

An important difference of our model from a majority of market models used in both the industry and the academia is that our model does {\bf not} assume a competitive market equilibrium. As discussed at length by Duffie \cite{Duffie}, this paradigm underlies three cornerstone Nobel prize-winning theories of modern Finance, which are used by many practitioners on both the sell and buy sides. 
On the other hand, George Soros, a famous guru of financial markets, called this paradigm an "absurd postulate"\footnote{"Economics ended up with the theory of
{\it rational expectations}, which maintains that there is a single optimum view of the future, that which corresponds to it, and eventually all
the market participants will converge around that view. This postulate is {\it absurd}, but it is
needed in order to allow economic theory to model itself on {\it Newtonian Physics}." (G.~Soros). We thank Vivek Kapoor for this reference.}. 

Our model can be interpreted as an attempt to reconcile such opposite views. Our suggested answer is that both sides are right in their own ways, but we offer a practical and easily computable {\it unifying} framework. This allows us to quantify Soros' critique and propose a simple model that can be used to describe 
markets in three different states: disequilibrium, quasi-equilibrium, and a perfect thermal equilibrium.  The latter scenario may only occur if there is no inflow of information in a market - hardly a realistic scenario. 

The last case of  a perfect thermal equilibrium corresponds  to assumptions of the competitive market equilibrium paradigm. 
While we believe that for financial markets the last limit is in a way 'non-physical'\footnote{It is non-physical in the sense that it contradicts the very existence of markets where market makers generate liquidity and speculators make profits by digesting new information - neither should exist in competitive market equilibrium models. This is because a perfect equilibrium is only possible for a closed system that does not exchange information with an outside world. Therefore, competitive market equilibrium models do not try to answer the question {\it why} markets exist, but rather simply postulate first-order optimality (equilibrium) conditions, and then explore the consequences \cite{Sornette_book}. In physics, a perfect thermodynamic equilibrium is achieved in the thermodynamic limit of a closed system, and corresponds to a state of a 'heat death of the Universe' \cite{Landau}.},  it is the limit described by competitive market equilibrium models such as the 
Modigliniani-Miller's capital structure irrelevance for the market value of a corporation,
the Capital Asset Pricing Model (CAPM) of William Sharpe (1964), and the Black-Scholes model of option pricing\footnote{The Black-Scholes model relies on a weaker form of competitive market equilibrium paradigm known as the no-arbitrage principle \cite{Duffie}.}\cite{Duffie}. 

These models consider market dynamics as equilibrium fluctuations around a perfectly thermodynamically equilibrium market state. Therefore, they implicitly assume that there is no inflow/outflow of money and information in a market as a whole, and the market is in a state of a maximum entropy.
 This may be a reasonable approximation in a steady/slow market, which may explain why these models work reasonably well under 'normal' market conditions.   

But this assumption of competitive market equilibrium also suggests that these models should behave progressively worse during periods of market instabilities, crises and market crashes - an observation that seems to be widely recognized in the literature.

The general reason for such model failures when they are needed most is that in all these cases, a view of a market as an equilibrium 
fluctuation around a stationary state where entropy is already maximized becomes inadequate to describe market dynamics, see also below on analogies with self-organizing systems and living organisms.           

The above remarks concern with potential theoretical implications of our framework. Irrespective, our model also attempts to address the needs of market {\it practitioners} that want to make a profit rather than do a theoretical research into the dynamics of the markets.
 
To this end, in addition to providing a multi-period extension of the Black-Litterman model\footnote{Note that because the latter is a one-period model, a question of a market in equilibrium vs non-equilibrium cannot even be formulated in this framework.}, our model produces "market-implied" values of market impact parameters and risk aversion parameter of an agent that {\it dynamically maintains} such market-optimal portfolio, as well as a "market-implied" optimal investment strategy, which can be viewed for monitoring of the market or individual players in the market (see below). Given an explicit formula produced by our model for a market-implied optimal strategy, expressions like 'a strategy that beats the market' can now be probably given a more quantitative meaning {\it ex-ante} rather than {\it ex-post}. 

Finally, one more interesting insight may be provided by the fact that one of parameters estimated by the model from market data is a parameter
$ \beta $ that describes a degree of rationality of the market (another name for $ \beta $ is the "inverse temperature" of the market). 
This suggests that market-implied value of $ \beta $ can be used as a monitoring tool and possibly a predictive signal to have an aggregative view of a market, a specific exchange, or even a specific dealer\footnote{The latter case corresponds to a possible application of the model developed in this paper for an individual investor rather than for the market as a whole, see also below.}. 

\subsection{Outlook of our approach}
\label{sect:Outlook}

The main intuitive idea behind the model can be introduced as follows. While the real market dynamics are highly complex as they are driven, to a large extent, by a very large number of individual rational or bounded-rational market participants, is is commonly known that market players exhibit a strong tendency to a herding behavior: when markets are up, everyone is buying, and when markets are down, everyone is selling. 

This suggests a concept of a {\it representative investor} whose objective is to optimize a given investment portfolio given some objective function. Such representative investor is otherwise known in the literature as an {\it agent}. In this view of the world, an environment, i.e. the market, is clearly {\it external} to the agent. 

But what if we take an {\it inverse} optimization view of this problem, as in Black-Litterman \cite{BL}? In this approach, the optimal portfolio is already 
{\it known}, it is the market portfolio itself. But then, who is an {\it agent} that {\it dynamically} maintains (rebalances) such market-optimal portfolio? 

We can identify such agent with a 'collective mode' of {\it all} individual traders involved in the market, that are guided in their decisions by a commonly agreed set of predictors $ {\bf z}_t $ which may include news, other market indicators and/or indexes, variables describing the current state of the limit order book, etc. Therefore, the first difference of our framework from  conventional utility-based models is that our agent is a {\it sum} of all investors, rather than their {\it average}, i.e. a 'representative' investor. 

Because such agent aggregates actions of a partly {\it homogeneous} and partly {\it heterogeneous} crowd of individual investors, it can not be a fully {\it rational} agent, but rather should be represented as an agent with {\it bounded rationality}. Bounded rationality, which will be explained in more details below, is a second key difference of our framework from a classical agent-based approach.
 
 Furthermore, because jointly all individual trades by {\it all} market participants 
amount to {\it actual} market moves that dynamically re-adjust the market-optimal 
portfolio, such agent 
can then be identified with a bounded-rational component of the market {\it itself}. 

If we adopt such view, the actual dynamics of market prices can now be mathematically modeled as 
a sequential decision-making process of such bounded-rational agent who is engaged in {\it self-learning via self-play} in a partly controllable and partly uncontrollable environment which is identified with the {\it rest} of the market. The first component identified with a RL agent can then be thought of as a 'mind' of such {\it self-organizing} market, that learns about its environment and itself via self-play in such open environment.

Our agent embodies an 'Invisible Hand' of the market, which is {\it goal-oriented} in our framework, as will be made more clear below.
The Invisible Hand is implemented in our model as a fictitious self-play of a bounded-rational RL agent.
Agent's self-play amounts to mimicking a risk-averse investor seeking a dynamic Markowitz-optimal portfolio, while actions of this investor are randomized by entropy. As will be shown below in Sect.~\ref{sect:Adversarial}, this is mathematically equivalent to portfolio optimization in a two-party game with 
an {\it adversarial} player, such that the original agent and its imaginary adversary form a Nash equilibrium.   
As a result, the agent simultaneously mimicks {\it all} traders as a bounded-rational 'mind' of a self-organizing market\footnote{
Equivalence between self-organization in dynamic systems and sequential decision making was emphasized by Yukalov and Sornette in \cite{YS_2014}.
A similar approach in neuroscience 
is a unified free-energy model of the brain of Friston \cite{Friston_2010}, see also \cite{Friston_2018} for recent applications of the free-energy principle to living systems. In short, this approach suggests that "all biological systems instantiate a hierarchical generative model of the world that implicitly minimizes its internal entropy by minimizing free energy" \cite{Friston_2018}.}.

This produces a dynamic model of market price dynamics, where a {\it total} price impact of all traders in the market, who try to construct Markowitz-optimal portfolios, amounts to a {\it dynamically} generated mean reversion in market-observed asset returns. 
The resulting model can be interpreted as a Geometric Mean Reversion model with external signals, where mean reversion arises dynamically, rather than being introduced by hands, as is done in descriptive models of market dynamics. The resulting model can be viewed as a non-linear factor model for returns that can be estimated using standard methods of statistics such as Maximum Likelihood. 
 
More than that, because our resulting asset return model is a {\it structural} model, mean reversion in our model has a 'story' behind it.
In our approach it is produced by a total bounded-rational action of all (bounded-rational or rational) agents in the market, that dynamically optimize their investment portfolios following mean-variance strategies. This can be compared with a mechanism for mean reversion due to zero-intelligence 'noise traders' suggested in 1988 by Poterba and Summers \cite{PS_1988}.  We have only one agent, but it is bounded-rational, unlike many noise traders with a zero 
total rationality/intelligence in the model of \cite{PS_1988}.    
 
\subsection{Possible insights from the model}
\label{Possible_insights}

Because we formulate the problem in a setting of inverse, rather than direct portfolio optimization, the objective of  
a bounded-rational agent can be viewed as the problem of rebalancing its own fictitious "shadow" portfolio, such that it is kept as close as possible to the market portfolio in such continuous self-play. Note that except for a bounded rationality of an agent assumed in such framework, it resembles the classical pole balancing problem of Reinforcement Learning (see e.g. \cite{SB}), where now it is the market portfolio that serves the role of a pole, and we {\it invert} the problem.   

Our model is also quite similar to an index tracking problem, except we set it as an {\it inverse} optimization problem to infer market views of its own dynamics, instead of solving a {\it forward} optimization problem of finding a good tracking portfolio for an index.
Note that data for such model formulation is readily available as level-1 limit order book (LOB) data (level-2 LOB data can be incorporated in the model via a set of external predictors $ \bf{z}_t $, see below). 

This is unlike a (mathematically identical) portfolio optimization problem for an {\it individual} trader, that would require trader's 
{\it proprietary} execution data for model estimation. If, however,  such trader's proprietary data {\it are} available, our framework can be used in the same way to construct a probabilistic model of the trader. This could be used, in particular, by regulators for monitoring activities of exchanges or individual traders. 

Note that in a single-period setting, our problem formulation brings us back to the BL model, where instead of multi-period trading strategies, we have 
 just single-period optimal portfolio allocations. 
 
 On the other hand, in a multi-period formulation, it extends the setting of the BL model in multiple ways, including a {\it probabilistic} model of observed actions, that takes into account effects that are absent in a single-period settings, such as dynamic market impacts and dynamically changing predictors.
As our model is probabilistic, i.e. {\it generative}, it can be used for {\it forward} simulation of dynamics.    

Also note that in a multi-period setting, it is a combination of non-linearity induced by market impacts and dynamical exogenous predictors $ \bf{z}_t $ that may produce potentially very rich dynamics that would be driven by a combinators of external signals $ {\bf z}_t $, non-linear system feedback via agent's trades, and incontrollable noise. 
As we will show below, our model is tractable in a quasi-equilibrium setting using conventional tools of constrained convex optimization, due to its simple structure with a {\it quadratic} non-linearity of dynamics. 

On the other hand, external signals $ {\bf z}_t $ have their own dynamics, and might operate at different frequencies from typical times of market responses to news, events, and other changes in predictors $ {\bf z}_t $. Therefore, due to its non-linearity, and depending on a relation between characteristic times of market responses and signal changes, the model can describe both an equilibrium and non-equilibrium settings with such non-linear dynamics. 
A combination of non-linearity of dynamics with particular patterns of external signals $ {\bf z}_t $ whose changes provide new information to the agent, can lead to potentially very rich dynamics. 

We will leave an exploration into {\it generative} properties of our model for a future research. 
The focus of the present paper is rather of a {\it batch-mode} (off-line) learning from {\it past} data. 
Such learning can be done using model-free or model-based Reinforcement Learning (see e.g. \cite{SB}) when rewards are observable, or Inverse Reinforcement Learning (IRL) when they are not. As in our case rewards (either of a single investor, or 'market-implied' rewards) are {\it not} observable, we rely on an IRL-based approach for learning in such setting. 

Our model, where rewards are not observed but rather {\it inferred} from data, belongs in a class of model-based IRL approaches with a parametrized
reward function and dynamics. The objective of modeling in this approach is to infer the {\it reward function} and {\it action policy} from data by 
tuning model parameters. As in our case we 
solve the dynamic inverse portfolio optimization problem for a {\it market-optimal} portfolio, our IRL approach infers 
{\it market-implied} reward function and optimal action policy.

Note that in typical applications of RL for financial decision making, an agent is typically a (representative or particular) trader or a financial institution
who is {\it external} to the market. In contrast, in our approach, an agent is the bounded-rational component of the market {\it itself}, as it is now 
{\it inseparable} from the market, so long as it {\it maintains} the market-optimal portfolio. 

Therefore, our model is a dynamic model of the market itself, rather than a model of an external representative investor in such market.  Our model is inspired by IRL, Information Theory, statistical physics and neuroscience, yet it is based on a simple parametric specification of a one-step reward, and a simple specification of dynamics. 
 
 The model is tractable as a non-linearity of dynamics is 'only' quadratic.
Furthermore, because we use a simple low-dimensional parametric specification of the 'actual' reward of the agent, the data requirements for the model are modest. The model does not need tens, hundreds, or thousands years of training data, even though both the state and action spaces in our problem are very high-dimensional. 

Computationally, the model amounts to a simple and transparent scheme, rather than being a black-box model in the spirit of 
Deep Reinforcement Learning. This is because a simple parametric specification of the model enables proceeding without sophisticated function approximations that are typically implemented in Deep Reinforcement Learning by deep neural networks.
The main computational tool employed by the model is (an iterative version of) the conventional Maximum Likelihood estimation method available via standard off-the-shelf numerical optimization software. This can be conveniently done with TensorFlow using its automatic differentiation functionality. 
        
The paper is organized as follows. In Sect.~\ref{sect:Related_work}, we review related work, and simultaneously provide further high-level
details of our framework.  
In Sect.~\ref{sect:Portfolio} we introduce our notation and describe an investment portfolio of stocks.
In Sect.~\ref{sect:RL}, we present a RL formulation of the model.
Sect.~\ref{sect:IRL} re-formulates the model in an IRL setting, and presents our solution to the problem of finding an 
optimal policy and reward function for a single investor case.
The IRL problem for the market as a whole is addressed in Sect.~\ref{sect:IRL_market}.
The same section introduces an effective market dynamics model that is obtained as a by-product of our IRL solution.
Experiments are presented in Sect.~\ref{sect:Experiments}.
Sect.~\ref{sect:Discussion} discusses our results and outlines future directions. A brief summary is given 
in Sect.~\ref{sect:Summary}.

\section{Related work}
\label{sect:Related_work}

Our model builds on several threads developed separately by the Quantitative Finance, Reinforcement Learning, Information Theory, Physics and Neuroscience communities. Here we provide a brief overview of related work in these different fields that have a close overlap with the model developed here, as well as explain their 
relation with our approach.

\subsection{What kind of equilibrium holds for markets?}

Quoting Duffie, "while there are important alternatives, a current basic paradigm for valuation, in both academia and in practice, is that of competitive market equilibrium" \cite{Duffie}. While this was said in 1997, this assessment remains true to this day.

Of course, deficiencies of standard financial models based on competitive market equilibrium and/or no-arbitrage paradigm were not left unnoticed both within 
the financial community, and among researchers in other disciplines, most notably physics and computer science. The latter disciplines contributed a number of interesting and fresh ideas to financial modeling \cite{Bouchaud_book}, \cite{Sornette_book}. In particular, agent-based models may provide interesting insight into how financial markets can operate when viewed as {\it evolving complex systems}, see e.g. \cite{Abergel}. 

The main challenge with agent-based models is that while they are capable of explaining some stylized facts of the market, they can hardly be turned, at least at the current stage, into practically useful tools - in part, due to their high computational complexity. While models such as CAPM or the Black-Scholes model may miss some important features of real markets, they also work reasonably well under certain market and trade conditions, and they are fast. 

Yet, to better model effects such as market liquidity, Amihud {\it et. al.}  suggested  that, instead of assuming competitive market equilibrium, researchers should assume an "equilibrium level of disequilibrium" \cite{Amihud_2005}. In physics, this is normally referred to as a non-equilibrium steady state. 

Viewing markets as an 'equilibrium disequilibrium' is beneficial if we are willing to consider them as evolving and self-organizing systems that may bear some similarities to living organisms. Boltzmann and Sch{\"o}dinger have emphasizes that activities of living organisms are impossible in thermal equilibrium, and necessarily depend on harnessing a pre-existing disequilibrium. In other words, as a consequence of the Second Law of thermodynamics, living organisms can only exist as processes {\it on the way} to a state of maximum entropy describing a thermal equilibrium, but {\it not} in this state itself \cite{Schrodinger}, \cite{Marsland_2017}.  

A demand-based option pricing model that does not rely on no-arbitrage assumptions was proposed by Garleanu {\it et. al.} \cite{Garleanu_2007}. 
A Reinforcement Learning based option pricing model that similarly does not rely on no-arbitrage but uses instead a model-free and data-driven Q-learning approach was proposed by one of the authors in \cite{IH_2017}\footnote{If so desired, the latter model can also be constructed as an arbitrage-free model, by using a suitable utility function, instead of a quadratic utility \cite{IH_2017}.}. Residual inefficiencies of markets resulting from multi-step strategies and market impact were studied by Esipov \cite{Esipov_2012}.

\subsection{Optimal portfolio execution} 

A close analog of a setting of our model is a problem of optimal execution in stock trading, one of the classical problems of Quantitative Finance.
The problem amounts to designing an optimal strategy (policy) for partitioning a large trade order to buy or sell a large block of a stock of some company into smaller chunks, and buy these chunks sequentially so that a potential market impact would be minimized, and respectively the total cost of implementing the trade will be minimized as a result. This is a problem solved many thousands times a day by brokers, as well as those asset managers and hedge funds that execute such trades themselves instead of passing trade orders to brokers.

The classical way to address such (forward) optimization problem is to start with building and calibrating models for stock dynamics and price impact.
Provided this is done, the next step is to define a {\it cost function} that specifies loss that will be observed upon taking certain actions in certain states.
If we focus for now on execution strategies that involve only market orders but not limit orders, then these market orders will be our {\it actions} $ {\bf a}_t $\footnote{This is sufficient if we look at aggregate actions of {\it all} traders, i.e. the market itself, which is the main setting of our model in this paper. 
If the model is applied to an {\it individual} investor, restricting a model to modeling only market orders may be a reasonable approximation for 
liquid stocks, while for stocks with limited liquidity optimal strategies may involve combinations of market and limit orders. Extensions of our framework to such setting of mixed market and limit orders for individual investors will be provided elsewhere.}. 

Assume that the trade order is to sell $ N $ shares of a given stock within time $ T $. {\it Optimal} actions $ {\bf a}_t^{\star} $ are obtained from (forward) optimization of total cumulative costs of execution, as determined by a {\it policy} $ \pi_t = \pi_t({\bf y}_t) $. Here $ t $ is current time and $ {\bf y}_t $ is    
a {\it state} vector of the system that includes the current mid-price of the stock $ S_t $, the number of stocks $ n_t $ currently held, and values of external signals $ {\bf z}_t $ that may include, in particular, predictors derived from properties of the limit order book (LOB). 

If $ \pi_t^{\star}({\bf y}_t)  $ is a (deterministic) optimal policy, then the optimal action $ {\bf a}_t^{\star} $ is simply the value 
$  {\bf a}_t^{\star}  =  \pi_t^{\star}({\bf y}_t) $.
The classical multi-period optimal execution problem was formulated in the dynamic programming (DP) setting by Bertsimas and Lo \cite{Bert_Lo}
for a risk-neutral investor, and then extended by Almrgen and Chriss \cite{AC}  to a risk-averse investor.  

\subsection{Inverse portfolio optimization}

In this paper, we consider three (related) modifications to the direct optimization problem described above. {\it First}, we take the view of dynamic {\it inverse} optimization, in the spirit of the Black-Litterman model \cite{BL} and its reformulation in \cite{Bertsimas_2012}, and  assume that such optimization problem was already {\it solved} by the market itself. Respectively, we look for market-implied optimal trading policies/strategies rather than trading/execution strategies of an individual investor.  However, our market-wise aggregate trader-agent does the same thing as nearly all traders in the market do, i.e. it 
dynamically optimizes its own investment portfolio. 

\subsection{Dynamic portfolio management with constrained convex optimization}

In our specification of single-step rewards, or negative costs of trading, we follow a large literature on multi-period mean-variance optimization.
An accessible review of a version of such mean-variance optimization is given by Boyd {\it et. al.} \cite{Boyd_2017}. 
We largely adopt the notation and assumption of the portfolio model suggested by Boyd {\it et. al.}, while in addition we explicitly introduce predictors and market impacts effects not considered in \cite{Boyd_2017}. Quadratic objective functions for multi-period portfolio optimization discussed at length 
in \cite{Boyd_2017} are formulated within the conventional DP approach that assumes a known model, including a known risk aversion parameter.

\subsection{Stochastic policies }

The {\it second} modification we make to the classical formulation of the optimal execution problem is that we consider {\it stochastic} (probabilistic), rather than {\it deterministic} policies $ \pi $. A stochastic policy $ \pi_t({\bf y}_t) $ describes a probability distribution, so that action $ {\bf a}_t $ becomes a sample from this distribution,
$ {\bf a}_t \sim \pi ( {\bf y}_t ) $, rather than a fixed number. Respectively, an {\it optimal} action would be a sample from an {\it optimal} policy, 
$  {\bf a}_t^{\star} \sim \pi^{\star} ( {\bf y}_t ) $. Deterministic policies can now be viewed as a special case of stochastic policies, where the action distribution is a Dirac delta-function $ \pi ( {\bf a}_t | {\bf y}_t ) = \delta \left( {\bf a}_t - {\bf a}_t^{\star}({\bf y}_t) \right) $ where $ {\bf a}_t^{\star}({\bf y}_t )$ is an optimal action 
for state $ {\bf y}_t $, which corresponds to a deterministic policy setting of the classical DP approach.

What would be the meaning of such probabilistic modeling of execution orders, given that at the end they amount to specific numbers, rather than probabilities?
Such choice can be justified for both the direct and inverse problems of optimal execution. 

Let's start with an argument why stochastic policies can be useful for {\it direct} optimization. Given that parameters defining optimal strategies are estimated from data, the resulting policy is always stochastic {\it de-facto}, even though this is {\it not} explicitly recognized in deterministic policy execution models such Bertsimas and Lo \cite{Bert_Lo} and Almgren and Chriss \cite{AC} models. 

Adapting stochastic policies as a principal modeling tool allows one to explicitly {\it control uncertainty} around an optimal action in each state of the world. The latter can be identified with a mode of the policy distribution, while uncertainty around this value would be specified by properties of this distribution, and measured, in a simplest case, by the variance of a predicted optimal action.
This argument is rather similar to an argument for stochastic, rather than deterministic, portfolio allocations for a one-period Markowitz-like portfolio optimization problem, which was put forward by Marschinski {\it et. al.} \cite{Rossi}.      

In the setting of {\it inverse} portfolio optimization adopted in this paper, the usefulness of stochastic policies becomes even more evident.
In this case, stochastic policies are needed in order to account for a possible {\it sub-optimality} of a policy used in generated data. Such events would be incompatible with an assumption of a strict optimality of {\it each} action in the data, leading to vanishing probabilities of observed execution paths.
Reliance on stochastic rather than deterministic policies allows one to cope with possible sub-optimality of historical data.

\subsection{Reinforcement Learning}
 
Deterministic policy optimization problem in a dynamic mean-variance optimization setting similar to Boyd {\it et. al.} \cite{Boyd_2017} was reformulated in a data-driven Reinforcement Learning (RL) way by Ritter \cite{Ritter}. Ritter considers the classical {\it on-line} Q-learning for the problem of multi-period portfolio optimization from data, using a quadratic risk-adjusted cost function. This translates the problem into a {\it data-driven} forward optimization that can  can be solved, given enough training data, by the famous Q-Learning of Watkins and Dayan \cite{Watkins}.

The difference between our approach and Ritter's is that we consider an {\it off-line} (batch-mode) learning, and we do {\it not} observe one step costs (or, equivalently, negative rewards). Therefore, our setting is that of IRL, while Ritter \cite{Ritter} considers an on-line RL formulation. Also, unlike \cite{Ritter} that uses a discretized state space, we use a continuous-state formulation. 
Furthermore, Ritter considers a general optimal portfolio investment problem for a given (representative?) investor, while here we focus on modeling an agent that represents a bounded-rational component of the market as a whole. This transforms our approach into a {\it market} model, unlike the case considered by Ritter, which is a {\it trader} model.

Quadratic risk-adjusted objective functions were considered in an apparently different problem of optimal option pricing and hedging using a model-free, data-driven approach in the work by one of the authors \cite{IH_2017, NuQLear}. The approach used in this work assumes {\it off-line}, batch-mode learning, that enables using data-efficient batch RL methods such as Fitted Q Iteration \cite{Ernst, Murphy}.

We use entropy-regularized Reinforcement Learning in the form suggested by Tishby and co-workers under the name of G-learning \cite{G-Learning}, 
as a way to do Reinforcement Learning in a noisy environment.
While \cite{G-Learning} assumed a tabulated discrete-state/discrete-action setting, in our case both the state and action spaces are high-dimensional continuous spaces. For a tutorial-style introduction to Information-constrained Markov Decision Processes, see Larsson {\it el. al.} \cite{Larsson_2017}.

\subsection{Inverse Reinforcement Learning}

A {\it third} modification we introduce to the classical portfolio optimization scheme is that we assume that some critical model parameters are {\it unknown}.
Note that forward optimization using Dynamic Programming always assumes that dynamics and model parameters are {\it known}, or estimated using independent models. In particular, market impact parameters or risk aversion parameters are not easy to mark down without using additional models to estimate them {\it before} using them in a direct execution optimization method. Moreover, traders do not necessarily even {\it think} in terms of {\it any} utility function, and respectively may not even know their {\it own} risk-aversion parameter $ \lambda $.

Unlike such DP approach, in our model we treat these parameters as {\it unknown}, and estimate them {\it simultaneously} with estimating optimal policy from historical trading data. What we obtain with such procedure can be interpreted as {\it implied} market impact and risk aversion parameters, similar 
to how implied volatilities are used to price and hedge options in option markets. In particular, even if traders may not {\it think} in terms of a quadratic utility function with some pre-determined value of $ \lambda $, their {\it observed behavior} might be consistent with such simple utility function, with some {\it data-implied} risk aversion rate $ \lambda = \lambda_{imp} $. 

Note that when risk aversion $ \lambda $ and parameters determining market impact are unknown, it also 
means that one-step {\it costs} (see below) are unknown as well.
Our data therefore consist of sequences of states and actions, but it does not reveal {\it costs} incurred by following these actions. Such problems of estimating costs (or rewards) from an observed behavior are solved using methods of Inverse Optimal Control (IOC) when dynamics are {\it known}, or using Inverse Reinforcement Learning (IRL) when dynamics are {\it unknown}.  

In this paper, we address this problem using model-based IRL. Our framework relies on a model for specification of one-step costs, market impact, and risk metrics (we will use quadratic risk measures going forward). 
On the side of IRL literature, our approach is based on Maximum Entropy IRL developed in \cite{Ziebart_2008}, and extended to continuous-space formulation in \cite{Monfort_Ziebart_2015}. A closely related method is Iterative Quadratic-Gaussian Regulator (IQGR) of Todorov and Li 
\cite{Todorov_2005}.    
 
 \subsection{Neuroscience and biology}
 
 Our approach is similar to a Free-Energy Principle (FEP) approach to living systems and  the brain function 
 developed by Friston and collaborators in \cite{Friston_2010, Friston_2018}. Under this formalism, "for an organism to resist dissipation and persist as an adaptive system that is a part of, coupled with, and 
 yet statistically independent from, the larger system in which it is embedded, it must embody a probabilistic model of the statistical interdependencies and regularity of its environment" \cite{Friston_2018}.
 
 Our model applies similar approach, based on ideas from statistical thermodynamics, to the market as a dynamic persistent and adaptive system that embodies a bounded-rational RL agent that imitates a 'mind' of the market as a {\it goal-directed} 'living organism' in an adversarial environment. 
We implement the above requirement that the agent should embody a probabilistic model of its environment by formulating this problem as Inverse Reinforcement Learning. The free energy arises in this approach either as a way to regularize (Inverse) Reinforcement Learning in a noisy environment by entropy, as in G-learning \cite{G-Learning}, or as a way to model bounded-rational decision-making of the agent by imposing constraints on information processing costs
\cite{Ortega_2012, Tishby_2012, Ortega_2015}, or equivalently as a way to account for an adversarial character of the environment, see the next two sections.

\subsection{Thermodynamics, Bounded Rationality and Information Theory}

Another, and mathematically equivalent way to introduce entropy and free energy into the problem of sequential decision-making, was formulated within an information-theoretic and physics-inspired approach in \cite{Ortega_2012, Tishby_2012, Ortega_2015}.  In particular, Ortega {\it et. al.}  \cite{Ortega_2012, Ortega_2015} emphasize that a regularization 'inverse temperature' parameter $ \beta $ that corresponds to a cost of information processing in a system, can also be interpreted as a {\it degree of rationality} of an agent that dynamically maximizes its free energy (i.e. an entropy-regularized value function). 

This interpretation is provided by noting that parameter $ \beta $ determines complexity of a search for a better policy starting from a given prior policy
\cite{Ortega_2012, Ortega_2015}. Agents with large $ \beta \rightarrow \infty $ can afford a highly complex (costly) search for a better policy, and therefore 
are more rational than agents that live in a world with a small value $ \beta \rightarrow 0 $. In this regime, an agent cannot afford to change from a prior policy, and therefore behaves as an irrational (entropy-dominated) agent. 
The information-theoretic approach thus provides a quantitative and tractable framework for a bounded-rational agent of Simon \cite{Simon_1956}. 

\subsection{Self-play, adversarial learning, and the free energy optimization}

An adversarial interpretation of Information Theoretic Bounded Rationality was suggested in \cite{Ortega_2014} where it was shown that a single-agent free energy optimization is equivalent to a fictitious game between an agent and an imaginary adversary. In our model, we have a similar setting, where an agent representing a bounded-rational component of the market optimizes its free energy. The optimization amounts to a dynamical optimization of
 agent's portfolio in a stochastic market environment with information processing costs. The latter are expressed as an entropy regularization of a value function, see below.
As will be shown in Sect.~\ref{sect:Adversarial}, using the method of \cite{Ortega_2014}, such self-play can be equivalently viewed as {\it adversarial} learning in a fictitious {\it two-party} game with an adversarial opponent. 

\subsection{Bounded Rational Information Theoretic IRL (BRIT-IRL)}

Our approach integrates ideas of Maximum Entropy IRL with a Bounded Rational Information-Theoretic interpretation of the process of learning, and applies them to make inferences of an 'Invisible Hand', in the spirit of the Black-Litterman model.
In splitting the market into its bounded-rational self and the rest, the model also has strong similarities with the free-energy approach to the brain and biological systems \cite{Friston_2010, Friston_2018}. In our approach, such view is applied to a financial market as a {\it dynamic self-organizing} system, with a focus on {\it inverse} rather than direct learning.   

As our setting is of {\it inverse} learning, instead of assuming some value of degree of rationality $ \beta $, we {\it infer} such parameter implied by the market data within our model. This produces a dynamic "market-implied" index of rationality $ \beta_t $ that can be used as a simple monitoring statistic, or possibly as a predictor of future events in the market. If the model is applied to an {\it individual} investor, provided corresponding proprietary trading data are available, it can produce an implied 'amount of rationality' of that particular trader.

\section{Investment portfolio}
\label{sect:Portfolio}   

We adopt the notation and assumption of the portfolio model suggested by Boyd {\it et. al.} \cite{Boyd_2017}.
In this model, dollar values of positions in $ n $ assets $ i = 1, \ldots, n $ are denoted as a vector $ {\bf x}_t $ with components 
$ \left( x_t \right)_i  $ for a dollar value of asset $ i $ at the beginning of period $ t $. 
In addition to assets $ {\bf x}_t $,  an investment portfolio includes a risk-free bank cash account $ b_t $ with a risk-free interest rate $ r_f $.
A short position in any asset  $ i $ then corresponds to a negative 
value $ \left( x_t \right)_i  <  0 $. The vector of mean of bid and ask prices of assets at the beginning of period $ t $ is denoted as $ {\bf p}_t $, with 
$ \left( p_t \right)_i > 0 $ being the price of asset $ i $. 
Trades $ {\bf u}_t $ are made at the beginning of interval $ t $, so that asset values $ {\bf x}_t^{+} $ immediately after trades are deterministic:
\beq
\label{h_t+}
{\bf x}_t^{+} = {\bf x}_t + {\bf u}_t
\eeq
The total portfolio value is 
\beq
\label{v_t}
v_t = {\bf 1}^T  {\bf x}_t + b_t
\eeq
where $ {\bf 1} $ is a vector of ones. The post-trade portfolio is therefore
\beq
\label{v_t+}
v_t^{+} =  {\bf 1}^T  {\bf x}_t + b_t^{+} = {\bf 1}^T  \left( {\bf x}_t + {\bf u}_t \right) + b_t^{+} = v_t + {\bf 1}^T   {\bf u}_t + b_t^{+} -  b_t
\eeq
We assume that all re-balancing of stock positions are financed from the bank cash account (additional cash cost related to the trade will be introduced below). This imposes the following 'self-financing' constraint: 
 \beq
 \label{self_fin}
 {\bf 1}^T   {\bf u}_t + b_t^{+} -  b_t = 0
 \eeq
 which simply means that the portfolio value remains unchanged upon an instantaneous re-shuffle of the wealth between the stock and cash:
 \beq
 \label{v_t_2}
v_t^{+} =  v_t 
\eeq
The post-trade portfolio $ v_t^{+} $ and cash are invested at the beginning of period $ t $ until the beginning of the next period. The return of asset $ i $ over 
period $ t $ is defined as
\beq
\label{r_t_1}
\left(r_t \right)_i = \frac{ \left( p_{t+1} \right)_i - \left( p_{t} \right)_i }{\left( p_{t} \right)_i}, \; \; i = 1, \ldots, n
\eeq
Asset positions at the next time period are then given by
\beq
\label{h_t_1}
{\bf x}_{t+1} = {\bf x}_t^{+} + {\bf r}_t \circ {\bf x}_t^{+}
\eeq
where $ \circ $ stands  for an element-wise (Hadamard) product, and $ {\bf r}_t \in \mathbb{R}^{n} $ is the vector of asset returns from period $ t $ to period $ t + 1 $. The next-period portfolio value is then obtained as follows:
\beq
\label{v_t_p_1}
v_{t+1} = {\bf 1}^T {\bf x}_{t+1} = ( 1 + {\bf r}_t )^T {\bf x}_t^{+} = ( 1 + {\bf r}_t )^T (  {\bf x}_t +  {\bf u}_t )
\eeq

Given a vector of returns $ {\bf r}_t $ in period $ t $, the change of the portfolio value in excess of a risk-free growth is
\bea
\label{dV_t}
\Delta v_t   
\hskip-0.5cm && \equiv v_{t+1} - (1+r_f) v_t 
= ({\bf 1}+ {\bf r}_t)^T ({\bf x}_t + {\bf u}_t) + (1+ r_f) b_t^{+}  - (1+ r_f) {\bf 1}^T {\bf x}_t - (1 + r_f) b_t  \nonumber \\
&& = ({\bf r}_r - r_f {\bf 1})^T (
{\bf x}_t + {\bf u}_t )
\eea
where in the second equation we used Eq.(\ref{self_fin}).  

\subsection{Terminal condition}

A terminal condition for the market portfolio is obtained from the requirement that at a planning horizon $ T $, all stock positions should be equal to the actual observed weights of stocks in the market index. This implies that $ 
{\bf x}_T = {\bf x}_{T}^{M} $ where $ {\bf x}_T^M $ are market cap weights in the S\&P 500 index at time $ T $. By Eq.(\ref{h_t+}), this fixes the action $ {\bf u}_T $ at the last time step:
\beq
\label{u_T} 
{\bf u}_T = {\bf x}_{T}^{M} - {\bf x}_{T-1} 
\eeq 
Therefore, action $ {\bf u}_T $ at the last step is deterministic 
and is not subject to optimization that should be applied to $ T $ remaining actions $ {\bf u}_{T-1}, \ldots, {\bf u}_0 $.
   
If the model is applied to an individual investor, the planning horizon $ T $ is an investment horizon for that investor, while the terminal condition
(\ref{u_T}) can be replaced by a similar terminal condition for the investor portfolio.  

\subsection{Asset returns model}

We assume the following linear specification of one-period excess asset returns:
\beq
\label{r_t}
{\bf r}_t - r_f {\bf 1} = {\bf W} {\bf z}_t - {\bf M}^T {\bf u}_t + \varepsilon_t 
\eeq
where $ {\bf z}_t $ is a vector of predictors with factor loading matrix  $ {\bf W} $, $ {\bf M} $ is a matrix of permanent market impacts with a linear impact specification, and $ \varepsilon_t $ is a vector of residuals with 
\beq
\label{residuals}
\mathbb{E} \left[ \varepsilon_t \right] = 0, \; \Var_t \left[ \varepsilon_t \right] = \Sigma_r
\eeq
Equation (\ref{r_t}) specifies stochastic returns $ {\bf r}_t $, or equivalently the next-step stock prices, as driven by external signals $ {\bf z}_t $, control (action) variables $ {\bf u}_t $, and 
uncontrollable noise $ \varepsilon_t  $. 

Though they enter 'symmetrically' in Eq.(\ref{r_t}), two drivers of returns $ {\bf z}_t $ and $ {\bf u}_t $ play entirely different roles. While
 signals $ {\bf z}_t $ are completely {\it external} for the agent, actions $ {\bf u}_t $ are {\it controlled} degrees of freedom.
In our approach, we will be looking for {\it optimal} controls  $ {\bf u}_t $ for the market-wise portfolio. When we 
set up a proper optimization problem, we solve for an optimal 
action $ {\bf u}_t $. As will be shown in this paper, this optimal control turns out to be a linear function of $ {\bf x}_t $, plus noise. 
Substituting it back into 
Eq.(\ref{r_t}), this produces effective {\it dynamically} generated dynamics that involve only stock prices, see  Eq.(\ref{GMR}) below in Sect.~\ref{sect:GMR_dynamics}\footnote{The reader interested only in the final asset return model resulting from our framework but not in its derivation can jump directly to Eq.(\ref{GMR}).}.
 
\subsection{Signal dynamics and state space}

For dynamics of signals ${\bf z}_t $,  similar to \cite{GP}, we will assume a simple 
multi-variate mean-reverting Ornstein-Uhlenbeck (OU) process for a $ K $-component vector $ {\bf z}_t $:
\beq
\label{OU_z}
{\bf z}_{t+1} = \left( {\bf I} - \Phi \right) \circ {\bf z}_t + \varepsilon_t^{z}
\eeq
where $ \varepsilon_t^{z} \sim \mathcal{N} \left(0, \Sigma_z \right) $ is the noise term, and $ \Phi $ is a diagonal matrix of mean reversion rates.

It is convenient to form an extended state vector $ {\bf y}_t $ of size $ N + K $ by concatenating vectors $ {\bf x}_t $ and $ {\bf z}_t $: 
\beq
\label{y_t}
{\bf y}_t = 
 \left[ \begin{array}{c}
  {\bf x}_t \\
  {\bf z}_t
 \end{array} \right]
 \eeq 
The extended vector $ {\bf y}_t $ describes a full state of the system for the agent that has some control of its $ x$-component, but no control of its 
$ z $-component. 

\subsection{One-period rewards}

We first consider an idealized case when there are no costs of taking action $ {\bf u}_t $ at time step $ t $. An instantaneous random reward received upon taking such action is obtained by substituting Eq.(\ref{r_t}) in Eq.(\ref{dV_t}):
\beq
\label{R_insta}
R_t^{(0)}({\bf y}_t, {\bf u}_t) = \left(  {\bf W} {\bf z}_t - {\bf M}^T {\bf u}_t + {\bf \varepsilon}_t \right)^T \left( {\bf x}_t + {\bf u}_t \right)
\eeq
In addition to this reward that would be obtained in an ideal friction-free world, we have to add (negative) rewards received due to instantaneous market impact and transaction fees\footnote{We assume no short sale positions in our setting, and therefore do not include borrowing costs.}. Furthermore, 
we have to include a negative reward due to risk in a newly created portfolio position at time $ t + 1 $. Similar to \cite{Boyd_2017}, we choose a simple quadratic measure of such risk penalty, as the variance of the instantaneous reward (\ref{R_insta}) conditional on the new state $ {\bf x}_t + {\bf u}_t $, 
multiplied by the risk aversion parameter $ \lambda $:
\beq
\label{R_risk}
 R_t^{(risk)}({\bf y}_t, {\bf u}_t) = - \lambda \Var_t \left[ \left. R_t^{(0)}({\bf y}_t, {\bf u}_t) \right| {\bf x}_t + {\bf u}_t \right] = - \lambda
 ( {\bf x}_t + {\bf u}_t )^T \Sigma_r  ( {\bf x}_t + {\bf u}_t )
 \eeq
 To specify negative rewards (costs) of an instantaneous market impact and transaction costs, it is convenient to represent 
each action $ u_{ti} $ 
as a difference of two non-negative action variables $  u_{ti}^{+}, u_{ti}^{-} \geq 0 $:
\beq
\label{z_pm}
u_{ti} = u_{ti}^{+}  - u_{ti}^{-} \, , \; \;  \left| u_{ti} \right| = u_{ti}^{+} + u_{ti}^{-} \, , \; \; u_{ti}^{+}, u_{ti}^{-} \geq 0 
\eeq
so that $ u_{ti} = u_{ti}^{+} $ if  $ u_{ti} > 0 $ and $ u_{ti} = - u_{ti}^{-} $ if  $ u_{ti} < 0 $. 
The instantaneous market impact and transaction costs are then given by the following expressions:
\bea
\label{R_impact_trade}
R_t^{(impact)}({\bf y}_t, {\bf u}_t)  
\hskip-0.5cm && = - {\bf x}_t^T  \Gamma^{+} {\bf u}_t^{+}  - {\bf x}_t^T\Gamma^{-} {\bf u}_t^{-}  - {\bf x}_t^T \Upsilon {\bf z}_t   \nonumber \\
R_t^{(fee)}({\bf y}_t, {\bf u}_t) 
\hskip-0.5cm && = - \nu^{+T} {\bf u}_t^{+} - \nu^{-T} {\bf u}_t^{-} 
\eea
Here $  \Gamma^{+}, \,  \Gamma^{-}, \,  \Upsilon $ and $  \nu^{+}, \,  \nu^{-} $ are, respectively, matrices-valued and vector-valued parameters that in a simplest case can be parametrized in terms of single scalars multiplied by unit vectors or matrices.
 
 Combining Eqs.(\ref{R_insta}, (\ref{R_risk}), (\ref{R_impact_trade}), we obtain our final specification of a risk- and cost-adjusted instantaneous reward function for the problem of optimal portfolio liquidation:
 \beq
 \label{R_adj}
  R_t({\bf y}_t, {\bf u}_t) = R_t^{(0)}({\bf y}_t, {\bf u}_t) +  R_t^{(risk)}({\bf y}_t, {\bf u}_t)  +  R_t^{(impact)}({\bf y}_t, {\bf u}_t) 
  +  R_t^{(fee)}({\bf y}_t, {\bf u}_t) 
 \eeq
 The {\it expected} one-step reward given action $ {\bf u}_t = {\bf u}_t^{+}  - {\bf u}_t^{-} $ is given by 
 \beq
 \label{R_adj_exp}
  \hat{R}_t({\bf y}_t, {\bf u}_t)  
   = \hat{R}_t^{(0)}({\bf y}_t, {\bf u}_t) +  R_t^{(risk)}({\bf y}_t, {\bf u}_t)  +  R_t^{(impact)}({\bf y}_t, {\bf u}_t) 
  +  R_t^{(fee)}({\bf y}_t, {\bf u}_t) 
 \eeq
where  
\beq
\label{exp_R_0} 
 \hat{R}_t^{(0)}({\bf y}_t, {\bf u}_t)  
 = \mathbb{E}_{t,u} \left[  R_t^{(0)}({\bf y}_t, {\bf u}_t) \right] =  \left(  {\bf W} {\bf z}_t - {\bf M}^T ( {\bf u}_t^{+} - 
 {\bf u}_t^{-} ) \right)^T \left( {\bf x}_t + {\bf u}_t^{+} - {\bf u}_t^{-} \right)
\eeq
where $ \mathbb{E}_{t,u} \left[ \cdot \right] = \mathbb{E} \left[ \cdot | {\bf y}_t, {\bf u}_t \right]  $ stands for 
averaging over next-periods realizations of market returns.  

Note that the one-step expected reward (\ref{R_adj_exp}) is a quadratic form of its inputs. 
We can write it more explicitly using vector notation:
\beq
\label{R_t_y}
\hat{R} ({\bf y}_t, {\bf a}_t) = 
{\bf y}_t^T {\bf R}_{yy} {\bf y}_t  + {\bf a}_t^T {\bf R}_{aa} {\bf a} 
 +  {\bf a}_t^T {\bf R}_{ay}{\bf y}_t  
 +   {\bf a}_t^T  {\bf R}_{a} 
 \eeq
 where
 \bea
 \label{R_y_param}
  && {\bf R}_{aa}=  
 \left[ \begin{array}{cc}
-  {\bf M}  -  \lambda  \Sigma_r &  {\bf M} +  \lambda  \Sigma_r \\
 {\bf M}  +  \lambda  \Sigma_r & -  {\bf M}  -  \lambda  \Sigma_r 
 \end{array} \right], \;  
 {\bf R}_{yy} = 
 \left[ \begin{array}{ll}
  - \lambda \Sigma_r  & {\bf W} - \Upsilon   \\
  0 & 0  
 \end{array} \right], \nonumber \\ 
 && {\bf R}_{ay} = 
 \left[
 \left[ \begin{array}{l}
-  {\bf M}  -  2 \lambda  \Sigma_r  -   \Gamma^{+}  \\
  {\bf M}  +  2 \lambda  \Sigma_r  -   \Gamma^{-} 
 \end{array} \right],
 \left[ \begin{array}{l}
 {\bf W} \\
 {\bf W} \end{array} \right]
 \right], \; \; 
 {\bf R}_{a} = 
 -  \left[ \begin{array}{l}
 \nu^{+}   \\
 \nu^{+}   
 \end{array} \right]
\eea

\subsection{Multi-period portfolio optimization}

Multi-period portfolio optimization is equivalently formulated either as maximization of risk- and cost-adjusted returns, as in 
the Markowitz portfolio model, or as minimization of risk- and cost-adjusted trading costs. The latter specification is usually 
used in problems of optimal portfolio liquidation.
 
A multi-period risk- and cost-adjusted reward maximization problem reads 
\bea
\label{opt_1_2}
&& \mbox{maximize $ \mathbb{E}_t \left[ \sum_{t'=t}^{T-1} \gamma^{t'-t}  \hat{R}_{t'}({\bf y}_{t'}, {\bf a}_{t'})   \right]  $}   \\
&& \mbox{ where $   \hat{R}_t({\bf y}_t, {\bf a}_t)  = {\bf y}_t^T {\bf R}_{yy} {\bf y}_t  +  {\bf a}_t^T {\bf R}_{aa} {\bf a} 
+  {\bf a}_t^T {\bf R}_{ay}{\bf y}_t  
 +   {\bf a}_t^T  {\bf R}_{a}    $}   \nonumber \\
&& \mbox{ w.r.t. $ 
{\bf a}_t = 
 \left( \begin{array}{c}
  {\bf u}_{t}^{+} \\
 {\bf u}_{t}^{-}   
\end{array} \right)
 \geq 0 $}, 
 \nonumber \\ 
&& \mbox{ subject to $  {\bf x}_t  + {\bf u}_t^{+} - {\bf u}_t^{-} \geq 0 $} \nonumber 
\eea
Here $ 0 < \gamma \leq 1 $ is a discount factor.
Note that the sum over future periods $ t'=[t, \ldots, T-1] $ does not include the last period $ t' = T $, because the last action is fixed by Eq.(\ref{u_T}).

An equivalent cost-focused formulation is obtained by flipping the sign of the above problem, and re-phrasing it as minimization of trading costs 
$  \hat{C}_t({\bf y}_t, {\bf a}_t) = -  \hat{R}_t({\bf y}_t, {\bf a}_t) $:
\bea
\label{opt_1_2_2}
&& \mbox{minimize $ \mathbb{E}_t \left[ \sum_{t'=t}^{T-1} \gamma^{t'-t}  \hat{C}_{t'}({\bf y}_{t'}, {\bf a}_{t'})   \right]  $}   \\
&& \mbox{ where $  
\hat{C}_t({\bf y}_t, {\bf a}_t) = -  \hat{R}_t({\bf y}_t, {\bf a}_t) $}
\eea
subject to the same constraints as in (\ref{opt_1_2}).

\subsection{Dynamic Inverse Portfolio Optimization} 

When the model dynamics are known (or independently estimated from data), the dynamic portfolio optimization problem 
of Eq.(\ref{opt_1_2}) can be formulated as a problem of Stochastic Optimal Control (SOC), also known as a Dynamic Programming approach.
This approach was pursued in Ref.~\cite{Boyd_2017} in a general setting of convex portfolio optimization, see also references there on previous work on this topic. In particular, 
one well-known example is a dynamic mean-variance model of Garleanu and Pedersen \cite{GP} with quadratic transaction costs.

We keep a convex multi-period portfolio formulation while adding to it modeling of market impact and external signals, and focusing 
on  a {\it inverse} optimization problem, rather than a forward optimization problem as in \cite{Boyd_2017}. 
We can refer to this problem as a Dynamic {\it Inverse} Portfolio Optimization (DIPO) problem. The word 'dynamic' here means that a learned optimal policy should be {\it adaptive} to predictors $ {\bf z}_t $.

In DIPO learning, we assume that an optimal portfolio strategy has been already {\it found}, perhaps not quite optimally, in the past by an {\it expert} trader. We assume that we have a record of $ N $ different runs of such nearly optimal strategy, each of length $ T $, performed by this expert trader.  
Following the common conventions of the RL/IRL literature, we can call this data samples expert demonstrations, or expert trajectories.
The problem is then to find the optimal execution policy from these data.
   
We may differentiate between two possible settings for such data-driven DIPO learning that  can be encountered in practice. First, in a setting of Reinforcement Learning we have access to historical data consisting of stock market prices, actions taken (i.e. portfolio trades), {\it and} risk-adjusted {\it rewards} received upon taking these actions (see below for details). In addition, the data consists of all predictive factors ("alpha-factors") 
that might be predictive of rewards.
The objective is to learn and improve a policy that was used in the data, so that the new improved policy can be used to generate higher rewards in the future.

The other setting is of Inverse Reinforcement Learning (IRL), where everything is the same as above, except we do {\it not} observe rewards anymore. The objective is to learn the reward function that leads to the observed behavior, and learn the policy as well.  

This is the setting of this paper, where we use an IRL framework to represent {\it all} traders in the market as {\it one} market-wise 'expert trader' who  
is mathematically modeled as a  bounded-rational RL agent. A reward function of this agent is learned from market data, plus whatever signals $ {\bf z}_t $ that are used by the model. The learned parameters include market-implied risk aversion $ \lambda $, market impact parameters $ \mu_i $, weights $ {\bf W} $
of predictors $ {\bf z}_t $, and market-implied 'rationality index' $ \beta $.

Note that if {\it proprietary} trading data from a particular trader or broker are available, the same framework can be applied to learn a reward function of that particular trader. Such setting might be interesting given that the value of a 'true' risk aversion parameter is often unknown to investors themselves,
as they may not base their decisions on a quadratic utility model. When applied to an individual investor, the model developed here may offer a probabilistic model of that {\it particular} investor, with parameters estimated on trading data of this investor, combined with the market data.

Regarding the policy optimization problem, as rewards are not observed in the IRL setting, this problem is in general both harder and less well-posed in comparison to the RL setting. In particular, unlike RL off-policy methods such as Q-learning that can learn, given enough data, even from data with purely random 
actions, IRL methods cannot proceed with data with entirely random actions. For IRL to work, data collected should correspond to some good, though not necessarily {\it optimal} policy. Probabilistic IRL methods are capable of learning when demonstrated data does not always correspond to optimal actions.     

While our main focus in this paper is on the IRL setting, we will start below with RL approaches to the problem.

\section{Reinforcement Learning of optimal trading}
\label{sect:RL}

In this section, we will discuss a data-driven Reinforcement Learning approach to multi-period portfolio optimization of Eq.(\ref{opt_1_2}).
We first introduce stochastic policies and a Bellman equation with stochastic policies, and then consider an entropy-regularized methods for MDP corresponding to Eq.(\ref{opt_1_2}).

\subsection{Stochastic policy}

Note that the multi-period  portfolio optimization problem
(\ref{opt_1_2}) assumes that an optimal policy that determines actions $ {\bf a}_t $ is a deterministic policy that can also be described as a delta-like probability distribution
\beq
\label{determin_policy}
\pi({\bf a}_t | {\bf y}_t) = \delta \left( {\bf a}_t - {\bf a}_t^{\star} ({\bf y}_t ) \right)
\eeq
where the optimal deterministic action $  {\bf a}_t^{\star} ({\bf y}_t )$ is obtained by 
maximization of the objective (\ref{opt_1_2}) with respect to controls $ {\bf a}_t $. 

But the actual trading data may be sub-optimal, or noisy at times, because of model mis-specifications, market timing lags, human errors etc.
Potential presence of such sub-optimal actions in data poses serious challenges, if we try to assume deterministic policy (\ref{determin_policy}) that assumes the the action chosen is {\it always} an optimal action. This is because such events should have zero probability under these model assumptions, and thus would produced vanishing path probabilities if observed in data.

Instead of assuming a deterministic policy (\ref{determin_policy}), {\it stochastic} policies described by {\it smoothed} distributions $ \pi({\bf a}_t | {\bf y}_t ) $, are more useful for inverse problems such as the problem of inverse portfolio optimization. In this approach, instead of maximization with respect to 
deterministic policy/action $  {\bf a}_t $, we re-formulate the problem as maximization over {\it probability distributions} 
$ \pi({\bf a}_t | {\bf y}_t ) $:
\bea
\label{opt_2}
&& \mbox{maximize $ \mathbb{E}_{q_{\pi}} \left[ \sum_{t'=t}^{T-1} \gamma^{t'-t}  \hat{R}_t({\bf y}_{t'}, {\bf a}_{t'})   \right]  $}   \\
&& \mbox{ where $  \hat{R} ({\bf y}_t, {\bf a}_t) = 
 {\bf y}_t^T {\bf R}_{yy} {\bf y}_t  + {\bf a}_t^T {\bf R}_{aa} {\bf a} 
+ {\bf a}_t^T {\bf R}_{ay}{\bf y}_t  
 +   {\bf a}_t^T  {\bf R}_{a} 
$}   \nonumber \\
&& \mbox{ w.r.t. $ 
q_{\pi}( \bar{x}, \bar{a} | {\bf y}_0) = \pi ({\bf a}_0 | {\bf y}_0 ) \prod_{t=1}^{T-1} \pi ({\bf a}_t | {\bf y}_t ) 
P \left( {\bf y}_{t+1} | {\bf y}_t, {\bf a}_t \right)   $} \nonumber \\ 
&& \mbox{ subject to $ \int d {\bf a}_t \, \pi \left( {\bf a}_t | {\bf y}_t \right) 
= 1 $}  \nonumber     
\eea 
Here $\mathbb{E}_{q_{\pi}} \left[ \cdot \right] $ stands for expectations with respect to path probabilities defined according to the third line in 
Eqs.(\ref{opt_2}).

Note that due to inclusion of a quadratic risk penalty in the risk-adjusted return $  \hat{R} ({\bf x}_t, {\bf a}_t ) $ the original problem of risk-adjusted return optimization is re-stated in Eq.(\ref{opt_2}) as maximizing the expected cumulative reward in the standard MDP setting, thus making the problem amenable to a standard risk-neutral approach of MDP models. Such simple risk adjustment based on one-step variance penalties was suggested in a non-financial context by Gosavi \cite{Gosavi}, and used in a Reinforcement Learning based approach to option pricing in \cite{IH_2017, NuQLear}.

Another comment that is due here is that a probabilistic approach to actions in portfolio trading appears, on many counts, a more natural way than a formalism based on deterministic policies. Indeed, even in a simplest one-period setting, because the Markowitz-optimal  solution for portfolio weights is a function of {\it estimated} stock means and covariances, they are in fact {\it random} variables. Yet the probabilistic nature of portfolio optimization is not recognized as such in Markowitz-type single-period or multi-period optimization  settings such as (\ref{opt_1_2}).
A probabilistic portfolio optimization formulation was suggested in a one-period setting by Marshinski {\it et. al.} \cite{Rossi}.

\subsection{Reference policy}
\label{sect:reference_policy}

We assume that we are given a probabilistic {\it reference} (or {\it prior}) policy $ \pi_0( {\bf a}_t | {\bf y}_t ) $ which should be decided upon prior to attempting 
the portfolio optimization (\ref{opt_2}). Such policy can be chosen based on a parametric model, past historic data, etc. 
We will use a simple Gaussian reference policy 
\beq
\label{pi_0}
\pi_0 ( {\bf a}_t | {\bf y}_t ) = \frac{1}{ \sqrt{ (2 \pi)^{N}  \left| \Sigma_p \right| }} \exp\left( - \frac{1}{2} \left( {\bf a}_t - \hat{ {\bf a}} ( {\bf y}_t) \right)^{T} \Sigma_p^{-1}  
\left( {\bf a}_t - \hat{{\bf a}}( {\bf y}_t) \right) \right)
\eeq
where $ \hat{ {\bf a}} ( {\bf y}_t)  $ can be a deterministic policy chosen to be a linear function of a state vector $ {\bf y}_t $:
\beq
\label{pi_0_mean}
\hat{ {\bf a}} ( {\bf y}_t)  
=  
\hat{\bf A}_0 +   \hat{{\bf A}}_{1}  {\bf y}_t  
\eeq 
A simple choice of parameters in (\ref{pi_0}) could be to specify them in terms of only two scalars $ \hat{a}_0, \, \; \hat{a}_1 $ as follows:
$  \hat{ {\bf A}}_0 = \hat{a}_0 {\bf 1}_{|A|} $ and $  \hat{ {\bf A}}_1 = \hat{a}_1 {\bf 1}_{|A| \times |A|} $ where $ |A| $ is the the size of vector $ {\bf a}_t $,  $  {\bf 1}_A $ and $  {\bf 1}_{A\times A} $ are, respectively, a vector and matrix made of ones.  The scalars  
$ \hat{a}_0 $  and 
$ \hat{a}_1 $ would then serve as hyper-parameters in our setting. Similarly, covariance matrix $ \Sigma_p $ for the prior policy 
can be taken to be a simple matrix with constant correlations $ \rho_p $ and constant variances $ \sigma_p $. 

As will be shown below, an {\it optimal} policy has the same Gaussian form as the prior policy (\ref{pi_0}), with updated parameters 
 $  \hat{{\bf A}}_0 $, $ \hat{ {\bf A}}_1 $ and 
$ \Sigma_p $. These updates will be computed iteratively starting with their initial values defining the prior (\ref{pi_0}). 
Respectively, updates at iteration $ k $ will be denoted by upper subscripts, e.g. $ \hat{ {\bf A}}_0^{(k)} $, $  \hat{{\bf A}}_1^{(k)} $.

Furthermore, it turns out that a linear dependence on $ {\bf y}_t $ at iteration $ k $, driven 
by the value of $  \hat{{\bf A}}_1^{(k)} $ arises even if we set $  \hat{ {\bf A}}_1
=  \hat{ {\bf A}}_1^{(0)} = 0 $ in the prior (\ref{pi_0}). Such choice of a state-independent prior $  \pi_0 ( {\bf a}_t | {\bf y}_t ) = \pi_0 ( {\bf a}_t ) $, although not very critical, reduces the number of free parameters in the model by two, as well as simplifies some of the analyses below, and hence will be assumed going forward. It also makes it unnecessary to specify the value of $ \bar{\bf {y}}_t $ in the prior (\ref{pi_0}) (equivalently, we can initialize it at zero).
The final set of hyper-parameters defining the prior (\ref{pi_0}) therefore includes only three values  of $  \hat{a}_0, \,   \rho_a, \,  \Sigma_p $.

\subsection{Bellman Optimality Equation}

Let 
\beq
\label{V_star}
V_t^{\star} ( {\bf y}_t) = \max_{ \pi(\cdot| y)} \mathbb{E} \left[ \left. \sum_{t'=t}^{T-1} \gamma^{t'-t} \hat{R}_{t'} ({\bf y}_{t'}, {\bf a}_{t'} ) \right| {\bf y}_t \right]
\eeq
The optimal state value function $ V_t^{\star} ( {\bf x}_t) $ satisfies the Bellman optimality equation (see e.g. \cite{SB})
\beq
\label{Bellman_V}
V_t^{\star} ( {\bf y}_t) = \max_{ {\bf a}_t } \, \hat{R}_t ({\bf y}_t, {\bf a}_t ) + \gamma \mathbb{E}_{ 
t, {\bf a}_t} \left[  V_{t+1}^{\star} ( {\bf y}_{t+1}) \right] 
\eeq 
The optimal policy $ \pi^{\star} $ can be obtained from $ V^{\star} $ as follows:
\beq
\label{pi_star_V_star}
 \pi_t^{\star} ({\bf a}_t | {\bf y}_t ) = \arg \max_{ {\bf a}_t} \,  \hat{R}_t ({\bf y}_t, {\bf a}_t) +  \gamma \mathbb{E}_{ 
t, {\bf a}_t } \left[  V_{t+1}^{\star} ( {\bf y}_{t+1}) \right] 
\eeq
The goal of Reinforcement Learning (RL) is to solve the Bellman optimality equation based on samples of data. Assuming that an optimal value function is found by means of RL, solving for the optimal policy $ \pi^{\star} $ takes another optimization problem as formulated in Eq.(\ref{pi_star_V_star}).  

\subsection{Entropy-regularized Bellman optimality equation}

Following \cite{Dai}, we start with reformulating the Bellman optimality equation using a Fenchel-type representation:
\beq
\label{V_star_Fenchel}
 V_t^{\star} ( {\bf y}_t)  = \max_{ \pi(\cdot| y) \in \mathcal{P}} \sum_{{\bf a}_t \in \mathcal{A}_t }  \pi ({\bf a}_t | {\bf y}_t )
\left(   \hat{R}_t ({\bf y}_t, {\bf a}_t) +  \gamma \mathbb{E}_{ 
t, {\bf a}_t } \left[  V_{t+1}^{\star} ( {\bf y}_{t+1}) \right] \right)
\eeq
Here $ \mathcal{P} = \left\{ \pi: \, \pi \geq 0, {\bf 1}^T \pi = 1 \right\} $ stands for a set of all valid distributions. Eq.(\ref{V_star_Fenchel}) is equivalent to the original Bellman optimality equation (\ref{V_star}), because for any $  x \in \mathbb{R}^n $, we have $ \max_{i \in \{ 1, \ldots, n \} } x_i = 
\max_{\pi \geq 0, || \pi || \leq 1 }  \pi^T x $.  Note that while we use discrete notations for simplicity of presentation, all formulas below can be equivalently expressed in continuous notations by replacing sums by integrals. For brevity, we will denote the expectation $ \mathbb{E}_{ 
{\bf y}_{t+1}| {\bf y}_t, {\bf a}_t } \left[ \cdot \right] $ as $ \mathbb{E}_{t, {\bf a}} \left[ \cdot \right] $ in what follows.

The one-step {\it information cost} of a learned policy $ \pi ( {\bf a}_t | {\bf y}_t) $ relative to a reference policy $ \pi_0( {\bf a}_t | {\bf y}_t) $  is defined as follows \cite{G-Learning}:
\beq
\label{info_cost}
g^{\pi} ({\bf y}, {\bf a} ) = \log \frac{  \pi ( {\bf a}_t | {\bf y}_t) }{ \pi_0 ( {\bf a}_t | {\bf y}_t) }
\eeq
Its expectation with respect to policy $ \pi $ is the Kullback-Leibler (KL) divergence of $ \pi(\cdot|  {\bf y}_t) $ and $ \pi_0( \cdot |  {\bf y}_t) $:
\beq
\label{Eg_KL}
\mathbb{E}_{\pi} \left[ \left. g^{\pi} ({\bf y}, {\bf a} ) \right| {\bf y}_t \right] = KL[ \pi || \pi_0] ({\bf y}_t) \equiv 
\sum_{{\bf a}_t } 
\pi ( {\bf a}_t | {\bf y}_t)  
\log \frac{ \pi ( {\bf a}_t | {\bf y}_t) }{\pi_0 ( {\bf a}_t | {\bf y}_t) } 
\eeq
The total discounted information cost for a trajectory is defined as follows:
\beq
\label{I_pi}
I^{\pi}({\bf y} ) = \sum_{t'=t}^{T} \gamma^{t'-t} \mathbb{E} \left[ \left. g^{\pi} ({\bf y}_{t'}, {\bf a}_{t'} ) \right| {\bf y} _t = {\bf y} \right]
\eeq
The {\it free energy} function $ F_t^{\pi} ({\bf y}_t) $ is defined as the value function (\ref{V_star_Fenchel}) augmented by the information cost penalty
(\ref{I_pi}): 
\bea
\label{F_pi}
F_t^{\pi}({\bf y}_t) 
\hskip-0.5cm && =   V_t^{\pi} ( {\bf y}_t) - \frac{1}{\beta}  I^{\pi} ( {\bf y}_t)  \nonumber \\
&& = \sum_{t'=t}^{T} \gamma^{t'-t} \mathbb{E} \left[  \hat{R}_{t'} ({\bf y}_{t'}, {\bf a}_{t'} ) - \frac{1}{\beta}  g^{\pi} ({\bf y}_{t'}, {\bf a}_{t'} ) 
\right]  
\eea
Note that $ \beta $  in Eq.(\ref{F_pi}) serves as the "inverse temperature" parameter
that controls a trade-off between reward optimization and proximity to the reference policy, see below. The free energy $ F_t^{\pi}({\bf y}_t)  $ 
is the entropy-regularized value function, where the amount of regularization can be tuned to better cope with noise in 
data\footnote{Note that in physics, as well as in the free-energy principle literature \cite{Friston_2010, Friston_2018}, free energy is defined with a negative sign relative to Eq.(\ref{F_pi}). This difference is purely a matter of a sign convention, as maximization of Eq.(\ref{F_pi}) can be re-stated as minimization of its negative. With our sign convention for the free energy function, we follow Reinforcement Learning and Information Theory literature \cite{Ortega_2012, Tishby_2012, Ortega_2015, Larsson_2017}.}. 
The reference
policy $ \pi_0 $ provides a "guiding hand" in the stochastic policy optimization process that we describe next.

A Bellman equation for the free energy function $ F_t^{\pi} ({\bf y}_t) $ is obtained from (\ref{F_pi}):
\beq
\label{Bellman_F}
F_t^{\pi}({\bf y}_t)  =    \mathbb{E}_{ {\bf a}|y}  \left[ \hat{R}_{t} ({\bf y}_{t}, {\bf a}_{t} )
 - \frac{1}{\beta}  g^{\pi} ({\bf y}_{t}, {\bf a}_{t})  + 
\gamma \mathbb{E}_{t, {\bf a}}  \left[ F_{t+1}^{\pi}({\bf y}_{t+1})  \right]  \right]
\eeq
For a finite-horizon setting, Eq.(\ref{Bellman_F}) should be supplemented by a terminal condition 
\beq
\label{F_pi_T}
F_T^{\pi}({\bf y}_T) =  \left. \hat{R}_{T} ({\bf y}_{T}, {\bf a}_{T} ) \right|_{ {\bf a}_T = - {\bf u}_{T-1}}
\eeq
(see Eq.(\ref{u_T})).
Eq.(\ref{Bellman_F}) can be viewed as a soft probabilistic relaxation of the Bellman optimality equation for the value function, with the KL information cost penalty 
(\ref{Eg_KL}) as a regularization controlled by the inverse temperature $ \beta $. In addition to such regularized value function (free energy), we will next introduce an entropy regularized Q-function.

\subsection{G-function: an entropy-regularized Q-function}

Similarly to the action-value function, we define the state-action free energy function $ G^{\pi} ( {\bf x}, {\bf a}) $ as \cite{G-Learning}
\bea
\label{G_fun}
G_t^{\pi} ( {\bf y}_t, {\bf a}_t) 
\hskip-0.5cm &&= \hat{R}_t ({\bf y}_{t}, {\bf a}_{t} ) 
 +  \gamma  \mathbb{E} \left[  \left. F_{t+1}^{\pi} ( {\bf y}_{t+1}) \right|  {\bf y}_t, {\bf a}_t \right]   \\  
&&=  \hat{R}_t ({\bf y}_{t}, {\bf a}_{t} ) 
   + \gamma  \mathbb{E}_{t, {\bf a}} \left[  \sum_{t'=t+1}^{T} \gamma^{t'-t-1} \left(   \hat{R}_{t'} ({\bf y}_{t'}, {\bf a}_{t'} ) - 
   \frac{1}{\beta} g^{\pi} ({\bf y}_{t'}, {\bf a}_{t'} ) \right)  \right] \nonumber \\
&&=  \mathbb{E}_{t, {\bf a}} \left[   \sum_{t'=t}^{T}  \gamma^{t'-t} \left(   \hat{R}_{t'} ({\bf y}_{t'}, {\bf a}_{t'} ) - 
   \frac{1}{\beta} g^{\pi} ({\bf y}_{t'}, {\bf a}_{t'} ) \right)  \right] \nonumber 
\eea
where in the last equation we used the fact that the first action $ {\bf a}_t $ 
in the G-function is fixed, and hence $ g^{\pi} ({\bf y}_{t}, {\bf a}_{t} ) = 0 $ when we condition on $ {\bf a}_t = {\bf a} $.

If we now compare this expression with Eq.(\ref{F_pi}), we obtain the relation between the G-function and the free energy
$ F_t^{\pi}({\bf y}_t ) $:
\beq
\label{G_F}
F_t^{\pi}({\bf y}_t ) = \sum_{ {\bf a}_t} \pi ( {\bf a}_t | {\bf y}_t) \left[ G_t^{\pi} ( {\bf y}_t, {\bf a}_t) -  \frac{1}{\beta}
\log \frac{  \pi ( {\bf a}_t | {\bf y}_t) }{ \pi_0 ( {\bf a}_t | {\bf y}_t) } \right]
\eeq
This functional is maximized by the following distribution $ \pi ( {\bf a}_t | {\bf y}_t) $:
\bea
\label{pi_from_F}
 && \pi ( {\bf a}_t | {\bf y}_t) = \frac{1}{Z_t} \pi_0 ( {\bf a}_t | {\bf y}_t) e^{ \beta G_t^{\pi} ( {\bf y}_t, {\bf a}_t) }  \\
 && Z_t = \sum_{{\bf a}_t} \pi_0 ( {\bf a}_t | {\bf y}_t) e^{ \beta G_t^{\pi} ( {\bf y}_t, {\bf a}_t) } \nonumber 
 \eea 
The free energy (\ref{G_F}) evaluated at the optimal solution (\ref{pi_from_F}) becomes
\beq
\label{F_opt}
F_t^{\pi}({\bf y}_t ) =  \frac{1}{\beta} \log Z_t =  \frac{1}{\beta} \log \sum_{{\bf a}_t} \pi_0 
( {\bf a}_t | {\bf y}_t) e^{ \beta G_t^{\pi} ( {\bf y}_t, {\bf a}_t)  }  
\eeq
Using Eq.(\ref{F_opt}), the optimal action policy (\ref{pi_from_F}) can be written as follows :
\beq
\label{pi_opt_F}
 \pi ( {\bf a}_t | {\bf y}_t) = \pi_0 ( {\bf a}_t | {\bf y}_t) e^{ \beta \left(G_t^{\pi} ( {\bf y}_t, {\bf a}_t) - F_t^{\pi}({\bf y}_t ) 
  \right) } 
 \eeq 
 Eqs.(\ref{F_opt}), (\ref{pi_opt_F}), along with the first form of Eq.(\ref{G_fun}) repeated here for convenience:
 \beq
 \label{G_from_F}
 G_t^{\pi} ( {\bf y}_t, {\bf a}_t) 
 = \hat{R}_{t}({\bf y}_{t}, {\bf a}_{t} ) 
 +  \gamma  \mathbb{E}_{t, {\bf a}} \left[  \left. F_{t+1}^{\pi} ( {\bf y}_{t+1}) \right|  {\bf y}_t, {\bf a}_t \right] 
 \eeq
 constitute a system of equations that should be solved self-consistently by backward recursion for $ t = T-1, \ldots, 0 $,
 with terminal conditions 
\bea
\label{terminal_G_F}
&& G_T^{\pi} ( {\bf y}_T, {\bf a}_T) 
 = \hat{R}_{T}({\bf y}_{T}, {\bf a}_{T} ) \\
 &&  F_T^{\pi}({\bf y}_T) = G_T^{\pi} ( {\bf y}_T, {\bf a}_T) 
 =  \hat{R}_{T}({\bf y}_{T}, {\bf a}_{T} ) \nonumber 
 \eea 
 The self-consistent scheme of Eqs.(\ref{F_opt}, \ref{pi_opt_F}, \ref{G_from_F}) \cite{G-Learning} can be used in both the RL setting, when rewards are {\it observed}, and in the IRL setting when they are {\it not}.  
 Before proceeding with these methods, we want to digress on an alternative interpretation of entropy regularization in Eq.(\ref{F_pi}), that can be useful for 
 clarifying the approach of this paper.

\subsection{Adversarial interpretation of entropy regularization}
\label{sect:Adversarial}

A useful alternative interpretation of the entropy regularization term in Eq.(\ref{F_pi}) can be suggested using its representation as a Legendre-Fenchel transform  of another function \cite{Ortega_2014}:
\beq
\label{Legendre_KL}
- \frac{1}{\beta} \sum_{{\bf a}_t}  \pi ( {\bf a}_t | {\bf y}_t) \log \frac{  \pi ( {\bf a}_t | {\bf y}_t) }{ \pi_0 ( {\bf a}_t | {\bf y}_t) }
= \min_{C({\bf a}_t , {\bf y}_t)} \sum_{{\bf a}_t}  \left( - \pi ( {\bf a}_t | {\bf y}_t) \left( 1 + C({\bf a}_t,  {\bf y}_t ) \right)  +  \pi_0 ( {\bf a}_t | {\bf y}_t) e^{ \beta  C({\bf a}_t , {\bf y}_t) } 
\right) 
\eeq
where $ C({\bf a}_t , {\bf y}_t) $ is an arbitrary function. Eq.(\ref{Legendre_KL}) can be verified by direct minimization of the right-hand side with respect to $ C({\bf a}_t , {\bf y}_t) $. 

Using this representation of the KL term, the free energy maximization problem (\ref{G_F}) can be re-stated as a max-min problem 
\beq
\label{G_F_max_min}
F_t^{\star}({\bf y}_t ) = \max_{ \pi} \min_{C}  \sum_{ {\bf a}_t} \pi ( {\bf a}_t | {\bf y}_t) \left[ G_t^{\pi} ( {\bf y}_t, {\bf a}_t) -
C({\bf a}_t,  {\bf y}_t ) -1 \right]
+ \pi_0 ( {\bf a}_t | {\bf y}_t) e^{ \beta  C({\bf a}_t , {\bf y}_t) } 
\eeq
The imaginary adversary's optimal cost obtained from (\ref{G_F_max_min}) is 
\beq
\label{opt_C}
C^{\star} ({\bf a}_t,  {\bf y}_t ) = \frac{1}{\beta} \log \frac{ \pi ( {\bf a}_t | {\bf y}_t)}{ \pi_0 ( {\bf a}_t | {\bf y}_t)}
\eeq 
Similarly to \cite{Ortega_2014}, one can check that this produces an {\it indifference} solution for the imaginary game between the agent and its adversarial environment where the total sum of the optimal G-function and the optimal adversarial cost
(\ref{opt_C}) is constant: $  G_t^{\star} ( {\bf y}_t, {\bf a}_t) + C^{\star} ({\bf a}_t,  {\bf y}_t )  = \mbox{const} $, which means that the game of the original agent and its adversary is in a Nash equilibrium. 

Therefore, portfolio optimization in a stochastic environment by a single agent that represents
a bounded-rational component of the market as a whole, as is done in our approach using the entropy-regularized free energy, is mathematically equivalent to 
studying a Nash equilibrium in a two-party game of our our agent with an adversarial counter-party with an exponential budget given by the last term 
in Eq.(\ref{G_F_max_min}).

\subsection{G-learning and F-learning}
\label{sect:G-learning} 
 
 In the RL setting when rewards are observed, the system Eqs.(\ref{F_opt}, \ref{pi_opt_F}, \ref{G_from_F}) can be reduced to one non-linear equation.
 Substituting the augmented free energy (\ref{F_opt}) into Eq.(\ref{G_fun}), we obtain
 \beq
 \label{soft_Q}
 G_t^{\pi} ( {\bf y}, {\bf a}) 
  =  \hat{R} ({\bf y}_{t}, {\bf a}_{t} ) +  \mathbb{E}_{t, {\bf a}} \left[  
   \frac{\gamma}{\beta}  \log \sum_{{\bf a}_{t+1}} \pi_0 
( {\bf a}_{t+1} | {\bf y}_{t+1}) e^{ \beta G_{t+1}^{\pi} ( {\bf y}_{t+1}, {\bf a}_{t+1})  } \right]  
\eeq
This equation provides a soft relaxation of the Bellman optimality equation for the action-value Q-function, with the G-function defined in
Eq.(\ref{G_fun}) being an entropy-regularized Q-function \cite{G-Learning}. 
 The "inverse-temperature" parameter $ \beta $ in Eq.(\ref{soft_Q}) 
 determines the strength of entropy regularization. In particular, if we take $ \beta \rightarrow \infty $, we recover the original Bellman optimality equation for the Q-funciton.  Because the last term in (\ref{soft_Q}) approximates the  $ \max(\cdot) $ function when $ \beta $ is large but finite, Eq.(\ref{soft_Q}) is known in the literature as soft Q-learning.
 
 For finite values $ \beta < \infty $, in a setting of Reinforcement Learning with observed rewards, Eq.(\ref{soft_Q}) can be used to specify {\it G-learning}
  \cite{G-Learning}: an off-policy time-difference (TD) algorithm that generalizes Q-learning to noisy environments where an entropy-based regularization can be needed. The G-learning algorithm of Ref.~\cite{G-Learning} was specified in a tabulated setting where both the state and action space are finite.
 In our case, we deal with high-dimensional state and action spaces, and in addition, we do not observe rewards, therefore we are in a setting of 
 Inverse Reinforcement Learning. 
 
 Another possible approach is to bypass the $ G $-function (i.e. the entropy-regulated Q-function) altogether, and proceed with the Bellman optimality equation for the free energy F-function (\ref{F_pi}). In this case, we have a pair of equations for $ F_t^{\pi}({\bf y}_t)  $ and $ \pi( {\bf a}_t | {\bf y}_t) $:
 \bea
 \label{F_learning}
&& F_t^{\pi}({\bf y}_t)  =    \mathbb{E}_{ {\bf a}|x}  \left[ \hat{R} ({\bf y}_{t}, {\bf a}_{t} )
 - \frac{1}{\beta}  g^{\pi} ({\bf y}_{t}, {\bf a}_{t})  + 
\gamma \mathbb{E}_{t, {\bf a}}  \left[ F_{t+1}^{\pi}({\bf y}_{t+1})  \right]  \right]  \nonumber \\
&& \pi ( {\bf a}_t | {\bf y}_t) = \frac{1}{Z_t} \pi_0 ( {\bf a}_t | {\bf y}_t) e^{ \hat{R} ({\bf y}_{t}, {\bf a}_{t} )  
+ \gamma \mathbb{E}_{t, {\bf a}} \left[  F_{t+1}^{\pi}({\bf y}_{t+1}) \right] } 
\eea 
Here the first equation is the Bellman equation (\ref{Bellman_F}) for the F-function, and the second equation is obtained by substitution of 
Eq.(\ref{G_from_F}) into Eq.(\ref{pi_from_F}). Also note that the normalization constant $ Z_t $ in Eq.(\ref{F_learning}) is in general different from the normalization constant in Eq.(\ref{pi_from_F}).
 
Eq.(\ref{F_learning}) shows that one-step rewards $  \hat{R} ({\bf y}_{t}, {\bf a}_{t} )  $ do {\it not} form on their own an alternative 
specification of single-step action probabilities 
$   \pi ( {\bf a}_t | {\bf y}_t) $. Rather, a specification of the sum  $ \hat{R} ({\bf y}_{t}, {\bf a}_{t} )  + 
\gamma \mathbb{E}_{t, {\bf a}} \left[  F_{t+1}^{\pi}({\bf y}_{t+1}) \right] $ is required \cite{Ortega_2015}. However, in a special case when dynamics are {\it linear} and rewards 
 $ \hat{R} ({\bf y}_{t}, {\bf a}_{t} ) $ are {\it quadratic}, 
the term $ \mathbb{E}_{t, {\bf a}} \left[  F_{t+1}^{\pi}({\bf y}_{t+1}) \right] $ has the same parametric form as the time-$t$ reward 
 $ \hat{R} ({\bf y}_{t}, {\bf a}_{t} ) $, therefore addition of this term amounts to a 'renormalization' of parameters of the one-step reward function (see below).
 Therefore, if the only objective of IRL is to learn a policy from data via modeling a reward function, a model can directly learn these 'renormalized' parameters from data. Splitting these values into a current-reward and expected future-reward part would be unnecessary in this case, 
 reducing the problem of finding an optimal policy in IRL to a standard Maximum Likelihood estimation. Such approach was considered e.g. in \cite{IRL_marketing} in a different context.

\section{Inverse Reinforcement Learning of optimal trading}
\label{sect:IRL}

In this section, we will simultaneously analyze two settings for our model: (i) a single investor IRL, and (ii) a market portfolio IRL. 
The main difference between these two cases is that while in the first case actions of an agent are observable, in the second case they are {\it not} directly observable, only their impact on market prices is observed. 

A second difference has to do with a planning horizon in the model. For a single investor case, we have a finite-horizon MDP problem where a task starts a given initial time $ t_0 $ and ends in $ T  $ steps at a specific time $ t_0 + T $. On the contrary, for the market portfolio IRL we do not have a well defined notion of a starting time $ t_0 $ and an end time $ T $. The only uncontroversial time-like parameter is the current time $ t $. 

A reasonable choice would be to get rid of an alleged time non-stationarity in a time homogeneous problem by setting $ t_0 = t $ 
(which means we start our task now), and to set $  T $ to infinity. The latter means changing the problem to a problem of an {\it infinite-horizon} IRL. 

On the other hand, as we will show below, computational algorithms for these two cases have many common or similar elements. In particular, an infinite-horizon setting can be numerically approximated by a fixed time horizon, while unobserved actions can be viewed as hidden variables that now become a part of inference of the model.

This implies that up to a certain point, inference of a market optimal portfolio and a single investor portfolio should involve many common elements.
In our setting, as our bounded-rational market-agent is a {\it sum} of all individual investors, state variables in these two formulations are linked in 
a very explicit way: what was a dollar amount of a single investor's investment in a given stock becomes a total market capitalization for this stock in the market portfolio case. 

Therefore, additivity of total individual investor's portfolios and actions into a single market-wise portfolio and a single action of a bounded-rational market-agent is built-in in our model by construction. This implies that the case of a market portfolio inference can be viewed as a generalization of a single investor case\footnote{It also opens a way to build a model of influential 'market movers" in a top-down manner by a probabilistic dissection ("thinning") of a market-optimal portfolio into sub-portfolios of individual major investors. We leave this for a future research.}. 

In this section, we will present a general solution to the problem of inference of optimal investment strategy from data made of observation of states, that 
works for both cases of a single investor and a market portfolio. This solution is based on a variational EM algorithm, and it can be used to find the original model parameters $ \Theta $. As will be shown in Sect.~\ref{sect:IRL_market}, for a specific case of a market portfolio, in addition to such general approach, our model can also be estimated in an alternative and simpler way, by re-formulating it as an {\it econometric} model of stock returns.   
Our presentation in the present section covers both cases of a single investor and market portfolio simultaneously when possible, and give separate analyses when it is not.       

\subsection{Likelihood functions}

We first consider the case of observable actions. Data in this case includes a set of $ D $ trajectories 
$ \zeta_i $ where $ i = 1, \ldots D $ of state-action pairs $ ( {\bf y}_t, {\bf a}_t ) $ where trajectory $ i $ starts at some time $ t_{0i} $ and runs until time
 $ T_i $. 
 
We consider a single trajectory $ \zeta $ where we set the start time $ t = 0 $ and the end time $ T $. As individual trajectories are considered independent, they 
will enter additively in the final log-likelihood of the problem. We assume that dynamics are Markov in the pair $ ( {\bf y}_t,  {\bf a}_t ) $.
 
The probability of complete data for trajectory $ \zeta $ is 
\beq
\label{P_c}
P_c \left( {\bf y}, {\bf a} | \Theta \right) =  p_{\theta}( {\bf y}_0) \prod_{t =0}^{T-1} \pi_{\theta}  
( {\bf a}_{t} | {\bf y}_{t}  ) p_{\theta} \left( {\bf y}_{t+1} | {\bf y}_t,  {\bf a}_t  \right) 
 \eeq
 Here $ p( {\bf y}_0) $ is a marginal probability of $ {\bf y}_t $ at the start of the $ i$-th demonstration, and  
 $ p_{\theta} \left( {\bf y}_{t+1} | {\bf y}_t {\bf a}_t  \right) $ is a probability of a new state $ {\bf y}_{t+1} $ conditional on the previous state $ {\bf y}_{t} $ and 
 action $ {\bf a}_t $ taken at this step. Note that the first action $ {\bf a}_0 $ is fixed, therefore we have $  \pi_{\theta}( {\bf a}_0 | {\bf y}_0 )  = 1 $.
Also note that in our model-based IRL setting, both the action policy $ \pi_{\theta} (\cdot | {\bf y}_t) $ and 
 transition probability  $ p_{\theta} \left( {\bf y}_{t+1} | {\bf y}_t {\bf a}_t  \right) $ depend on the same set of parameters. The joint distribution 
 $ p_{\theta} ({\bf y}_{t+1}, {\bf a}_t  |  {\bf y}_t) =  \pi_{\theta} ( {\bf a}_t | {\bf y}_t)  p_{\theta} \left( {\bf y}_{t+1} | {\bf y}_t {\bf a}_t  \right) $ is a 
 generative model in our framework. 
 
 For a {\it complete data} (i.e. when both $ {\bf y}_t $ and $ {\bf a}_t $ are observable), 
 we obtain the following log-likelihood
 \beq
 \label{full_LL}
 L_c ( {\bf \theta})  =  
\log P_c \left( {\bf y}, {\bf a} | \Theta \right) = \log  p_{\theta}( {\bf y}_0) + \sum_{t \in \zeta} \left( \log  \pi_{\theta} ( {\bf a}_t | {\bf y}_t) + 
\log p_{\theta} \left( {\bf y}_{t+1} | {\bf y}_t,  {\bf a}_t  \right) \right)
\eeq
 where $ {\bf y}_t $ and $ {\bf a}_t $ stand for values observed in data.  
 Given some simple parametric forms for the policy and transition probability functions, maximization of such complete data log-likelihood is rather straightforward. 
 Such inference problem with complete data corresponds to a single investor IRL in our model.
 
 A different situation arises for IRL of the market portfolio. In this case, actions $ {\bf a}_t $ of the agent are no longer observable. Respectively, we treat them as 
 {\it hidden} variables and integrate over all values of $ {\bf a}_t $ in the product over $ t $ in Eq.(\ref{P_c}). 
This produces the {\it expected} complete log-likelihood of data 
\beq
\label{L_e}
 L_e ( \theta ) =  \log  p_{\theta}( {\bf y}_0 )  + \sum_{t =0}^{T-1}  \log 
 \int d {\bf a}_t  
 \pi_{\theta}( {\bf a}_{t} | {\bf y}_{t} ) 
  p_{\theta} \left( {\bf y}_{t+1} | {\bf y}_t , {\bf a}_t \right)  
\eeq
As the log-likelihood function involves an integral over $ {\bf a}_t $, it is in general intractable in high dimensional action spaces.
Therefore, we will next address an approximate approach to evaluation of log-likelihood (\ref{L_e}). Furthermore, as Eq.(\ref{L_e}) is additive 
in time steps, in what follows we focus on practical ways to compute the integral over  $ {\bf a}_t $ in a single term entering the sum over $ t $ in 
(\ref{L_e}).

\subsection{EM algorithm}
\label{sect:EM}

Expectation Maximization (EM) algorithm is a powerful method for estimating parameters of models with incomplete observations and/or hidden variables. In our Eq.(\ref{L_e}), the role of hidden variables is played by actions $ {\bf a}_t $. In addition, we might introduce additional hidden variables for tractability 
of a resulting approximate likelihood. 

Let $ q( {\bf a}_t | {\bf y} ) $ be some distribution for actions $ {\bf a}_t $ that can depend on the data $ {\bf y} = ({\bf y}_{t}, {\bf y}_{t+1}) $. We can use it to write the 
expected one-step log-likelihood $ L_t $ for time step $ [t, t + 1] $ as follows: 
 \bea
 \label{L_e_q}
  L_t (\theta) 
\hskip-0.5cm && \equiv  \log  \int d {\bf a}_t \,  p_{\theta} \left( {\bf y}_{t+1},  {\bf a}_t  | {\bf y}_t \right) 
 =  \log  \int d {\bf a}_t \,  q( {\bf a}_t | {\bf y} ) \frac{p_{\theta} \left( {\bf y}_{t+1},  {\bf a}_t  | {\bf y}_t \right) }{q( {\bf a}_t | {\bf y} )} \nonumber \\
\hskip-0.5cm  && \geq  \int d {\bf a}_t \,  q( {\bf a}_t | {\bf y} ) \log  \frac{p_{\theta} \left( {\bf y}_{t+1},  {\bf a}_t  | {\bf y}_t \right) }{q( {\bf a}_t | {\bf y} )} 
 \eea
 where in the second line we used Jensen's inequality.  This produces the following low bound for expected log-likelihood of data:
 \bea
 \label{F_q_theta}
 \mathcal{F}(q, \theta) 
 \hskip-0.5cm && \equiv \int d {\bf a}_t \,  q( {\bf a}_t | {\bf y} ) \log  \frac{p_{\theta} \left( {\bf y}_{t+1},  {\bf a}_t  | {\bf y}_t \right) }{q( {\bf a}_t | {\bf y} )} 
 = \mathbb{E}_{q} \left[ \log p_{\theta} \left( {\bf y}_{t+1},  {\bf a}_t  | {\bf y}_t \right) \right] + H \left[ q \right] \nonumber \\
  \hskip-0.5cm && = - KL \left[ q( {\bf a}_t | {\bf y} )  || p_{\theta} \left( {\bf y}_{t+1},  {\bf a}_t  | {\bf y}_t \right) \right] 
 \eea
 where $ H \left[ q \right] = -  \int d {\bf a}_t \,  q( {\bf a}_t | {\bf y} ) \log  q( {\bf a}_t | {\bf y} ) $ is the entropy of distribution $  q( {\bf a}_t | {\bf y} ) $.
 The low bound (\ref{F_q_theta}) can be interpreted as a free energy with the 'energy function' $ \log p_{\theta} \left( {\bf y}_{t+1},  {\bf a}_t  | {\bf y}_t \right) $
 \cite{Neal_Hinton}. 
 
 The classical EM algorithm \cite{EM_1977} amounts to iterative maximization of the free energy (\ref{F_q_theta}) with respect to the distribution $ q $ and model parameters $ \theta $:
\bea
\label{EM_classical}
&& \mbox{ {\bf E step}: }  q^{(k+1)} = \underset{{\bf q}}{\mathrm{argmax}} \;  \mathcal{F} ( q, \theta^{(k)} ) \nonumber \\
&& \mbox{ {\bf M step}: } \theta^{(k+1)} =  \underset{\theta}{\mathrm{argmax}} \; \mathcal{F} ( q^{(k+1)},  \theta ) 
\eea 
Note that the E-step can formally be done analytically by noting that the last form of the free energy $ \mathcal{F}(q, \theta) $ in 
Eq.(\ref{F_q_theta}) indicates that its maximum as a function of $ q $ is attained when 
$  q( {\bf a}_t | {\bf y} ) = C p_{\theta} \left( {\bf y}_{t+1},  {\bf a}_t  | {\bf y}_t \right) $, where $ C $ is a normalization constant, which should be equal to 
$ 1/p_{\theta} \left( {\bf y}_{t+1} | {\bf y}_t \right) $ to have the right normalization of $ q( {\bf a}_t | {\bf y} ) $. Together this produces the following analytical result for the E-step:
\beq
\label{E-step-analytical}
 q^{(k+1)} = \frac{p_{\theta} \left( {\bf y}_{t+1},  {\bf a}_t  | {\bf y}_t, \theta^{(k-1)} \right)}{p_{\theta} \left( {\bf y}_{t+1} | {\bf y}_t , \theta^{(k-1)} \right)} = 
 p_{\theta} \left(  {\bf a}_t  |  {\bf y}_{t+1} , {\bf y}_t, \theta^{(k)} \right)
 \eeq
 so that $ q $ for the $ k$-th step is just the posterior distribution of $ {\bf a}_t $ computed with the model parameters from the previous iteration. The M-step 
 in Eq.(\ref{EM_classical}) then amounts to maximization of the expectation of the 'energy' $ \log p_{\theta} \left( {\bf y}_{t+1},  {\bf a}_t  | {\bf y}_t \right) $
 in parameters $ \theta $. This procedure guarantees a monotonous convergence to a local maximum of the free energy (\ref{F_q_theta}) 
 \cite{EM_1977, Neal_Hinton}.
 
 \subsection{Variational EM}
 
 As the M-step of the classical EM algorithm is intractable in our setting, we use the variational EM method where instead of a non-parametric specification 
 of the approximating distribution $ q $ leading to a non-parametric optimal solution for the E-step, we use a model-based specification $ q_{w} (\cdot ) $ parametrized by a set of 'recognition model' parameters $ \omega $. The E-step then amounts to maximization with respect to parameters $  \omega  $, while 
 the M-step is performed with an expectation defined by the distribution $ q^{(k+1)}(\cdot)  $.
 
A variational EM algorithm thus iteratively updates the recognition model parameters  $  \omega $ and the generative model parameters $  \theta $:
\bea
\label{EM}
&& \mbox{ {\bf E step}: } \omega^{(k+1)} = \underset{ \omega}{\mathrm{argmax}} \;  \mathcal{F} ( \omega, \theta^{(k)} ) \nonumber \\
&& \mbox{ {\bf M step}: } \theta^{(k+1)} =  \underset{\theta}{\mathrm{argmax}} \; \mathcal{F} ( \omega^{(k+1)},  \theta ) 
\eea  
While a variational version of the EM algorithm does not guarantees a monotonous increase of a log-likelihood at each step, it guarantees that the log-likelihood is non-decreasing (i.e. it either increases or stays constant) at each iteration.

To produce a practical computational scheme, we consider the following specification of a variational distribution $ q_{\omega}(\cdot) $ as a joint distribution of {\it four} hidden variables $ {\bf a}_t, \, \bar{\bf a}_t,  \,  \bar{\bf y}_t,  \, \bar{\bf y}_{t+1} $:
\beq
\label{q_marginal}
q_{\omega} ({\bf a}_t  | {\bf y})  =  \int d \bar{\bf a}_t  d  \bar{\bf y}_t  d \bar{\bf y}_{t+1}  q_{\omega} ({\bf a}_t,  \bar{\bf a}_t, \bar{\bf y}_t, \bar{\bf y}_{t+1}   | {\bf y})  
=  \int d \bar{\bf a}_t  d  \bar{\bf y}   q_{\bar{a} \bar{y}}(  
 \bar{\bf a}_t,\bar{\bf y}  | {\bf y}, \omega ) q_{a}(  {\bf a}_t  | \bar{\bf a}_t, \omega )  
\eeq
 where $  {\bf y} = ({\bf y}_t, {\bf y}_{t+1}) $ and $  \bar{\bf y} = (\bar{\bf y}_t, \bar{\bf y}_{t+1}) $. The hidden variables $ \bar{\bf a}_t,  \bar{\bf y} $ will serve below for linearization of dynamics, similar to the Robust Controllable Embedding (RCE) method of \cite{RCE}.

Using this distribution in Eq.(\ref{F_q_theta}), we obtain the following variational EM bound on the log-likelihood of observed data:
\bea
\label{F_Var_EM}
 \mathcal{F}(\omega, \theta) 
 \hskip-0.5cm && = 
 \int d \bar{\bf a}_t   d  \bar{\bf y}  \, 
 q_{\bar{a} \bar{y}}( \bar{\bf a}_t,\bar{\bf y}  | {\bf y},  \omega)
 \int d {\bf a}_t \,   q_{a}(  {\bf a}_t   |  \bar{\bf a}_t, \omega  ) \log  
 \frac{p_{\theta} \left( {\bf y}_{t+1},  {\bf a}_t  | {\bf y}_t \right) }{q_{\omega}( {\bf a}_t, \bar{\bf a}_t,  \bar{\bf y}  | {\bf y} )} \nonumber \\
 \hskip-0.5cm && \equiv  
 \int d \bar{\bf a}_t  d  \bar{\bf y}  \,  
 q_{\bar{a} \bar{y}}( \bar{\bf a}_t,\bar{\bf y} | {\bf y},  \omega ) 
 \mathcal{F}_a ( \omega, \theta, \bar{\bf a}_t, \bar{\bf y})  
\eea
where $  \mathcal{F}_a (\omega, \theta, \bar{\bf a}_t, \bar{\bf y}) $ is 
a conditional variational free energy :
\beq
\label{cond_Free_energy}
 \mathcal{F}_a ( \omega, \theta, \bar{\bf a}_t, \bar{\bf y}) = \int d  {\bf a}_t 
 q_{a}(  {\bf a}_t | \bar{\bf a}_t , \omega )  \log  
 \frac{ \pi_{\theta} \left( {\bf a}_t |  {\bf y}_t,  \bar{\bf a}_t, \bar{\bf y}_t \right) 
  p_{\theta} \left( {\bf y}_{t+1} | {\bf y}_t,  {\bf a}_t,  \bar{\bf a}_t , \bar{\bf y}  \right) }{q_{\omega}( {\bf a}_t, \bar{\bf a}_t,  \bar{\bf y}   | {\bf y}  )} 
 \eeq
where 
$  q_{\omega}( {\bf a}_t  | {\bf y} ) $ in the logarithm is computed as per Eq.(\ref{q_marginal}). Eqs.(\ref{F_Var_EM}) and (\ref{cond_Free_energy}) thus give a variational low bound on the likelihood of data for inference of a market portfolio, while for the case of an individual investor, we have to omit the inner integral over 
$ {\bf a}_t $ in Eq.(\ref{F_Var_EM}).

Note that in Eq.(\ref{cond_Free_energy}) we explicitly introduced the hidden variables into the generative model 
 $ p_{\theta}({\bf y}_{t+1}, {\bf a}_t | {\bf y}_{t}) $. 
 As will be shown below, these hidden variables are introduced to make two calculations involved in Eq.(\ref{cond_Free_energy}) tractable: computing 
 the integral in (\ref{cond_Free_energy}), and computing the policy  $  \pi_{\theta}  $ that this integral depends on.  

These two tasks are clearly sequential. We will first use conditioning on hidden variables to find a tractable representation of the action 
policy $  \pi_{\theta}  $, and then use this representation to compute the integral over $ {\bf a}_t $.   
Eq.(\ref{cond_Free_energy}) suggests that 
if the distribution $  q_{a}(  {\bf a}_t | \bar{\bf a}_t , \omega ) $ is sharply peaked around $ {\bf a}_t = \bar{\bf a}_t $, then  the conditional free energy 
$  \mathcal{F}_a (\omega, \theta, \bar{\bf a}_t, \bar{\bf y})  $ can be computed using  a saddle-point (Laplace) 
approximation. The remaining integral in Eq.(\ref{F_Var_EM}) over the conditioning hidden variables $ \bar{\bf a}_t, \, \bar{\bf y}_t , \, \bar{\bf y}_{t+1} $  
can then be computed using another 
saddle point approximation. 
This scheme will be presented in details below after we specify the variational policy distribution $ q_{w} $ and the generative model $ p_{\theta} $. 

\subsection{Variational distribution  $ q_{w} $}

Our variational model $ q_{\omega} $ is defined as follows:
\bea
\label{gen_var}
q_{\omega} \left( {\bf a}_t, \bar{\bf a}_t, \bar{\bf y} | {\bf y} \right) 
\hskip-0.5cm && = 
q_{\bar{a} \bar{y}} \left( \bar{\bf a}_t,\bar{\bf y} | {\bf y} \right) q_{a}(  {\bf a}_t   |  \bar{\bf a}_t, \omega  ) \nonumber \\
\hskip-0.5cm && =   
q_{\phi} \left( \bar{\bf y}_{t+1} | {\bf y}_{t+1} \right)
q_{\varphi} \left( \bar{\bf y}_{t} | {\bf y}_{t},\bar{\bf y}_{t+1} \right) 
q_{\bar{a}}(  \bar{\bf a}_t | {\bf y}_t, \omega ) 
 q_{a}(  {\bf a}_t   |  \bar{\bf a}_t, \omega  ) 
\eea
Here $ q_{\phi} $ and $ q_{\varphi} $ are variational forward and backward encoders, respectively \cite{RCE}.
As we assume time homogeneity, a functional form of the encoder $ q_{\phi} \left( \bar{\bf y}_{t+1} | {\bf y}_{t+1} \right) $ should be the same as of 
$ q_{\phi} \left( \bar{\bf y}_{t} | {\bf y}_{t} \right) $. 
 
We use Gaussian specifications for four marginals of the variational policy $ q_w $:
\bea
\label{Gaussian_q}
&&  q_{\bar{a}}(  \bar{\bf a}_t | {\bf y}_t, \omega ) = \mathcal{N} \left(  \bar{\bf a}_t | \mu_{a} ({\bf y}_t), \Sigma_{a} \right) , \; \; 
\nonumber \\
&& q_{\phi} \left( \bar{\bf y}_{t} | {\bf y}_{t} \right) = \mathcal{N} \left(   \bar{\bf y}_{t} |  \mu_{\phi} ({\bf y}_{t}), \Sigma_{\phi} \right) \nonumber \\
&& q_{\varphi} \left( \bar{\bf y}_{t} | {\bf y}_{t},\bar{\bf y}_{t+1} \right)  =  \mathcal{N} \left(   \bar{\bf y}_{t+1} |  
\mu_{\varphi} ( {\bf y}_{t},\bar{\bf y}_{t+1} ), \Sigma_{\varphi} \right) \\
&&  q_{a}(  {\bf a}_t | \bar{\bf a}_t , \omega ) = \mathcal{N} \left(  {\bf a}_t | \bar{\bf a}_t, \Sigma_{\delta} \right)   \nonumber 
\eea 
with constant covariance matrices and linear mean functions:
\bea
\label{mean_variational_q}
&& \mu_{a} ({\bf y}_t) =    \mu_{a}   +  \Lambda_{a} {\bf y}_t \nonumber \\
&&  \mu_{\phi} ({\bf y}_{t+1}) = \mu_{\phi}  + \Lambda_{\phi}  {\bf y}_{t+1} \nonumber \\
&&  \mu_{\varphi} ({\bf y}_{t},\bar{\bf y}_{t+1}) =   \mu_{\varphi}   +  \Lambda_{\varphi}^{(1)} {\bf y}_t +  \Lambda_{\varphi}^{(2)} \bar{\bf y}_{t+1}
\eea 
An alternative to these simple linear specifications could be non-linear means and covariances implemented by neural networks as 
in Ref.\cite{RCE}, or using some other universal function approximations such as Gaussian mixtures or trees. In this paper, we stick to simple 
linear Gaussian forms (\ref{Gaussian_q}), (\ref{mean_variational_q}).
 
The vector $ \omega $ of parameters of the variational distribution $ q_{\omega} $  thus includes three vectors $ {\bf \mu}_{a}, \,  \mu_{\phi},   \mu_{\varphi} $, four
'slope' matrices $  \Lambda_{a},  \, \Lambda_{\phi}, \, \Lambda_{\varphi}^{(1)}, \, \Lambda_{\varphi}^{(2)} $, and four covariance matrices
 $ \Sigma_{a}, \, \Sigma_{\phi} , \, \Sigma_{\varphi} , \, \Sigma_{\delta} $.  
For the marginalized distribution $ q_{\omega} ({\bf a}_t | {\bf y}_t ) $ in Eq.(\ref{q_marginal}), we obtain
\beq
\label{marg_q_w_Gauss}
q_{\omega} ({\bf a}_t | {\bf y}_t ) =   \int d \bar{\bf a}  \, q_{\bar{a}}(  \bar{\bf a} | {\bf y}_t )
 q_{a}({\bf a}_t   | \bar{\bf a} ) = \mathcal{N} \left( {\bf a}_t | \mu_{a} ({\bf y}_t), 
\Sigma_w \right), \; \; \Sigma_w =  \Sigma_{a} +   \Sigma_{\delta} 
\eeq
We can also marginalize over $ \bar{\bf y}_{t+1} $:
\beq
\label{marg_q_a}
q_{\bar{y}} \left( \bar{\bf y}_{t} | {\bf y}_{t}, {\bf y}_{t+1} \right)  = \int d  \bar{\bf y}_{t+1} \, 
q_{\phi} \left( \bar{\bf y}_{t+1} | {\bf y}_{t+1} \right) 
q_{\varphi} \left( \bar{\bf y}_{t} | {\bf y}_{t},\bar{\bf y}_{t+1} \right) = \mathcal{N} \left( \bar{\bf y}_t | \mu_{h} ({\bf y}_t, {\bf y}_{t+1}),  
\Sigma_h \right)
\eeq
where 
\bea
\label{sigma_h}
&& \mu_{h} ({\bf y}_t, {\bf y}_{t+1}) = \Lambda_{\varphi}^{(2)}  \left( \mu_{\phi}  + \Lambda_{\phi}  {\bf y}_{t+1} \right)  +  \Lambda_{\varphi}^{(1)} {\bf y}_t + 
 \mu_{\varphi}  \nonumber \\
&& \Sigma_h =   \Sigma_{\varphi} + \Lambda_{\varphi}^{(2)} \Sigma_{\phi} \left( \Lambda_{\varphi}^{(2)} \right)^T
\eea
Finally, the joint distribution $ q_h( \bar{\bf y}_t, \bar{\bf y}_{t+1} | {\bf y} ) $ is a Gaussian with the following inverse covariance matrix:
\beq
\label{inv_covar_joint}
\Sigma_j^{-1} = 
 \left[ \begin{array}{ll}
  \Sigma_{\phi}^{-1} +  \Lambda_{\varphi}^{(2)}  \Sigma_{\varphi}^{-1} \left( \Lambda_{\varphi}^{(2)}   \right)^T & 
  - \Lambda_{\varphi}^{(2)}  \Sigma_{\varphi}^{-1}  \\
 -  \Sigma_{\varphi}^{-1} \Lambda_{\varphi}^{(2)} &   \Sigma_{\varphi}^{-1} 
 \end{array} \right]
\eeq

\subsection{Calculation of conditional free energy $  \mathcal{F}_a $}


Let us write the conditional free energy (\ref{cond_Free_energy}) as follows:
\bea
\label{cond_F_a}
\mathcal{F}_a (\omega, \theta, \bar{\bf a}_t)
 \hskip-0.5cm && = \mathbb{E}_{ q_{a}} \left[  \log \pi_{\theta} \left( {\bf a}_t  |  {\bf y}_t \right) 
  p_{\theta} \left( {\bf y}_{t+1} | {\bf y}_t,  {\bf a}_t  \right) \right] - 
\mathbb{E}_{ q_{a}} \left[  \log q_{\omega}( {\bf a}_t, \bar{\bf a}_t, \bar{\bf y}   | {\bf y} ) \right] \nonumber \\
 \hskip-0.5cm && \equiv  \mathcal{E}_a (\omega, \theta, \bar{\bf a}_t) + \mathcal{H}_a   
\eea
The second term in this expression is given by the following expression:
\beq
\label{mathcal_H}
\mathcal{H}_a  \equiv - \log q_{\phi}( \bar{\bf y}_{t+1} | {\bf y}_{t+1}) - \log q_{\varphi}( \bar{\bf y}_{t} | {\bf y}_t, \bar{\bf y}_{t+1}) - 
\log q_{\bar{a}}( \bar{\bf a}_{t} | {\bf y}_{t}) + H \left[ q_{a} ({\bf a}_t | \bar{\bf a}_t) \right]
\eeq
where $  H \left[ q_{a} ({\bf a}_t | \bar{\bf a}_t) \right] $ is the entropy of the marginal $ q_{a} ({\bf a}_t | \bar{\bf a}_t) $:
\beq
\label{entropy_q_a}
 H \left[ q_{a} ({\bf a}_t | \bar{\bf a}_t) \right]  = - \int d {\bf a}_t q_{a} ({\bf a}_t | \bar{\bf a}_t)  \log q_{a} ({\bf a}_t | \bar{\bf a}_t) = 
 \frac{1}{2} \log \left\{ \left( 2 \pi e \right)^N \left| \Sigma_{\delta} \right| \right\}
 \eeq
 Using specifications of marginals in Eq.(\ref{Gaussian_q}), we obtain a closed-form expression for $ \mathcal{H}_a $:
 \bea
 \label{H_a_closed_form}
 \mathcal{H}_a 
  \hskip-0.5cm && = - \frac{1}{2} \left( \bar{\bf y}_{t+1} - \mu_{\phi} \right)^T \Sigma_{\phi}^{-1}   \left( \bar{\bf y}_{t+1} - \mu_{\phi} \right)
  - \frac{1}{2} \left( \bar{\bf y}_{t} - \mu_{\varphi} \right)^T \Sigma_{\varphi}^{-1}   \left( \bar{\bf y}_{t} - \mu_{\varphi} \right) \nonumber \\
  \hskip-0.5cm &&  - \frac{1}{2} \left( \bar{\bf a}_{t} - \mu_{a} \right)^T \Sigma_{a}^{-1}   \left( \bar{\bf a}_{t} - \mu_{a} \right) 
  +  \frac{1}{2} \log \left\{ \left( 2 \pi e \right)^N \left| \Sigma_{\delta} \right| \right\}
  \nonumber \\
  \hskip-0.5cm &&  - \frac{1}{2} \log \left| \Sigma_{\phi} \right|  - \frac{1}{2} \log \left| \Sigma_{\varphi} \right|  - \frac{1}{2} \log \left| \Sigma_{a} \right| 
  - \frac{1}{2} \left(2 N + N_{a} \right) \log 2 \pi 
 \eea
 where $ N $ and $ N_a $ stand for dimensions of vectors $ \bar{\bf y}_t $  and $ \bar{\bf a}_t $, respectively.
  
On the other hand, the first 'energy' term $  \mathcal{E}_a (\omega, \theta, \bar{\bf a}_t)  $ in the conditional free energy (\ref{cond_F_a}) cannot be computed in closed form.
Changing the integration variable $    {\bf a}_t \rightarrow   \delta {\bf  a}_t = {\bf a}_t  - \bar{\bf a}_t  $, we write this term as follows:
\beq
\label{energy_a}
 \mathcal{E}_a (\omega, \theta, \bar{\bf a}_t)  =  \int d  \delta {\bf a}_t \,   q_{a}(  \bar{\bf a}_t +  \delta {\bf a}_t | \bar{\bf a}_t , \omega ) \log  \left[
  \pi_{\theta} \left(\bar{\bf a}_t +  \delta {\bf a}_t |  {\bf y}_t \right) 
  p_{\theta} \left( {\bf y}_{t+1} | {\bf y}_t,  \bar{\bf a}_t +  \delta {\bf a}_t  \right) \right]
 \eeq
As the distribution $  q_{a}(  {\bf a}_t | \bar{\bf a}_t , \omega ) $ is sharply peaked around $ {\bf a}_t = \bar{\bf a}_t $ (as long as $ \Sigma_{\delta} $ is small enough), 
we can calculate this integral using a saddle point approximation. To this end, we need to compute  
$ \pi_{\theta} \left(\bar{\bf a}_t +  \delta {\bf a}_t |  {\bf y}_t \right) $ and $  p_{\theta} \left( {\bf y}_{t+1} | {\bf y}_t,  \bar{\bf a}_t +  \delta {\bf a}_t  \right) $ for small
values of $ \delta {\bf a}_t  $.

Let us start with a calculation of $  p_{\theta} \left( {\bf y}_{t+1} | {\bf y}_t,  \bar{\bf a}_t +  \delta {\bf a}_t  \right) $. 
A full transition probability for the state vector $ {\bf y}_t = [ {\bf x}_t, {\bf z}_t ]^{T} $ is given by the following expression:
\beq
\label{trans_prob_y}
p_{\theta} \left( {\bf y}_{t+1} | {\bf y}_t, \bar{\bf a}_t +  \delta {\bf a}_t  \right)  = p_z ( {\bf z}_{t+1} |  {\bf z}_t ) \,
p_{\theta}\left( {\bf x}_{t+1} | {\bf x}_t,\bar{\bf a}_t +  \delta {\bf a}_t \right)
\eeq
where 
\beq
\label{trans_prob_z}
p_z ( {\bf z}_{t+1} |  {\bf z}_t )  = \frac{1}{\sqrt{ \left( 2 \pi \right)^{K} \left| \Sigma_z \right|}} e^{ - \frac{1}{2} \left(  {\bf z}_{t+1} - \left( {\bf I} - 
\Phi \right) \circ {\bf z}_t  \right)^{T} \Sigma_z^{-1} 
\left(  {\bf z}_{t+1} - \left( {\bf I} - 
\Phi \right) \circ {\bf z}_t  \right)  }
\eeq
(see Eq.(\ref{OU_z})), where $ K $ is the number of component in the vector of predictors $ {\bf z}_t $. This term is independent of $ \delta {\bf a}_t $ and serves 
as a constant term in Eq.(\ref{energy_a}).

The second conditional transition probability $ p_{\theta}\left( {\bf x}_{t+1} | {\bf x}_t, \bar{\bf a}_t +  \delta {\bf a}_t \right) $ in (\ref{trans_prob_y}) can be 
computed as follows.  First, we obtain the dynamics of the portfolio vector $ {\bf x}_t $ using Eqs. (\ref{h_t_1}) and (\ref{r_t}):
\bea
\label{x_t_1}
{\bf x}_{t+1}
\hskip-0.5cm && = {\bf x}_t + {\bf u}_t  + {\bf r}_t \circ \left( {\bf x}_t +  {\bf u}_t \right) \nonumber \\
&&= 
 {\bf x}_t + {\bf u}_t +  \left(r_f {\bf 1} + {\bf W} {\bf z}_t -  {\bf M}^{T} {\bf u}_t + \varepsilon_t \right)  \circ  \left( {\bf x}_t +  {\bf u}_t \right) 
 \\
 && = 
 (1 + r_f) ( {\bf x}_t + {\bf u}_t ) + \mbox{diag} \left( {\bf W} {\bf z}_t  - {\bf M} {\bf u}_t \right) ( {\bf x}_t + {\bf u}_t ) 
 + \varepsilon ({\bf x}_t, {\bf u}_t)  \nonumber
\eea
Here we assumed that the matrix $ M$ of market impacts is diagonal with elements $ \mu_i $, and set 
\beq
\label{ADvar}
  {\bf M}  = \mbox{diag} \left( \mu_i \right) , \; 
 \varepsilon ({\bf x}_t, {\bf u}_t) \equiv  \varepsilon_t \circ  \left( {\bf x}_t +  {\bf u}_t \right) 
 \eeq
Eq.(\ref{x_t_1}) shows that the dynamics are non-linear in controls $ {\bf u}_t $ due to the market impact $ \sim {\bf M} $. 
Expanding the action $ {\bf u}_t $ as follows:
\[ 
{\bf u}_t =  [{\bf 1}, - {\bf 1}] {\bf a}_t  =  [{\bf 1}, - {\bf 1}] \bar{\bf a}_t + [{\bf 1}, - {\bf 1}] \delta {\bf a}_t  
\equiv \bar{\bf u}_t + \delta {\bf u}_t 
\]
so that $  \delta {\bf u}_t  =  [{\bf 1}, - {\bf 1}] \delta {\bf a}_t  =  {\bf 1}_{-1}^T   \delta {\bf a}_t  $ where $  {\bf 1}_{-1} \equiv [ {\bf 1}, - {\bf 1} ]^T  $,
a one-step conditional transition probability 
for $ {\bf x}_t $ reads
\beq
\label{one_step_p_x}
p_{\theta} \left( {\bf x}_{t+1} | {\bf x}_t, \bar{\bf a}_t +  \delta {\bf a}_t \right) = \frac{1}{\sqrt{ \left( 2 \pi \right)^{N} \left| \Sigma_r \right|}} e^{ - \frac{1}{2} \Delta_{t }^{T} \Sigma_r^{-1} 
\Delta_{t }}
\eeq
where 
\bea
\label{d_t}
\Delta_{t } 
&& \hskip-0.5 cm \equiv \frac{ {\bf x}_{t+1}}{ {\bf x}_t +  \bar{\bf u}_t + \delta {\bf u}_t  } - 1 - r_f - {\bf W} {\bf z}_t + {\bf M}^T
 \left(  \bar{\bf u}_t + \delta {\bf u}_t \right) \nonumber \\
&& \hskip-0.5 cm = {\bf d}_0 (\bar{\bf a}_t) +    {\bf d}_1 (\bar{\bf a}_t) \delta {\bf a}_t  +  {\bf d}_2 (\bar{\bf a}_t)  \left( \delta {\bf a}_t \right)^2 + \ldots
\eea
Here
\bea
\label{coeffs_d_t}
&&  {\bf d}_0 (\bar{\bf a}_t) =   \frac{ {\bf x}_{t+1}}{ {\bf x}_t +  {\bf 1}_{-1}^T \bar{\bf a}_t  } - 1 - r_f - {\bf W} {\bf z}_t  +  {\bf M}^T  {\bf 1}_{-1}^T \bar{\bf a}_t \nonumber \\
&&  {\bf d}_1 (\bar{\bf a}_t)  =   - \mbox{diag} \left(\frac{ {\bf x}_{t+1}}{ \left({\bf x}_t +   {\bf 1}_{-1}^T \bar{\bf a}_t \right)^2 }  \right)  {\bf 1}_{-1}^T 
+  {\bf M}^T {\bf 1}_{-1}^T    
\\
&&   {\bf d}_2 (\bar{\bf a}_t)  =    \mbox{diag} \left( \frac{ {\bf x}_{t+1}}{ \left({\bf x}_t +   {\bf 1}_{-1}^T \bar{\bf a}_t \right)^3 }  \right)  [{\bf 1},  {\bf 1}]         
\nonumber 
\eea
These expressions depend non-linearly on $ \bar{\bf a}_t $, and within a saddle point approximation, values  $ \bar{\bf a}_t $  in these expressions will be 
replaced by their mean values according to the distribution $  q_{\bar{a}} $ defined in Eq.(\ref{Gaussian_q}). On the other hand, other arguments of these expressions, namely  $ {\bf x}_t $ and $ \bf {x}_{t+1} $ (and $ {\bf z}_t $) are values directly observed in the variational likelihood (\ref{F_Var_EM}), as well as in the full likelihood (\ref{full_LL}).
   
Next we have to compute the action policy $ \pi_{\theta} \left(\bar{\bf a}_t +  \delta {\bf a}_t |  {\bf y}_t \right) $ for small
values of $ \delta {\bf a}_t  $. To this end, we write the state vector as $  {\bf y}_t = \bar{\bf y}_t + \delta   {\bf y}_t $ (the meaning of this decomposition will be 
explained below), and introduce a locally-quadratic parametrization for the G-function:
 \beq
 \label{G_dx_du}
  G_t^{\pi} ( {\bf y}_t , \bar{\bf a}_t +  \delta {\bf a}_t) = 
  \delta  {\bf a}_t^T {\bf G}_{aa} \delta {\bf a}_t + \delta {\bf y}_t^T {\bf G}_{yy} \delta {\bf y}_t  + 
 \delta  {\bf a}_t^T  {\bf G}_{ay} \delta {\bf y}_t + \delta  {\bf a}_t^T {\bf G}_{a} + 
  \delta {\bf y}_t^{T} {\bf G}_{y} +  g (\bar{{\bf y}}_t, \bar{{\bf a}}_t) 
  \eeq
As the optimal action policy is given by Eq.(\ref{pi_opt_F}), we have (where now $ {\bf y}_t = \bar{\bf y}_t + \delta   {\bf y}_t $)
\beq
\label{pi_opt_F_1}
 \pi (\bar{\bf a}_t +  \delta {\bf a}_t | {\bf y}_t) = \pi_0 (\bar{\bf a}_t +  \delta {\bf a}_t| {\bf y}_t) 
 e^{ \beta \left(G_t^{\pi} ( {\bf y}_t, \bar{\bf a}_t +  \delta {\bf a}_t) - F_t^{\pi}({\bf y}_t ) 
  \right) } 
 \eeq  
Substituting these expressions in Eq.(\ref{energy_a}) and retaining only quadratic terms in $ \delta {\bf a}_t $ in  in the 
$ \log p_{\theta} \left( {\bf x}_{t+1} | {\bf x}_t, \bar{\bf a}_t +  \delta {\bf a}_t \right) $ term (see Eq.(\ref{d_t})), we obtain
\beq
\label{cond_energy_result}
 \mathcal{E}_a (\omega, \theta, \bar{\bf a}_t)  =  \mathcal{E}_a^{(0)} (\omega, \theta)  + \mathcal{E}_a^{(1)} (\omega, \theta, \bar{\bf a}_t)
 \eeq
 where
 \bea
 \label{cond_energy_coeffs}
  \mathcal{E}_a^{(0)} (\omega, \theta) 
 && \hskip-0.5cm =   
 - \frac{1}{2} \left( \bar{\bf a}_t - \hat{A}_0 - \hat{A}_1 {\bf y}_t \right)^{T} \Sigma_p^{-1}  \left( \bar{\bf a}_t - \hat{A}_0 - \hat{A}_1 {\bf y}_t \right) - 
 \frac{1}{2} {\bf d}_0^T \Sigma_r^{-1} {\bf d}_0    \nonumber \\
 && \hskip-0.5cm + \log p_z ( {\bf z}_{t+1} |  {\bf z}_t ) - \frac{1}{2} \mbox{Tr} \left[ \Sigma_{\delta}  {\bf d}_1^T \Sigma_{r}^{-1} {\bf d}_1
 \right] - \mbox{Tr} \left[ \mbox{diag} \left( \Sigma_{\delta} \right)  {\bf d}_0^T \Sigma_{r}^{-1}  {\bf d}_2 \right]  \nonumber \\
 && \hskip-0.5cm  - \frac{1}{2} \mbox{Tr} \left[ \Sigma_{\delta}  \Sigma_{p}^{-1}  \right] 
 - \frac{1}{2} \log \left| \Sigma_p \right| - \frac{1}{2} \log  \left| \Sigma_r \right|- \frac{N}{2} \log (2 \pi) \nonumber \\
  \mathcal{E}_a^{(1)} (\omega, \theta, \bar{\bf a}_t) 
 && \hskip-0.5cm =  \beta \left( g (\bar{\bf y}_t, \bar{{\bf a}}_t)  - F_t^{\pi}({\bf y}_t) +  \delta{\bf y}_t^T {\bf G}_{yy} \delta {\bf y}_t 
 +   \delta {\bf y}_t^{T} {\bf G}_{y}  + \mbox{Tr} \left[  \Sigma_{\delta} {\bf G}_{aa} \right] \right) 
 \eea
 where we omitted for compactness the dependence of $ {\bf d}_0, \, {\bf d}_1 $ and $ {\bf d}_2 $ on $ \bar{\bf a}_t $, see Eq.(\ref{coeffs_d_t}), and 
 $ {\bf y}_t $ in $  \mathcal{E}_a^{(0)} (\omega, \theta) $ stands for the observed state vector at time $ t $. 
 The second expression $ \mathcal{E}_a^{(1)} (\omega, \theta, \bar{\bf a}_t) $ in Eq.(\ref{cond_energy_result}) thus collects all terms that depend on  
 the G- and F-functions, while terms independent of these functions are combined in  $ \mathcal{E}_a^{(0)} (\omega, \theta) $. 
 
To summarize so far, Eqs.(\ref{cond_energy_result}), (\ref{cond_energy_coeffs}), (\ref{mathcal_H}) jointly specify the conditional variational free energy (\ref{cond_F_a}), provided model parameters as well as the G-function  (\ref{G_dx_du}) and the F-function are known.  Once the conditional free energy 
$ \mathcal{E}_a (\omega, \theta, \bar{\bf a}_t)  $ is computed, the unconditional variational free energy (\ref{F_Var_EM}) can be calculated using another saddle point approximation for the integral over $ \bar{\bf a}_t $. This calculation will be presented next, while the following sections will describe the method of finding 
the policy $ \pi_{\theta} $ and the G-function (\ref{G_dx_du}) and a corresponding F-function by linearization around  $  \bar{\bf a}_t,  \bar{\bf y} $.
%

\subsection{Calculation of variational free energy $ \mathcal{F} $} 

Recall that in Eq.(\ref{G_dx_du}) we used the representation of the state vector $  {\bf y}_t = \bar{\bf y}_t + \delta {\bf y}_t $. This decomposes the {\it observable} 
vector $  {\bf y}_t $ into a sum of two {\it unobservable} quantities $ \bar{\bf y}_t  $ and $ \delta  {\bf y}_t $. When we condition on the linearization variable 
$ \bar{\bf y}_t $, we can write $ \delta {\bf y}_t =  {\bf y}_t - \bar{\bf y}_t $ when performing integration over the outer hidden variables  
$  \bar{\bf a}_t,  \bar{\bf y} $. 

The advantage of such decomposition of the observable $ {\bf y}_t $ into two unobservables $ \bar{\bf y}_t, \,  \delta {\bf y}_t $ is that now we can assume that the 
F-function is locally quadratic around a random hidden conditioning (linearization) value $  \bar{\bf a}_t,  \bar{\bf y} $, and parametrize it as follows:
 \beq
 \label{F_parametrization}
 F_t^{\pi}({\bf y}_t) = \delta  {\bf y}_t^T {\bf F}_{yy}  \delta  {\bf y}_t  
 +   \delta  {\bf y}_t^{T} {\bf F}_{y} + F_0 (\bar{{\bf y}}_t, \bar{{\bf a}}_t) 
 \eeq 
 Here 
 \beq
 \label{F_yyt}
  {\bf F}_{yy} = 
 \left[ \begin{array}{ll}
 {\bf F}_{xx}  & {\bf F}_{xz}  \\
 {\bf F}_{zx}  & {\bf F}_{zz}   
 \end{array} \right], \; \; 
 {\bf F}_{y} = 
 \left[ \begin{array}{l}
 {\bf F}_{x}    \\
 {\bf F}_{z}     
 \end{array} \right], \; \; 
\eeq
In a finite-horizon setting, parameters $  {\bf F}_{yy}, \,   {\bf F}_{y} , \,  F_0 $  become time-dependent, while in an infinite-horizon setting they do not 
explicitly depend on time. As will be shown below, the last term $ F_0 (\bar{{\bf y}}_t, \bar{{\bf a}}_t) $ in (\ref{F_parametrization}) is a quadratic functional
of $ (\bar{{\bf y}}_t, \bar{{\bf a}}_t) $.

Using Eq.(\ref{F_parametrization}) in (\ref{cond_energy_coeffs}), we have the following decomposition of the unconditional free energy (\ref{F_Var_EM}):
\bea
\label{F_Var_EM_decomp}
\mathcal{F}(\omega, \theta) 
 \hskip-0.5cm && =  
 \int d \bar{\bf a}_t  d  \bar{\bf y}  \,  
 q_{\bar{a} \bar{y}}( \bar{\bf a}_t,\bar{\bf y} | {\bf y}, \omega ) \left( \mathcal{H}_a + \mathcal{E}_a^{(0)} (\omega, \theta) + 
  \mathcal{E}_a^{(1)} (\omega, \theta, \bar{\bf a}_t)  \right) \nonumber \\
  \hskip-0.5cm && \equiv  \mathcal{H}  + \mathcal{F}^{(0)} (\omega, \theta) +  \mathcal{F}^{(1)} (\omega, \theta) 
 \eea
Here the first term can be computed analytically:
\bea
\label{H}
\mathcal{H} 
 \hskip-0.5cm && = - \int d \bar{\bf y}_t d \bar{\bf y}_{t+1}  q_h( \bar{\bf y}_t, \bar{\bf y}_{t+1} | {\bf y} ) \log q_h( \bar{\bf y}_t, \bar{\bf y}_{t+1} | {\bf y} ) + 
 H \left[ q_{\bar{a}} ( \bar{\bf a}_t | {\bf y}_t ) \right] \nonumber \\
  \hskip-0.5cm && =  \frac{1}{2} \log \left\{ \left( 2 \pi e \right)^{2N} \left| \Sigma_{j} \right| \right\} + 
   \frac{1}{2} \log \left\{ \left( 2 \pi e \right)^{N_a} \left| \Sigma_{a} \right| \right\}
  \eea
where the joint covariance matrix $ \Sigma_j $ is defined in Eq.(\ref{inv_covar_joint}).

The second term $  \mathcal{F}^{(0)} (\omega, \theta) $ in Eq.(\ref{F_Var_EM_decomp}) involves the integral of $ \mathcal{E}_a^{(0)} (\omega, \theta)  $ that collects all terms that are independent of the G- and F-functions. 
With a saddle point approximation, we replace $  \bar{\bf a}_t $ in coefficients (\ref{coeffs_d_t}) by its mean value  $ \langle  \bar{\bf a}_t \rangle = 
\mu_{a} ({\bf y}_t) $. Therefore, with this approximation, the remaining dependence of $ \mathcal{E}_a^{(0)} (\omega, \theta)  $ on $ \bar{\bf a}_t $ is quadratic due
to the first term. Integrating this expression with the Gaussian distribution $ q_{a} $ given by Eq.(\ref{Gaussian_q}), we obtain
\bea
\label{mathcal_F_0}
\mathcal{F}^{(0)} (\omega, \theta) 
\hskip-0.5cm && =  
 \int d \bar{\bf a}_t  d  \bar{\bf y}  \,  
 q_{\bar{a} \bar{y}}( \bar{\bf a}_t,\bar{\bf y} | {\bf y}, \omega )  \mathcal{E}_a^{(0)} (\omega, \theta)  
 =  \int d \bar{\bf a}_t \, q_{\bar{a}}(  \bar{\bf a}_t | {\bf y}_t, \omega )  \mathcal{E}_a^{(0)} (\omega, \theta)   \nonumber \\
 \hskip-0.5cm && =     
 - \frac{1}{2} \left( \mu_a - \hat{A}_0 + \left( \Lambda_a - \hat{A}_1 \right) {\bf y}_t \right)^{T} \Sigma_p^{-1} 
  \left( \mu_a - \hat{A}_0 + \left( \Lambda_a - \hat{A}_1 \right) {\bf y}_t \right) - 
 \frac{1}{2} {\bf d}_0^T \Sigma_r^{-1} {\bf d}_0    \nonumber \\
 \hskip-0.5cm  && + \log p_z ( {\bf z}_{t+1} |  {\bf z}_t ) - \frac{1}{2} \mbox{Tr} \left[ \Sigma_{\delta}  {\bf d}_1^T \Sigma_{r}^{-1} {\bf d}_1
 \right] - \mbox{Tr} \left[ \mbox{diag} \left( \Sigma_{\delta} \right)  {\bf d}_0^T \Sigma_{r}^{-1}  {\bf d}_2 \right]  \nonumber \\
 \hskip-0.5cm  && - \frac{1}{2} \mbox{Tr} \left[ \Sigma_{\delta}  \Sigma_{p}^{-1}  \right] - 
  \frac{1}{2} \mbox{Tr} \left[ \Sigma_{a}  \Sigma_{p}^{-1}  \right] 
 - \frac{1}{2} \log \left| \Sigma_p \right| - \frac{1}{2} \log  \left| \Sigma_r \right|- \frac{N}{2} \log (2 \pi) 
\eea
Lastly, we consider the third term in Eq.(\ref{F_Var_EM_decomp}) that depends on the G-function (\ref{G_dx_du}) and the F-function (\ref{F_parametrization}).
Using these expressions, we can write the integrand  $  \mathcal{E}_a^{(1)} $  of this term defined in the second of Eqs.(\ref{cond_energy_coeffs}) as follows
\bea
\label{mathcal_F_1}
 \mathcal{E}_a^{(1)} (\omega, \theta, \bar{\bf a}_t) 
 \hskip-0.5cm && =   \beta \left( g (\bar{\bf y}_t, \bar{{\bf a}}_t)  - F_t^{\pi}({\bf y}_t) +  \delta{\bf y}_t^T {\bf G}_{yy} \delta {\bf y}_t 
 +   \delta {\bf y}_t^{T} {\bf G}_{y}  + \mbox{Tr} \left[  \Sigma_{\delta} {\bf G}_{aa} \right] \right)  \\
  \hskip-0.5cm && =   \beta \left[ \delta{\bf y}_t^T \left( {\bf G}_{yy} -  {\bf F}_{yy} \right) \delta{\bf y}_t
 +  \delta {\bf y}_t^{T}  \left( {\bf G}_{y} -  {\bf F}_{y} \right) 
   + g (\bar{\bf y}_t, \bar{{\bf a}}_t)  - F_0 (\bar{{\bf y}}_t, \bar{{\bf a}}_t)  
 + \mbox{Tr} \left[  \Sigma_{\delta} {\bf G}_{aa} \right] \right]  \nonumber 
\eea
Relations between parameters of the G-function and F-function are derived in Appendix A in Sect.~\ref{sect:Backward_obs_rewards}, see Eqs.(\ref{F_coeffs}).
Using the following auxiliary quantities  (as defined below in Eq.(\ref{aux}) and repeated here for convenience)
\bea
\label{aux_1}
&& {\bf b}_t =  \bar{ {\bf a}}_t - \hat{ {\bf A}}_0  - \hat{ {\bf A}}_1 \bar{{\bf y}}_t   , \; 
 \tilde{\Sigma}_p = \Sigma_p^{-1} - 2 \beta {\bf G}_{aa},   \nonumber \\
 && \Gamma_{\beta} = \frac{1}{\beta} \left( {\bf I} - \left(\Sigma_p^{-1} \right)^T \tilde{\Sigma}_p^{-1} \right)  \Sigma_p^{-1}, \; 
 \Upsilon_{\beta} =   \tilde{\Sigma}_p^{-1} \Sigma_p^{-1}  \nonumber \\
 && {\bf E}_{ay} = \Upsilon_{\beta} \hat{\bf A}_{1} + \frac{1}{2} \beta  \tilde{\Sigma}_p^{-1}  {\bf G}_{ay}, \; 
 {\bf D}_{ay} =  {\bf G}_{ay}^{T} \Upsilon_{\beta} - \hat{ {\bf A}}_1^{T} \Gamma_{\beta}  \nonumber \\
 && {\bf E}_{a} = \hat{\bf A}_{1}^T \Upsilon_{\beta} {\bf G}_a + \beta {\bf G}_{ay}^T \tilde{\Sigma}_p^{-1} {\bf G}_a , \; 
 \mathcal{L}_{\beta} = \frac{1}{2 \beta} \left( \log  \left| \Sigma_p \right|  + \log \left| \tilde{\Sigma}_p \right|  \right) 
 \nonumber 
\eea
we obtain
\bea
\label{mathcal_F_1_1}
&&  {\bf F}_{yy} = {\bf G}_{yy} +  {\bf G}_{ay}^T  {\bf E}_{ay}
- \frac{1}{2}  \hat{ {\bf A}}_1^{T} \Gamma_{\beta}  \hat{ {\bf A}}_1 \nonumber \\
&&   {\bf F}_{y}  =  {\bf G}_{y}  - {\bf D}_{ay}  {\bf b}_t  
+ \hat{ {\bf A}}_1^{T} \Upsilon_{\beta} {\bf G}_a + \beta  {\bf G}_{ay}^T  \tilde{\Sigma}_p^{-1}  {\bf G}_a \\
&&  F_0 (\bar{{\bf y}}_t, \bar{{\bf a}}_t) = g (\bar{{\bf y}}_t, \bar{{\bf a}}_t)   
- \frac{1}{2} {\bf b}_t^T \Gamma_{\beta} {\bf b}_t -    {\bf G}_a^T \Upsilon_{\beta} {\bf b}_t + \frac{ \beta}{2}  {\bf G}_a^T \tilde{\Sigma}_p^{-1} {\bf G}_a
-  \mathcal{L}_{\beta} 
\nonumber  
 \eea 
 These relations suggest the following dependencies of different terms in free energy (\ref{F_parametrization}) on hidden variables  $ \bar{\bf a}_t $ and 
  $ \bar{\bf y}_t $. First, the quadratic term $ \delta  {\bf y}_t^T {\bf F}_{yy}  \delta  {\bf y}_t  $ is quadratic 
  in $ \bar{\bf y}_t $ (as $ \delta {\bf y}_t = {\bf y}_t - \bar{\bf y}_t $), and independent of $ \bar{\bf a}_t $. 
 The second term $ \delta {\bf y}_t^T  {\bf F}_{y}  $ is quadratic in  $ \bar{\bf y}_t $ and linear in $ \bar{\bf a}_t $. The free term 
 $  F_0 (\bar{{\bf y}}_t, \bar{{\bf a}}_t) $ is given by a sum of the term $ g (\bar{{\bf x}}_t, \bar{{\bf a}}_t) $ that cancels out in Eq.(\ref{mathcal_F_1}), and 
 a quadratic form in both  $ \bar{\bf y}_t $ and $ \bar{\bf a}_t $, as indicated by the last of Eqs.(\ref{mathcal_F_1_1}).

The integral of this expression can therefore be computed in closed form with Gaussian hidden variable distributions (\ref{Gaussian_q}).
Using Eqs.(\ref{mathcal_F_1}) and (\ref{mathcal_F_1_1}), we obtain the following results for expectations $ \mathbb{E}_{ \bar{\bf a}_t,\bar{\bf y}}[\cdot ]  $ of 
three terms in Eq.(\ref{F_parametrization}) under the variational distribution $ q_{\bar{a} \bar{y}} \left( \bar{\bf a}_t,\bar{\bf y} | {\bf y} \right) $:
\bea
\label{E_1_terms}
 \mathcal{E}_{yy}^{(1)} (\omega, \theta, \bar{\bf a}_t)
  \hskip-0.5cm && \equiv
 \mathbb{E}_{ \bar{\bf a}_t,\bar{\bf y}} \left[ \beta \delta{\bf y}_t^T \left( {\bf G}_{yy} -  {\bf F}_{yy} \right) \delta{\bf y}_t \right] 
= 
\beta \mbox{Tr} \left[ \Sigma_h^{-1} \left( \frac{1}{2}  \hat{ {\bf A}}_1^{T} \Gamma_{\beta}  \hat{ {\bf A}}_1 - {\bf G}_{ay}^T  {\bf E}_{ay} \right) \right]
\nonumber \\
 \hskip-0.5cm && + \beta 
 \left( {\bf y}_t - \mu_h({\bf y}) \right)^T \left( \frac{1}{2}  \hat{ {\bf A}}_1^{T} \Gamma_{\beta}  \hat{ {\bf A}}_1 - {\bf G}_{ay}^T  {\bf E}_{ay} \right) 
\left( {\bf y}_t - \mu_h({\bf y}) \right)  \nonumber \\
\mathcal{E}_{y}^{(1)} (\omega, \theta, \bar{\bf a}_t)
 \hskip-0.5cm && \equiv 
  \mathbb{E}_{ \bar{\bf a}_t,\bar{\bf y}} \left[ \beta \delta{\bf y}_t^T \left( {\bf G}_{y} -  {\bf F}_{y} \right) \right] 
 = \beta  \mbox{Tr} \left( \Sigma_h^{-1} {\bf D}_{ay} \hat{\bf A}_1 \right) 
 - \beta \mu_h({\bf y}) \hat{\bf A}_1^T {\bf D}_{ay}^T  \nonumber \\
  \hskip-0.5cm && + \beta \left( {\bf y}_t - \mu_h({\bf y}) \right)^T \left( {\bf E}_{a} + {\bf D}_{ay} \left( \mu_a ({\bf y}_t) - \hat{\bf A}_0 \right) \right) 
   \\ 
\mathcal{E}_{0}^{(1)} (\omega, \theta, \bar{\bf a}_t) 
 \hskip-0.5cm && \equiv 
 \mathbb{E}_{ \bar{\bf a}_t,\bar{\bf y}} \left[ \beta \left( g (\bar{\bf y}_t, \bar{{\bf a}}_t)  - F_0 (\bar{{\bf y}}_t, \bar{{\bf a}}_t)  
 + \mbox{Tr} \left[  \Sigma_{\delta} {\bf G}_{aa} \right]  \right) \right] =   
 \frac{\beta}{2}  \mbox{Tr} \left[  \Sigma_a \Gamma_{\beta} + \Sigma_h  \hat{\bf A}_1^T \Gamma_{\beta}  \hat{\bf A}_1 \right] 
 \nonumber \\
 \hskip-0.5cm && +  \frac{\beta}{2} \hat{\bf A}_0^T \Gamma_{\beta} \hat{\bf A}_0 
 - \beta  \hat{\bf A}_0^T \Gamma_{\beta} \mu_a ({\bf y}_t) - \beta \left( \mu_a({\bf y}_t) -  \hat{\bf A}_0 \right)^T 
 \Gamma_{\beta}  \hat{\bf A}_1 \mu_h({\bf y}) \nonumber \\
  \hskip-0.5cm && + \beta {\bf G}_a^T \Upsilon_{\beta} \left( \mu_a({\bf y}_t) -  \hat{\bf A}_0 -  \hat{\bf A}_1 \mu_h({\bf y}) \right) 
  - \frac{\beta^2}{2} {\bf G}_a^T  \tilde{\Sigma}_p^{-1}  {\bf G}_a  + \beta  \mbox{Tr} \left[  \Sigma_{\delta} {\bf G}_{aa} \right] + \beta \mathcal{L}_{\beta} 
  \nonumber 
 \eea
 where linear Gaussian mean functions $ \mu_a({\bf y}_t) $ and $ \mu_h({\bf y}) $ are defined in Eqs.(\ref{mean_variational_q}) and (\ref{sigma_h}), respectively.
 
 The final {\it closed form} result for the variational free energy (\ref{F_Var_EM_decomp}) is therefore given by the 
 sum of equations (\ref{H}), (\ref{mathcal_F_0}) and (\ref{E_1_terms}):
 \beq
 \label{F_var_final}
 \mathcal{F}(\omega, \theta, \pi_{\theta}) 
 =   \mathcal{H}  + \mathcal{F}^{(0)} (\omega, \theta) +  \mathcal{F}^{(1)} (\omega, \theta, \pi_{\theta}) 
 \eeq
 Here we added the policy $ \pi_{\theta} $ as an argument to $ \mathcal{F}(\omega, \theta, \pi_{\theta})  $ 
 to emphasize that the latter depends on three sets of inputs: variational parameters $ \omega $, 
 generative model parameters $ \Theta $, and the optimal policy $ \pi_{\theta} $. The variational free energy (\ref{F_var_final}) depends on the policy 
 $ \pi_{\theta} $ via its dependence on the parameter $ {\bf G}_{aa}, \bf{G}_{ay} $ etc. that determine the locally-quadratic representation (\ref{G_dx_du}) of 
 the optimal G-function (i.e. the optimal entropy-regularized Q-function).
 
  The variational EM algorithm amounts to iterative maximization of Eq.(\ref{F_var_final}). 
 As the whole expression for the variational free energy (\ref{F_var_final}) is analytical, both the E-step and the M-step of the algorithm are computationally light. 
  In the E-step, we maximize it with respect to variational parameters 
  $ \omega $ while keeping parameters $ \Theta $ and the G-function from the previous iteration. In the M-step, we maximize it with respect to generative model parameters $ \Theta $ and policy $ \pi_{\theta} $. The outputs of the M-step are updated values of parameters $ \Theta $ and updated 
  values of parameters of G-function (\ref{G_dx_du}).  We will now consider the M-step in more details.

\subsection{M-step: policy optimization}

In the M-step, updates of G-functions are done using Eqs.(\ref{R_hat}), (\ref{F_vect_aa}), (\ref{recursive_G}) derived in Appendix A.
These equations provide a practical implementation of the general self-consistent system of equations (\ref{F_opt}), (\ref{pi_opt_F}), (\ref{G_from_F}) in our setting 
of locally-quadratic expansion for the G-function. In this setting, all integrations in these equations are performed 
analytically, thus providing a tractable version of this approach in our highly dimensional continuous state-action setting. Note that the original version of G-learning was only explored in \cite{G-Learning} in a low-dimensional discrete state setting. 

As discussed in Appendix A, Eqs.(\ref{R_hat}), (\ref{F_vect_aa}), (\ref{recursive_G}) can be used for either a single investor or a market portfolio.
In the former case, the update is performed backward in time, starting with a terminal time $ T $ and a specific terminal condition on the F-function or/and 
G-function. In the latter case of a market portfolio, these equations can be used in a time-stationary setting as update rules for time-independent coefficients of the G-function. 

When coefficients of the Q-functions are computed in this way for time step $ t $, the optimal action distribution for $ \delta{\bf a}_t $
is computed using Eq.(\ref{pi_opt_F_1}) which we repeat here for convenience:
\beq
\label{pi_opt_F_1_1}
 \pi_{\theta} (\bar{\bf a}_t +  \delta {\bf a}_t | {\bf y}_t) = \pi_0 (\bar{\bf a}_t +  \delta {\bf a}_t| {\bf y}_t) 
 e^{ \beta \left(G_t^{\pi} ( {\bf y}_t, \bar{\bf a}_t +  \delta {\bf a}_t) - F_t^{\pi}({\bf y}_t ) 
  \right) } 
 \eeq 
 When $ \bar{\bf a}_t $ is fixed by conditioning, we view the distribution as a Gaussian distribution for $ \delta {\bf a}_t $ with the mean
 $  \widehat{\delta{\bf a}}_t =   \hat{ {\bf A}}_0  + \hat{ {\bf A}}_1 {\bf y}_t  - \bar{ {\bf a}}_t $.    
As the reference distribution $ \pi_0 $ is Gaussian and the Q-function is quadratic, the optimal action policy $ \pi $ is again Gaussian with 
a new mean and covariance:
\beq
\label{pi_post}
 \pi_{\theta} (\delta{\bf a}_t| {\bf y}_t) = \pi_0 ( \delta{\bf a}_t| {\bf y}_t) e^{ \beta \left(G_t^{\pi} ( {\bf y}_t, {\bf a}_t) - F_t^{\pi}({\bf y}_t ) 
  \right) } 
  = \mathcal{N} \left( \delta{\bf a}_t | \widehat{\delta{\bf a}}_t', \Sigma_p' \right)
 \eeq 
where $ \mathcal{N} (\cdot) $ is a multivariate Gaussian distribution with the following mean and covariance matrix:
\bea
\label{new_mean_var}
&&  \widehat{\delta{\bf a}}_t' =  \Sigma_p' \left(  \Sigma_p^{-1}  \widehat{\delta{\bf a}}_t + \beta {\bf G}_{ay}  \delta  {\bf y}_t + 
\beta {\bf G}_{a} \right) \nonumber \\
&&  \Sigma_p' = \left[  \Sigma_p^{-1} - 2 \beta {\bf G}_{aa} \right]^{-1} 
\eea
These relations can be viewed as Bayesian updates for the current iteration mean $  \widehat{\delta{\bf a}}_t $ (see Eq.(\ref{prior_mean})) and 
variance $ \Sigma_p $ of the optimal action policy relative to their values for  the "prior" reference policy (\ref{pi_0_1}). Note that in the limit $ \beta \rightarrow 0 $,
Eq.(\ref{new_mean_var}) produces no update, $  \widehat{\delta{\bf a}}_t' =  \widehat{\delta{\bf a}}_t $. This is as expected, as in this 'high-temperature' limit the agent only maximizes the negative of the KL entropy but not rewards.  

 They can be also expressed as updates for the action policy (\ref{pi_0}) in terms of original policy variables. As
  $  \widehat{\delta{\bf a}}_t =   \hat{ {\bf A}}_0  + \hat{ {\bf A}}_1 {\bf y}_t  - \bar{ {\bf a}}_t $, the update (\ref{new_mean_var}) 
  of the mean $ \widehat{\delta{\bf a}}_t  $ implies an update of parameters $ \hat{ {\bf A}}_0 $ and $  \hat{ {\bf A}}_1 $. Substituting this expression into 
  Eq.(\ref{new_mean_var}) and 
 comparing an intercept and linear terms in this equation produces an update for the mean of the policy  (\ref{pi_0}):
\bea
\label{update_A_hat}
&& \Sigma_p^{(k+1)}  =  \left[  \left(\Sigma_p^{(k)} \right)^{-1} - 2 \beta {\bf G}_{aa}^{(k)} \right]^{-1}  \nonumber \\
&& \hat{ {\bf A}}_{0}^{(k+1)}  = \bar{\bf a}_t +  \Sigma_p^{(k+1)}  \left(\Sigma_p^{(k)} \right)^{-1} \left( \hat{ {\bf A}}_{0}^{(k)} - \bar{\bf a}_t \right)
 +   \beta \Sigma_p^{(k+1)}   \left( {\bf G}_{a}^{(k)} -  {\bf G}_{ay}^{(k)} \bar{\bf y}_t \right)  \nonumber \\ 
&& \hat{ {\bf A}}_1^{(k+1)}  = \Sigma_p^{(k+1)}  \left( \left(\Sigma_p^{(k)} \right)^{-1} \hat{ {\bf A}}_1^{(k)} +  \beta   {\bf G}_{ay}^{(k)}    \right)
\eea
where we use values of parameters $ {\bf G}_{aa} $ etc. corresponding to the current iteration of the algorithm.
Again, these updates degenerate and become identities in the high temperature limit $ \beta \rightarrow 0 $. 
On the other hand, in the opposite limit $ \beta \rightarrow \infty $ we obtain finite and non-trivial updates. 

Note that in a finite-horizon setting of a single investor, 
parameters $ {\bf G}_{aa},  {\bf G}_{ay} $ etc. are time-dependent, therefore coefficients $  \hat{ {\bf A}}_1 $ will be also be time-dependent. 
On the other hand, for a market portfolio inference, parameters of the G-function are time-independent, thus parameters  $  \hat{ {\bf A}}_0 $ 
and $  \hat{ {\bf A}}_1 $ would also be time-independent\footnote{An apparent dependence of $ \hat{ {\bf A}}_{0} $ on $  \bar{\bf a}_t ,   \bar{\bf y}_t $ 
is a result of our conditioning on these values in the outside integral in Eq.(\ref{F_Var_EM}). While {\it updates} of $  \hat{ {\bf A}}_{0} $
may depend on the conditioning/linearization variables $ \bar{\bf a}_t,   \bar{\bf y}_t $ as in Eq.(\ref{update_A_hat}), a final fixed-point value of  
$  \hat{ {\bf A}}_{0} $ obtained with this method is a constant parameter  that is independent of  $ \bar{\bf a}_t,  \bar{\bf y}_t  $.}. 
 
The updated policy for step $ k + 1 $ now takes the form
\beq
\label{pi_post_M}
 \pi^{(k+1)} ( {\bf a}_t| {\bf y}_t) 
  = \mathcal{N} \left( {\bf a}_t|  \hat{ {\bf A}}_{0}^{(k+1)} +  \hat{ {\bf A}}_1^{(k+1)}  {\bf y}_t,  \Sigma_p^{(k+1)} \right)
 \eeq  
Equations (\ref{update_A_hat})  and (\ref{pi_post_M}) represent one of our main results. The point is that the last of 
Eqs.(\ref{update_A_hat}) shows that a non-zero coefficients $ \hat{\bf A}_1^{(k+1)} $ is obtained even if its value at the previous iteration was zero.
Applying this for $ k = 0 $, it means that this coefficient  (which induces the dependence of the optimal policy on the state $ {\bf y}_t $) becomes 
non-zero even if we start with  $ \hat{ {\bf A}}_1^{(0)} $ in the policy prior (\ref{pi_0}). 

Furthermore, it implies that at convergence, updates 
(\ref{pi_post_M}) produce some fixed values $ \hat{ {\bf A}}_{0}, \, \hat{ {\bf A}}_{1} $ of policy parameters. 
Our model therefore predicts that the optimal investment policy is Gaussian whose mean is is {\it linear} in the state 
variable $ {\bf y}_t = \left[  {\bf x}_t,  {\bf z}_t \right] $, as in the Iterative Linear-Quadratic Gaussian (iLQG) regulator of Todorov and Li \cite{Todorov_2005}.

When $ {\bf x}_t $ is identified with a market portfolio and an agent is our bounded-rational market-agent, 
Eq.(\ref{pi_post_M}) (used with such fixed-point values  $ \hat{ {\bf A}}_{0}, \, \hat{ {\bf A}}_{1} $) 
defines an optimal "market-implied" action policy. This provides a probabilistic and multi-period extension of a market-optimal static portfolio in a 
one-period setting of the Black-Litterman model \cite{BL} and inverse portfolio optimization approach of Bertsimas {\it et. al.} \cite{Bertsimas_2012}. 

On the other hand, as we mentioned above, the same framework can be applied to an individual investor provided we have access to proprietary 
trading data of that particular investor. In this case, actions $ {\bf a}_t $ will be actions of that investor. If these actions are observable,  
Eq.(\ref{pi_post_M}) can be 
directly used within a Maximum Likelihood estimation. We discuss this as a special case of our model in Appendix B, while here we proceed with the case when actions (of either a market-agent or an individual investor) are unobservable.

While the main focus of this paper is on inference of a market-wide bounded-rational agent, the algorithm can also be used for a single large investor 
whose trades impact the market but cannot be directly observed. Such setting may be of interest for intraday trading when the 
market moves have stronger causality relations with impacts of individual large trades. While for this case variables $ {\bf x}_t $ correspond to the dollar 
values of positions in different stocks, they become total capitalizations of all firms in a market portfolio for inference of a market.      

\subsection{IRL for a market portfolio vs IRL for a single investor}
\label{sect:IRL_market_vs_single_investor}

Up to this point in the paper, our mathematical formulation for a single-investor and market portfolio was nearly uniform. 
In both cases, the optimal investment policy is given by Eq.(\ref{pi_post_M}), and in both cases, inference can be made using variational 
EM algorithm with a single-step variational free energy given by Eq.(\ref{F_var_final}). Now we come to differences between these two cases.

The first difference is in computational procedures for computing parameters
entering these equations. For a single investor case, if actions are unobserved, coefficients in Eqs.(\ref{pi_post_M}) and (.(\ref{F_var_final}) are time-dependent, and 
should be computed by a backward recursion starting from a terminal date $ t = T $, as described in Appendix A\footnote{A single investor case with unobserved actions may probably be less common than a scenario with observable actions, but the latter is a straightforward case as it does not need hidden variables at 
all, see Appendix B.}. For a market portfolio case, the problem is stationary, as there is no single unique horizon  $ T $ for planning in the market. 

This means that coefficients are now time-independent.  
The self-consistent set of Eqs.(\ref{F_opt}), (\ref{pi_opt_F}), (\ref{G_from_F}) for the stationary case reads
\bea
\label{F_opt_G_opt_stationary}
&& F^{\pi}({\bf y}_t ) =   \frac{1}{\beta} \log \sum_{{\bf a}_t} \pi_0 
( {\bf a}_t | {\bf y}_t) e^{ \beta G^{\pi} ( {\bf y}_t, {\bf a}_t)  }   \nonumber \\
&& G^{\pi} ( {\bf y}_t, {\bf a}_t) 
 = \hat{R}({\bf y}_{t}, {\bf a}_{t} ) 
 +  \gamma  \mathbb{E}_{t, {\bf a}} \left[  \left. F^{\pi} ( {\bf y}_{t+1}) \right|  {\bf y}_t, {\bf a}_t \right]  \\
&&   \pi ( {\bf a}_t | {\bf y}_t) = \pi_0 ( {\bf a}_t | {\bf y}_t) e^{ \beta \left(G^{\pi} ( {\bf y}_t, {\bf a}_t) - F^{\pi}({\bf y}_t ) 
  \right) }  \nonumber 
\eea
Computationally, this formulation amounts to solving the self-consistent system Eqs.(\ref{F_opt_G_opt_stationary}) as fixed point equations for time stationary G-function, F-function, and policy $ \pi_{\theta} $. 
In this setting, equations (\ref{F_coeffs}) become fixed point matrix equations, because now they relate matrix coefficients of the F-function 
(\ref{F_parametrization_2}) with themselves, rather than with their next-period values, as was the case in a finite-horizon specification. In the stationary setting, 
these equations can be used as update rules for parameters of the F-function by reading them from the right to the left, the same way as they are used in each step of a time-depending case.

A second major difference of IRL for the market portfolio from the single investor case is that while states $ {\bf y}_t $ are directly observable in this settings, actions  $ {\bf a}_t $ are {\it not}. They {\it might} be made observable in
a multi-agent version of the model, where the objective would be to model market-beating strategies, rather than just market-fitting strategies. However, in the inverse optimization IRL setting of this paper, we have only one agent representing a bounded-rational component of the market itself, thus it cannot trade stocks with other agents.

Therefore, its actions $ {\bf a}_t $ cannot be observed or interpreted as changes in {\it numbers} of stocks in the portfolio. 
Our agent does only a fictitious self-play of its trading decisions, but does not trade directly with any other counter-party. The only observable effects of actions of the agent are price changes resulting from heating the market via the trading impact mechanism.

We are now ready to formulate our final variational EM algoritm 
for inference of either an individual investor or a market optimal portfolio. A 
different and simpler algorithm for a special (and the most interesting) case of a market optimal portfolio will be presented in Sect.~\ref{sect:IRL_market}.

\subsection{Invisible Hand Inference with Free energy (IH-IF) algorithm}
\label{sect:IH-IF}

The complete IRL algorithm for learning the optimal policy of a bounded-rational agent (either a market-agent or a single investor) whose actions are unobservable that we call the Invisible Hand Inference with Free energy (IH-IF) is given by  Algorithm~\ref{Algorithm I}. 

 Our algorithm is a variational EM algorithm that amounts to iterative maximization of Eq.(\ref{F_var_final}). In the E-step, we maximize it with respect to variational parameters 
  $ \omega $ while keeping parameters $ \theta = (\lambda, \mu_i, 
\beta, {\bf W}, \Gamma, \Upsilon), \hat{ {\bf A}}_0, \, \hat{ {\bf A}}_1, \,  \Sigma_p $ and the G-function from the previous iteration. In the M-step, we maximize it with respect to generative model parameters $ \Theta $ and policy $ \pi_{\theta} $. The outputs of the M-step are updated values of parameters $ \theta $ and updated 
  values of parameters of G-function (\ref{G_dx_du}).  

In more details, at each iteration, we sample a new random mini-batch of $ N_b $ T-step trajectories $ ({\bf y}_1, \ldots, {\bf y}_{t+T} ) $. 
For the case of a market portfolio, we can take $ T = 1 $, so that a mini-batch has $ N_b $ one-step transitions  $ ({\bf y}_1, {\bf y}_{t+1} ) $.
For inference of a single large investor, $ T $ should be set to be a finite planning horizon of the investor. 

All subsequent calculations in a given iteration of the algorithm are done for this mini-batch. We define the free energy of a mini-batch as 
\beq
\label{F_minib}
 \mathcal{F}_b(\omega, \theta)  = \sum_{b=1}^{N_b} \sum_{t=0}^{T}  \mathcal{F}(\omega, \theta, t)
 \eeq
 where  $ \mathcal{F}(\omega, \theta, t) $ is defined in Eq.(\ref{F_var_final}), while here we add a third argument to emphasize the time dependence in observations.

In the E-step, we maximize $ \mathcal{F}_b(\omega, \theta) $ with respect to variational parameters $ \omega $.
In the M-step, we compute updates of parameters of the G-function, policy $ \pi_{\theta} $ as functions of $ \theta $, and then use these expressions to compute 
 $ \mathcal{F}_b(\omega, \theta) $  as a  function of $ \theta $. 
 
 This is done as follows.
 In step 1, the expectation of the next-time F-function is computed with Eq.(\ref{F_next_mat}) used as an update equation for parameters of the 
model, or within a backward recursion that starts with a fixed terminal condition at time $ t = T $, for IRL of an individual investor. 
In step 2, we compute the reward using Eq.(\ref{R_r_dx}). 
In step 3, an update of the Q-function is performed 
using Eq.(\ref{recursive_G}). The time-$t$ F-function is computed in step 4 using Eq.(\ref{F_coeffs}). 
Finally, in step 5, 
the optimal 
policy as a function of $ \theta $ is recomputed using Eq.(\ref{pi_post_M}).
Computing these quantities for all transitions in the mini-batch, we obtain the free energy (\ref{F_minib}) for the mini-batch. 
 This is used to produce an update of the current estimation of $ \theta $ using a 
learning rate  $ \alpha_{\theta} $. 
The new updated values of $ \theta $  are then used to update parameters  $ \hat{ {\bf A}}_1^{(k)}, \, \hat{ {\bf A}}_1^{(k)}, \,  \Sigma_p^{(k)} $ of the policy $ \pi_{\theta} $.
Then the algorithm proceeds to the next iteration.

\vskip0.5cm
\begin{algorithm}[H]
\label{Algorithm I}
    \SetAlgoLined
    \KwData{a sequence of states and signals}
    \KwResult{the reward function, optimal policy, and value function }
     Set  the learning rates $ \alpha_{\theta}, \, \alpha_{\omega} $, batch size $ N_b $, initial parameters $ \theta^{(0)}, \,  \omega^{(0)},  
     \hat{ {\bf A}}_0^{(0)}, \, \hat{ {\bf A}}_1^{(0)}, \,  \Sigma_p^{(0)} $  \\
    Set  $ k = 1 $ \\
    \While{not converged}{
    	Draw a new mini-batch of $ N_b $  $T$-step trajectories $ ( {\bf y}_t, \ldots, {\bf y}_{t+T} ) $ (can set $ T = 1 $ for a market portfolio) \\ 
    	{\it E-step}: \\
	 Compute the free energy $ \mathcal{F}_b(\omega, \theta^{(k-1)}) $  of the mini-batch using Eq.(\ref{F_minib}) \\
	Update recognition model parameters $ \omega^{(k)} =  (1-\alpha_{\omega})  \omega^{(k-1)} 
	+ \alpha_{\omega} \frac{\partial}{\partial \omega}  \mathcal{F}_b(\omega, \theta^{(k-1)}) $  \\
    	{\it M-step}:  Maximize $ \mathcal{F}_b(\omega^{(k)}, \theta) $ as a function of $ \theta $: \\
	\For{ \mbox{each transition} $ ( {\bf y}_t, {\bf y}_{t+1} ) $ ( for a single investor, take $ t = T-1, \ldots, 0 $)}{
	    	1. Compute the expected value at time $ t $ of the F-function at time $ t + 1 $. \\
		2. Compute the reward as a function of $ \theta $. \\
		3. Use steps 1 and 2 to update the Q-function at time $ t $ \\
		4. Compute the value of the F-function at time $ t $. \\
		5. Recompute the policy distribution  $ \pi_{\theta} ( {\bf a}_t| t, {\bf y}_t) $ as a function of $ \theta $ by updating its mean and variance.   \\     
        }
        Compute the free energy $ \mathcal{F}_b(\omega^{(k)}, \theta) $  of the mini-batch using Eq.(\ref{F_minib}) \\
        Update the parameter vector $ \theta^{(k)} = (1-\alpha_{\theta})  \theta^{(k-1)} + \alpha_{\theta} 
        \frac{\partial}{\partial \theta} \mathcal{F}_B(\omega^{(k)}, \theta)  $ \\
        Use the new value $ \theta^{(k)} $ to compute $  \hat{ {\bf A}}_1^{(k)}, \, \hat{ {\bf A}}_1^{(k)}, \,  \Sigma_p^{(k)} $ \\
        Increment $ k = k + 1 $
    }
\caption{The Invisible Hand Inference with the Free energy (IH-IF) variational EM IRL algorithm that learns the reward function, optimal policy and value function from a history of prices and signals, for either a market portfolio or a single investor.}
\end{algorithm}

\section{IRL for the market portfolio}
\label{sect:IRL_market}

When actions are unobserved or unobservable, the variational EM formulation (\ref{F_var_final}) provides a general and tractable 
algorithm to estimate the original model 
parameters $ \Theta $ from observed trajectories of stock capitalizations. The price one has to pay to solve the problem in this way is a need to specify a variational distribution with its own parameters $ \omega $, and estimate these parameters jointly with $ \Theta $ in a way specified by a variational EM algorithm. 

As we will show next, an alternative and simpler method of estimation model can be obtained simply by plugging Eq.(\ref{pi_post_M}) into the market return model 
(\ref{r_t}). To this end, we note that that once we obtained Eq.(\ref{pi_post_M}), we can 'forget' how it was derived using RL, IRL, neuroscience etc., and simply treat it as a model with free tunable parameters  $ \hat{ {\bf A}}_{0}, \, \hat{ {\bf A}}_{1} $  and $ \Sigma_p $. Substituting Eq.(\ref{pi_post_M}) into Eq.(\ref{r_t}) 
gives rise to a purely {\it econometric} model of 
market returns, which can be viewed (and estimated) as a model on its own. 
As will be shown below, this produces a model that predicts {\it mean reversion} in stock returns.

\subsection{Market dynamics: dynamically generated mean reversion}
\label{sect:GMR_dynamics}

Recall that for a vector of $ N $ stocks, we introduced a size $ 2 N $-action vector 
$ {\bf a}_t = [{\bf u}_t^{(+)}, {\bf u}_t^{(-)}] $, so that an action $ {\bf u}_t $ was defined as a difference of two non-negative numbers 
$ {\bf u}_t = {\bf u}_t^{(+)} -  {\bf u}_t^{(-)} = [{\bf 1}, - {\bf 1}] {\bf a}_t \equiv {\bf 1}_{-1}^{T} {\bf a}_t $.

Therefore, the joint distribution of $ {\bf a}_t = [{\bf u}_t^{(+)}, {\bf u}_t^{(-)} ] $ is given by our Gaussian policy
$  \pi_{\theta}({\bf a}_t |{\bf y}_t ) $. This means that the distribution of 
$ {\bf u}_t = {\bf u}_t^{(+)} -  {\bf u}_t^{(-)} $ is also Gaussian. Let us write it therefore as follows:
\beq
\label{pi_theta_u}
\pi_{\theta}({\bf u}_t |{\bf y}_t ) =   \mathcal{N}\left({\bf u}_t | \bf{U}_0 + \bf{U}_1 {\bf y}_t, \Sigma_u \right) 
\eeq
Here $ \bf{U}_0 = {\bf 1}_{-1}^{T}  \bf{A}_0 $ and $ \bf{U}_1 =  {\bf 1}_{-1}^{T}  \bf{A}_1 $.

Eq.(\ref{pi_theta_u}) means that $ {\bf u}_t $ is a Gaussian random variable that we can write as follows:
\beq
\label{u_SDE}
{\bf u}_t = \bf{U}_0 + \bf{U}_1 {\bf y}_t + \varepsilon_t^{(u)}  = \bf{U}_0 + \bf{U}_1^{(x)} {\bf x}_t + \bf{U}_1^{(z)} {\bf z}_t + \varepsilon_t^{(u)} 
\eeq
where $ \varepsilon_t^{(u)} \sim \mathcal{N}(0,\Sigma_u) $ is a Gaussian random noise.  

The most important feature of this expression that we need going forward is is linear dependence on the state $ {\bf x}_t $.  As can be seen in  
Eqs.(\ref{update_A_hat}) and (\ref{pi_post_M}), the variational EM algorithm developed above suggests that a coefficient of such dependence should be non-vanishing. 

This is the only result from the model developed in this paper that we will use in this section in order to construct a simple dynamic market model resulting from our approach. In order to end up with non-negative market prices in the model, we use a deterministic limit of Eq.(\ref{u_SDE}), where in addition we set $ \bf{U}_0 = \bf{U}_1^{(z)} = 0 $, and replace $ \bf{U}_1^{(x)} \rightarrow \phi  $
to simplify the notation. We thus obtain a simple deterministic policy
\beq
\label{determ_u}
{\bf u}_t =  \phi  {\bf x}_t 
\eeq
Next, let us recall Eqs.(\ref{h_t_1}) and (\ref{r_t}), which we repeat were with a substitution ${\bf W} \rightarrow {\bf w} $ and $ {\bf M} \rightarrow \mu $ :
\bea
\label{r_t_one_more}
&& {\bf x}_{t+1} = (1 + r_t) \circ ( {\bf x}_t + {\bf u}_t )  \nonumber \\
&& {\bf r}_t - r_f {\bf 1} = {\bf w} {\bf z}_t -  \mu {\bf u}_t + \varepsilon_t^{(r)} 
\eea
where $ r_f $ is a risk-free rate, $ {\bf z}_t $ is a vector of predictors with factor loading matrix  $ {\bf w} $, $ \mu $ is a matrix of permanent market impacts with a linear impact specification, and $ \varepsilon_t^{(r)} $ is a vector of residuals with 
$ \mathbb{E} \left[ \varepsilon_t^{(r)} \right] = 0 $ and 
$ \mbox{Var}_t \left[ \varepsilon_t^{(r)} \right] = \Sigma_r $.

In general case, the second equation in (\ref{r_t_one_more}) assumes a single vector of predictor $ {\bf z}_t $ for all stocks in a market portfolio.
If we have $ K $ individual predictors $ {\bf z}_{t}^{(i)} = [ z_{t1}^{(i)}, \ldots, z_{tK}^{(i)} ] $  for each stock $ i $,  we can stack them together as 
$ {\bf z}_t = [{\bf z}_{t}^{(1)}, \ldots, {\bf z}_{t}^{(N)} ]^T $, so that $ {\bf z}_t $ has length $ K N $. Respectively, matrix $ {\bf w} $ will have the size $ N \times K N $.
Each row $ i $ in this matrix will only have $ K $ non-zero elements in positions $ i, \ldots, i + K $ (so that to only include $ i$'s name predictors). 
This results in $ K N $ free parameters in matrix $ {\bf w} $. If desired or needed, the number of free parameters can be reduced if we enforce some symmetries, e.g. enforce a requirement that factor loadings for all names in a given sector should have the same value.  

Substituting Eq.(\ref{determ_u}) into Eqs.(\ref{r_t_one_more}) and simplifying, we obtain 
\beq
\label{GMR_0}
\Delta {\bf x}_t = \mu \circ \phi \circ ( 1 + \phi ) \circ {\bf x}_t \circ \left(  \frac{ \phi + (1 + \phi) (r_f + {\bf w} {\bf z}_t )}{ \mu \phi (1+ \phi)}  - {\bf x}_t \right) + ( 1 + \phi) \circ {\bf x}_t \circ \varepsilon_t^{(r)} 
\eeq
Introducing parameters 
\beq
\label{params}
\kappa \Delta t   =   \mu   \circ \phi \circ ( 1 + \phi ), \;  \theta ({\bf z}_t)  =     \frac{ \phi + (1 + \phi) (r_f + {\bf w} {\bf z}_t )}{ \mu \phi (1+ \phi)} , \; 
 \sigma ({\bf x}_t ) \sqrt{ \Delta t} =  ( 1 + \phi) \circ {\bf x}_t  
\eeq   
(here $ \Delta t $ is a time step) and replacing $ \varepsilon_t^{(r)} \rightarrow \varepsilon_t $, we can write Eq.(\ref{GMR_0}) more suggestively as 
\beq
\label{GMR}
\Delta {\bf x}_t = \kappa \circ  {\bf x}_t \circ 
\left( \theta ({\bf z}_t )  - {\bf x}_t \right)  \Delta t +  \sigma ({\bf x}_t) \sqrt{ \Delta t}   \circ \varepsilon_t
\eeq
In this equation, $ \circ $ stands for an element-wise (Hadamard) product. Note that this equation has a {\it quadratic} mean reversion. It is quite different from models with {\it linear} mean reversion such as the Ornstein-Uhlenbeck (OU) process. Eq.(\ref{GMR}) is the second main result of this paper.

%
 Equation (\ref{GMR}) describes mean reverting dynamics with a signal-driven mean reversion level $ \theta ({\bf z}_t ) $, and 
 a mean reversion speed $ \kappa $ proportional to market impact parameter vector $  {\mu} $. 
%
 It is easy to 
 see that in the limit of vanishing market impact $  {\mu}  \rightarrow 0, \, \phi \rightarrow 0 $, Eq.(\ref{GMR}) reduces to 
 the log-normal return model given by Eq.(\ref{r_t}) without the action term $ {\bf u}_t $:
 \beq
 \label{no_mu}
 \frac{\Delta {\bf x}_t}{ {\bf x}_t}  =  
 r_f  +    {\bf w}  {\bf z}_t    +  \varepsilon_t
\eeq
Therefore, the conventional log-normal return dynamics (with signals) is reproduced in our framework in the limit $  {\mu}  \rightarrow 0, \, \phi \rightarrow 0 $.
However, when parameters $  {\mu}, \,  \phi $ are small but non-zero,  Eqs. (\ref{no_mu}) and (\ref{GMR}) describe  {\it qualitatively different} dynamics.
While Eq.(\ref{no_mu}) is scale-invariant with respect to scale transformations $ {\bf x}_t \rightarrow \alpha {\bf x}_t $ with $ \alpha $ being a scaling parameter, the non-linear mean reverting dynamics (\ref{GMR}) are {\it not} scale invariant.

This is of course due to the fact that our market-wide agent aggregates all agents in the market.  As  their individual trade impacts induce 
a dependence of dynamics on a dimensional market impact parameter $ \mu $, scale invariance is broken in the resulting market dynamics (\ref{GMR}).

Therefore, even if parameters $ \kappa $, $ \phi $  are small but non-vanishing, Eq.(\ref{GMR}) produces a potentially highly complex non-linear dynamics with broken scale invariance and ensuing multi-period auto-correlations.  
   
 These non-linear dynamics with a {\it dynamically} generated mean reversion level $ \theta( {\bf z}_t)  $ are produced from simple linear dynamics 
 (\ref{r_t}) with a Linear-Quadratic-Gaussian (LQG) control $ {\bf u}_t $. 
 A peculiar feature of our model is that it has very clear origins for both the {\it level} and the {\it speed} of mean reversion.
 As can be seen from Eqs.(\ref{GMR}), the level $ \theta({\bf z}_t ) $ is driven by external signals $ {\bf z}_t $, which makes an intuitive sense.
 On the other hand, the {\it speed} of reverting to such 'target' price values is proportional to the market impact parameter vector 
 $ {\mu} $, that also intuitively makes sense.

It is important to note here that our model demonstrates some features that are typical for self-organizing systems, such as non-linear mean reversion effects, long-term correlations resulting from such mean reversion, and a dynamic adaptivity to external signals $ {\bf z}_t $. Therefore, our construction of self-learning by a fictitious self-play by an agent, that imitates simultaneously all traders in the market, provides a specific 
illustration of equivalence between self-organization and decision-making that was suggested in \cite{YS_2014}.

Another important comment has to do with time scales in the problem. There are a few of them in our model. First, we have a vector of external signals 
$ {\bf z}_t $. Each one of them has its own relaxation time $ \tau_{zk} $ where $ k = 1, \ldots, K $ is a number of signals. 

Assume for simplicity that we have only one scalar signal $ z_t $ with a characteristic relaxation time $ \tau_z \sim 1/\kappa_z $ where $ \kappa_z $ 
is the mean reversion speed of the signal. This can be compared with the characteristic relaxation time of the {\it system} $ \tau_x \sim 1/ \kappa $.
The setting of this paper implicitly assumes that $ \tau_x \leq \tau_z $, that is, $  \kappa \geq  \kappa_z $, so that the market is close to a non-equilibrium steady state, and it manages to digest a new information in signals $ {\bf z}_t $ at each step, and fully adjust market prices (at the price of the
 information cost $ g_t $, see Eq.(\ref{info_cost})).
 
 On the other hand, we might have a very different dynamics if $  \kappa \leq  \kappa_z $. In this case, the market would be in non-equilibrium transient state without a steady state. 
 Yet a different scenario may occur when a large jump in $ {\bf z}_t $ occurs at time $ t $ relative to its previous value (following e.g. a major financial, economic or political event), and then continues to fluctuate only mildly around a new level. In this case, the mean stock price level $\theta({\bf z}_t ) $
 that adjusted at time $ t $ to the {\it previous} value of the signals, becomes not the true dynamic optimum, but only a {\it metastable} state. 
 Further comments on such scenario will be given in Sect.~\ref{sect:Discussion}.

 
 In a one-dimensional (1D) case with a constant mean reversion level $  \theta({\bf z}_t )  = \theta $, Eq.(\ref{GMR}) produces the
  following dynamics for a re-scaled variable $ s_t = x_t/ \theta $: 
  \beq
 \label{logistic_noise}
 \Delta s_t = \mu s_t ( 1 - s_t) + \sigma  \sqrt{ \Delta t}  s_t \varepsilon_t, \; \; \mu \equiv \kappa \theta \Delta t 
  \eeq
 Dynamics described by Eq.(\ref{logistic_noise}) or its noiseless limit $ \sigma \rightarrow 0 $ are widely encountered or used in physics and biology. 
 In particular, the limit $ \sigma \rightarrow 0 $ of Eq.(\ref{logistic_noise}) describes the logistic map dynamics, that arises e.g. 
 in the Malthus-Verhulst model of population growth (see e.g. \cite{vanKampen}), or in Feigenbaum bifurcations in the logistic map chaos, 
 that arise when $ 3 \leq \mu < 4 $ in Eq.(\ref{logistic_noise}), see e.g. \cite{Sternberg}. When $ \sigma > 0 $, Eq.(\ref{logistic_noise}) describes a logistic map with a multiplicative thermal noise, which may produce highly complex dynamics \cite{Baldovin}. 
 
 We can also consider a continuous-time limit of 1D dynamics implied by Eq.(\ref{GMR}):
 \beq
 \label{GMR_1D}
 d x_t = \kappa   x_t \left( \theta   -  x_t \right) dt +  \sigma   x_t  \, d W_t
\eeq
 where $ W_t $ is a standard Brownian motion. This 1D process is known in the  economics and finance literature as a Geometric Mean 
 Reversion (GMR) process. Equivalently, we can introduce a scaled variable $ s_t = \kappa x_t $, for which we obtain
 \beq
 \label{GMR_physics}
 d s_t = \left( \lambda_t s_t - s_t^2 \right) dt + \sigma s_t \, dW_t , \; \; \lambda_t \equiv \kappa \theta_t 
 \eeq
 which is a form mostly used in physics literature \cite{Horsthemke}.  As discussed in \cite{Horsthemke}, if we keep parameter $ \lambda_t \equiv \kappa \theta_t 
 $ constant in time, i.e. $ \lambda_t  \rightarrow \lambda $ and look at the behavior of the system in the limit $ \sigma \rightarrow 0 $, the system exhibits a second-order phase transition at $ \lambda = 0 $.
 
When $ \sigma >  0 $ while $ \theta_t = \theta $ is kept fixed, Eq.(\ref{GMR_physics}) has one or two transition points corresponding to two extrema of its 
stationary distribution:
\beq
\label{extrema_GMR}
s_1 = 0, \; \; s_2 = \kappa \theta - \nu \frac{\sigma^2}{2}
\eeq
where $ \nu = 2 $ and $ \nu = 1 $ for the Ito and Stratonovich interpretation of SDE (\ref{GMR_physics}), respectively. The second transition point exists only if 
$    \kappa \theta >  \nu \frac{\sigma^2}{2} $. When this constraint is satisfied, the system (\ref{GMR_physics}) undergoes a noise-induced transition 
\cite{Horsthemke}.
 
 We can produce a few equivalent descriptions of the dynamics described by Eq.(\ref{GMR_1D}) by using changes of variable in this equation. In particular, if we define $ s_t = 1/x_t $, then the stochastic differential equation for $ s_t $ using Ito's prescription reads
 \beq
 \label{1_over_x}
 d s_t = \left( \kappa - (\kappa \theta - \sigma^2) s_t \right) dt - \sigma s_t  d W_t 
 \eeq
 where now the drift becomes linear in the transformed variable $ s_t = 1/x_t $. 
 
 Another useful form is obtained if instead we define  $ s_t = \log x_t/c $ where
 $ c > 0 $ is a fixed number having dimension of the currency of the market portfolio (e.g. the USD) that we need to introduce on the grounds of dimensionality analysis. For example, we can choose $ c = \langle x \rangle $ to be a time-average value of $ x_t $ within an observation period.
 Using Ito's prescription with this choice of $ c $, the SDE for $ s_t = \log x_t/ \langle x \rangle $ reads
 \beq
 \label{log_x_eq}
 d s_t  = \kappa \left( \theta - \frac{\sigma^2}{2 \kappa} -  \langle x \rangle e^{s_t} \right) dt + \sigma d W_t 
 \eeq
 Note that with this form, the noise becomes additive rather than multiplicative as in Eqs.(\ref{GMR_1D})  or (\ref{1_over_x}). On the other hand, the drift becomes exponential. It is easy to see that Eq.(\ref{log_x_eq}) requires the condition $ 2 \kappa \theta > \sigma^2 $ in order for Eq.(\ref{log_x_eq}) to have a stationary distribution.  
 
 Note that because $ x_t $ is a total market capitalization of a firm (or all firms in the index, depending on how we use the 1D setting here), $ \log x_t $ will be given by a log-stock price plus a log of total number of shares outstanding. When the latter is constant, $ s_t   = \log x_t/c $ is equal to the log-price of the stock plus a constant term.
   
The GMR model (\ref{GMR_1D}) was used by Dixit and Pindyck \cite{Dixit}, and its properties were further studied by Ewald and Yang \cite{EY} who have shown that this process is bounded, non-negative, and has a stationary distribution  under the constraint $ 2 \kappa \theta > \sigma^2 $. 
 Rather than introducing such mean-reverting dynamics phenomenologically, our model {\it derives} them (in a multi-variate setting) from an underlying dynamic optimization problem of a bounded-rational agent. 
 
 The non-stationary multivariate Geometric Mean reverting process (\ref{GMR}) can be interpreted as either an equilibrium or quasi-equilibrium statistical process (which is the case usually assumed in econometric and financial models), or as an non-equilibrium Langevin  process \cite{vanKampen}. In the rest of this section, we assume the former setting, while some further comments on the latter case will be provided in Sect.~\ref{sect:crashes}.
 
\subsection{IRL by Maximum Likelihood: market portfolio}
\label{sect:IRL_market_portfolio}

Here we assume a quasi-equilibrium setting when the market manages to attain an equilibrium distribution (\ref{pi_post_M}) in each period, following changes
in signals $ {\bf z}_t $. In this case, standard statistical methods, such as Maximum Likelihood, can be applied to estimate the model.  
The negative log-likelihood function for observable data with this model reads
\beq
\label{log_like_market}
LL_M (\Theta) = - \log \prod_{t=0}^{T-1} 
\frac{1}{ \sqrt{ (2 \pi)^{N}  \left| \Sigma_x \right| }} 
e^{ - \frac{1}{2} \left(   {\bf v}_t
 \right)^{T} 
\Sigma_x^{-1}  
\left(  {\bf v}_t \right)} , \; \;  {\bf v}_t \equiv \frac{{\bf x}_{t+1} -  {\bf x}_{t}}{{\bf x}_{t}}  
-  \kappa \circ \left( \theta( {\bf z}_t)   - {\bf x}_t \right)  \Delta t 
\eeq
where $ {\bf x}_t $  now stands for observed stock market prices, and $ \Sigma_x = \sqrt{\Delta t} \Sigma_r $. Note that because the model is Markov, the product over $ t = 0, \ldots, T-1 $ does not 
necessarily mean a product of transitions along the same trajectory, but can be viewed as a product of $ T $ one-step transitions that do not correspond to 
consecutive time moments. 

Parameters that can be estimated from data are therefore the vector of mean reversion speed 
parameters $ \kappa $, factor loading matrix $ {\bf w} $, and covariance matrix $ \Sigma_x $. 

Note that instead of defining the likelihood in terms of the original variables $ {\bf x}_t $, we could defined it instead in terms of a transformed variable $ {\bf s}_t =  \log {\bf x}_t/ \langle x \rangle $. The negative log-likelihood, when re-expressed in terms of the original observables $ {\bf x}_t $ would then be of the same Gaussian form as in Eq.(\ref{log_like_market}) where the variable $ {\bf v}_t $ would be defined as 
\beq
\label{v_t_1}
{\bf v}_t = \log \frac{{\bf x}_{t+1}}{ {\bf x}_t } - \kappa  \circ  \left(  \theta( {\bf z}_t) - \frac{\sigma^2}{2 \kappa}  - {\bf x}_t \right) \Delta t 
\eeq

\section{Experiments}
\label{sect:Experiments}

In this section we describe our experiments with the market model Eq.(\ref{GMR}). Further details for calibrated model parameters are provided 
in Appendix C.

To show detailed results, we use the DJI index instead of the S\&P500 index that is more commonly used as a market portfolio.
We analyze the daily data on market caps of all firms in the DJI index from 2010 to the end of 2017. We use the current composition of DJI that includes Apple that was added in 2016.
 We re-scale all data points by dividing by the average total market cap of the index for the whole period, which is approximately equal to \$160Bn for our dataset.

Similar to \cite{Boyd_2017}, our approach takes signals $ {\bf z}_t $ as given, and assumes that they are obtained through 
a search for 'alpha' that is beyond the scope of our framework.
Calibrated model parameters will necessarily depend on the choice of predictors $ {\bf z}_t $. One of our objectives here is to 
illustrate such dependence on the choice of signals.

To this end, we test our model using two sets of experiments with two different sets of predictors  $ {\bf z}_t $. We build both sets as predictors of market caps (or equivalently prices) rather than predictors of returns. 

The first set of predictors includes two predictors for each stock: a perfect signal and a random signal. 
The perfect (oracle) signal is obtained as a (demeaned) realized next-day return. This test can serve as a sanity/implementation test for the model. It is expected to provide a stable calibration of parameters, nearly zero volatility, and an order of magnitude of difference between estimated weights of the perfect signal and the random signal. The results are as expected, see
tables \ref{Tab:kappa_perfect} and \ref{Tab:sigma_perfect} in Appendix C where we show calibrated parameters for separate annual runs (we do not report weights to save space).

The second set of predictors are given by a pair of demeaned exponential moving averages of the (re-scaled) market caps. The two signals use parameters $ \gamma = 0.9 $ and $ \gamma = 0.96 $ of the exponential moving averaging, corresponding to the lookback windows of 7 days and 15 days, respectively.

In both sets of experiment, we estimate the resulting model parameters by minimizing the negative log-likelihood (\ref{log_like_market}) subject to constraints of non-negativity of weights $ w_1, \, w_2 $ of two predictors, and adding a regularization term $ \lambda (w_1 + w_2 -1)^2 $. While the results are only weakly dependent on the value of regularization parameter in the range $ \lambda \sim 10^{-3} - 10^{-2} $, we report the results for the value of $ \lambda = 10^{-2} $. 
The covariance matrix $ \Sigma_x $ is taken to be diagonal $ \Sigma_x = \mbox{diag}(\sigma_i^2) $. We set $ \Delta t = 1 $, thus we report daily rather than annualized values of  $ \kappa $ and $ \sigma^2 $.

Calibrated parameters $ \kappa $ and $ \sigma^2 $ for the exponential moving average signals are shown in tables \ref{Tab:kappa_exp} and 
\ref{Tab:sigma_exp} in Appendix C. As could be expected, the resulting parameters are substantially different from those obtained with the first set of signals.
Calibrated parameters are less stable, which is unsurprising given that moving averages are not very good predictors of future prices. In particular, we observe occasional negative values of $ \kappa $ that suggest a local divergence from a predicted value, rather than a convergence to this value. In Figs. 1, 2, 3 we show the market cap vs a fitted mean level 
for the IBM, JPM and XOM stocks for a two-month period in 2017. Results obtained for other stocks and other periods are similar.

Note that it would be wrong to try to estimate parameter $ \kappa $ by simply running a regression of $ \Delta x_t $ on $ x_t $ and treating the signals $ {\bf z}_t $  as a part of a noise term in such regression. As $ {\bf z}_t $ is a random process itself, such procedure would violate the {\it i.i.d.} assumption for a noise term in 
such regression.

\begin{figure}[ht]
\begin{center}
\includegraphics[scale=0.4]{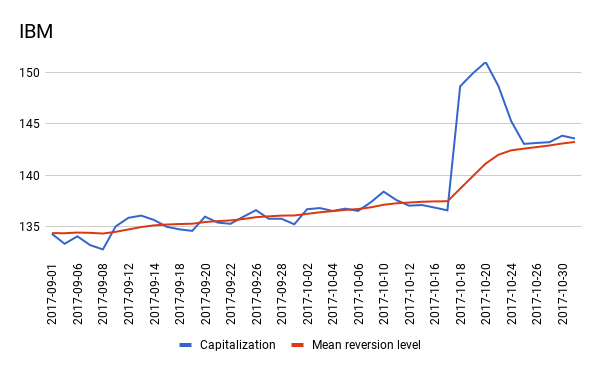}
\label{fig:mv_IBM}
\caption{Market cap vs estimated mean level: IBM}
\end{center}
\end{figure}

\begin{figure}[ht]
\begin{center}
\includegraphics[scale=0.4]{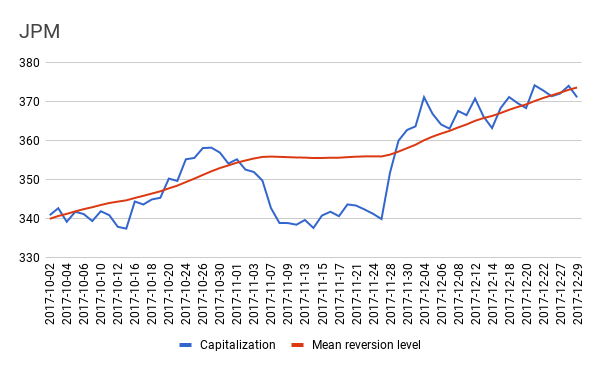}
\label{fig:mv_JPM}
\caption{Market cap vs estimated mean level: JPM}
\end{center}
\end{figure}

\begin{figure}[ht]
\begin{center}
\includegraphics[scale=0.4]{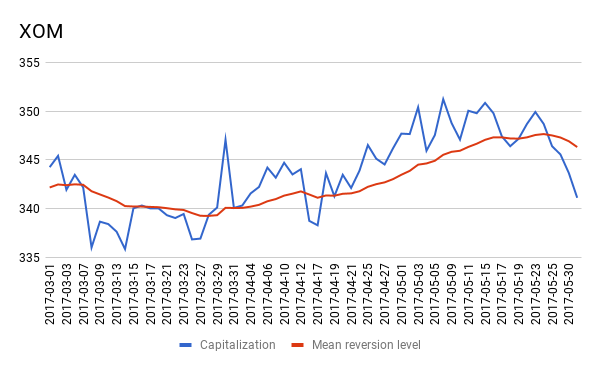}
\label{fig:mv_XOM}
\caption{Market cap vs estimated mean level: XOM}
\end{center}
\end{figure}

\section{Discussion and future directions}
\label{sect:Discussion}

\subsection{Mean reversion in asset returns}

One of the most interesting implications of our model is its prediction of a non-linear mean-reverting behavior of asset returns.
While mean reversion in intraday data for stock markets is a well established fact, its presence in longer-horizon returns is a topic of a long discussion in the literature. The latter started with Poterba and Summers who argued for mean reversion in stock returns as resulting from actions of 'noise traders' that do not have any objectives in trading, i.e. have zero intelligence \cite{PS_1988}.  Implications of mean reversion for a long-term optimal asset management were discussed in \cite{SB_2012}. 

In our model, mean reversion in asset returns has a very transparent origin. It results from a total market impact from traders that optimize their investment portfolios following mean-variance Markowitz-type optimization strategies by adapting to changing signals and changes in the market.  
The resulting stock price dynamics are of {\it non-linear mean reversion} type (a multi-variate Geometric Mean Reversion process with external factors), even though we start with a simple Gaussian policy $ \pi $ for the agent in our model. Non-linearity of the dynamics in our model is both a manifestation and a result of a feedback loop via the price impact mechanism.

Interestingly, the {\it dynamically} generated mean reversion in our model produces time-decaying auto-correlations in the system, i.e. multi-period effects that were absent in the original formulation\footnote{In particular, our model did not originally include any "permanent impact" effect which may not {\it a priori} be an well-defined notion in an MDP setting.}. Note that presence of slowly time-decaying auto-correlations and adaptivity to external signals are typical for self-organizing systems, see e.g. \cite{YS_2014}. Therefore, our model demonstrates some features of a self-organizing behavior by the dynamic generation of a mean reversion level for stocks. 


\subsection{Non-equilibrium behavior and market crashes}
\label{sect:crashes}

As was discussed in Sect.~\ref{sect:GMR_dynamics}, our setting above assumes that changes of external signals are slow enough, so that market
has sufficient time to adjust to new information in signals $ {\bf z}_t $ from one period to another. If external signals were just constant in time, the 
system would eventually settle in a stationary equilibrium state.

A different situation can occur if signals $ {\bf z}_t $ exhibits a large jump at time $ t $ relative to their value at time $ t - 1 $. In this case, the system can find itself trapped in a meta-stable state - a previously globally optimal state that becomes a local minimum following a jump of $ {\bf z}_t $ to a new value.
A meta-stability, rather than stability of this state is ensured by the presence of a potential barrier separating the global and local minima. 
A transition from a metastable state to a new dynamically optimal stable state would be activated by noise $ \varepsilon_t^{(x)} $, see e.g. \cite{vanKampen} on how such transitions are modeled in physics. 
In the financial setting, decays of such meta-stable states via a thermally-activated diffusion can describe market crashes. Such transitions can be studied either numerically using simulations, or theoretically using methods of \cite{vanKampen}. Non-equilibrium phase transitions induced by multiplicative 
noise as in Eq.(\ref{logistic_noise}) were studied in \cite{Van_Broeck_1997}.  Statistical physics of driven non-equilibrium dynamics of system with thermal fluctuations is studied in \cite{England_2016}. 

\subsection{Multi-agent formulations: market-fitting or market-beating strategies?}

As the objective of this paper was to make inference of a Bounded-Rational 'Invisible Hand' that drives the market as a whole using Inverse Reinforcement Learning, we used a single agent setting. In our formulation, this single agent self-learns by self-playing. As we showed in this paper, this formulation, though may appear somewhat abstract or even 'theological', gives rise to quite specific observable and computable consequences such as the prediction of mean reversion in asset returns, implied rationality and risks aversion parameters, and a market-implied optimal strategy.

On the other hand, it would be interesting to extend the setting of this model to a multi-agent formulation. On-line multi-agent Reinforcement Learning, where 
two or more bounded-rational agents implement Markowitz-like, or possibly more advanced investment 
strategies in a noisy market environment with external signals, can create potentially rich market-{\it beating} strategies.

\subsection{Implied rationality of the market}

Recall that the inverse temperature parameter $ \beta $ controls the degree of rationality of the RL agent that dynamically replicates the market portfolio by minimizing its trading cost. We have showed the result of calibration of the market model (\ref{GMR}) implied by our framework. In this setting, the original 
model parameters are embedded in parameters defining Eq.(\ref{GMR}), see  Eqs.(\ref{params}). The latter parameters are calibrated to market data.

To infer the original parameters of the model including  $ \beta $, one can instead use the IH-IF algorithm from Sect.~\ref{sect:IH-IF}. Inference of market-implied 
rationality parameter $ \beta $ and risk aversion $ \lambda $ will be addressed elsewhere.

\subsection{The market as an information perception-action system}

Analysis of the RL agent representing a coherent bounded-rational component of the market that we developed above included analysis of information costs of actions.
This analysis can be extended by including information costs of information {\it extraction} \cite{Tishby_2012, Ortega_2015, Genewein_2015, Still_2017}. 

The value of this extension is in its focus on the external signals $ {\bf z}_t $. In our model, we took them as given, effectively leaving the information 
costs of their {\it extraction} outside of the scope of the model. Analysis along the lines of \cite{Tishby_2012, Ortega_2015, Genewein_2015, Still_2017}
allows one to assess the value of these signals for the {\it full} perception-action cycle. Note that traditionally, signals are accessed based on their ability to predict the future, e.g. their own future. 

However, this is not the same as the ultimate goal of these signals, which is to improve rewards.
A perception-action cycle analysis in \cite{Tishby_2012, Ortega_2015, Genewein_2015} specifies {\it useful} information in signals, as opposed to {\it useless} information that should be discarded as its use amounts to a dissipated energy (heat) instead of an increase of the free energy. 
Extensions of the model developed in this paper along these lines of analysis of the perception-action cycle of financial markets will be presented elsewhere.


\section{Summary}
\label{sect:Summary}

As was discussed e.g. by Sornette in \cite{Sornette_book}, economic models  differ from models in the physical sciences in that economic agents anticipate the future and act accordingly, thus impacting the present. A value in finance depends on views of market participants on the future. This is very different from 
physics where quantities such as e.g. the mass of a proton are clearly independent of public views on the future. Such observations led many researchers to suggest that ideas from biology and genetics can be useful for financial modeling \cite{Sornette_book}.

As we discussed in Sect.~\ref{sect:Related_work}, our model shares a number of similarities with models in biology, e.g. \cite{Friston_2010}, \cite{Friston_2018}. Our bounded-rational market-wide agent aggregates all traders in the market who anticipate the future in their trading decisions. Optimal actions of the agent are those that maximize its free energy, similar to models of \cite{Friston_2010}, \cite{Friston_2018}.

Our model provides a computational scheme based on Inverse Reinforcement Learning and the variational EM algorithm to infer parameters of the model.
As in our model the market-wide agent that implements the 'Invisible Hand' is a {\it sum} of all agents, it provides a unifying framework for inference of either a market portfolio or a single investor. Furthermore, for the most interesting case of a dynamic inference of the market portfolio, our model provides a multi-period extension of the Black-Litterman model \cite{BL}. 
Finally, our approach suggests a non-stationary multivariate Geometric Mean Reversion (GMR) process (\ref{GMR}) as a model for 
market dynamics.

\def\thesection{A}	
\setcounter{equation}{0}
\def\theequation{\thesection.\arabic{equation}}

\section*{Appendix A: Optimal action and optimal G-function with locally-quadratic expansion}
\label{sect:Appendix_A}

\subsection{Linearization of dynamics}

Here we develop a tractable computational scheme based on conditioning on  the linearization variables $ \bar{\bf a}_t, \;  \bar{\bf y}_t  $, and 
expanding the dynamics and functions of interest (the G-function and the action policy $ \pi_{\theta} $) in Taylor series in small deviations from these values.

In this section we use the symbols $  {\bar {\bf a}}_t $, $  {\bar {\bf y}}_t $ as fixed conditioning values in 
calculation of conditional variational free energy  (\ref{cond_F_a}), or equivalently as {\it realizations} of random hidden variables 
$  {\bar {\bf a}}_t $, $  {\bar {\bf y}}_t $. Note that when these values are fixed, we also have fixed values of a related pair 
$ ( {\bar {\bf u}}_t,  {\bar {\bf x}}_t)  \equiv  ( {\bf 1}_{-1}^T {\bar {\bf a}}_t,  {\bf 1}_{0}^T {\bar {\bf y}}_t) $, where 
$ {\bf 1}_{0} = [ {\bf 1}, {\bf 0} ]^{T} $ and $ {\bf 1}_{-1} = [ {\bf 1}, - {\bf 1} ]^{T} $. 

Let us come back to Eq.(\ref{x_t_1}) that shows that the dynamics are non-linear in controls $ {\bf u}_t $ and the state vector $ {\bf y}_t $. 
Define deviations $ \delta {\bf x}_t $ and $ \delta {\bf u}_t $ from linearization points in the $ ({\bf x}, {\bf u}) $ space:
\beq
\label{x_hat_u_hat}
{\bf x}_t  = {\bar {\bf x}}_t  + \delta {\bf x}_t , \;  {\bf u}_t   = {\bar {\bf u}}_t +  \delta {\bf u}_t
\eeq
  
We linearize the dynamics equation (\ref{x_t_1}) by keeping linear terms in deviations $ \delta {\bf x}_t, \,  \delta {\bf u}_t $.
This yields
\beq
\label{delta_x_t}
 \delta {\bf x}_{t+1} = \Omega_0 + \Omega_x \delta {\bf x}_{t} +\Omega_u \delta {\bf u}_{t}  +   \Omega_z  \delta {\bf z}_{t}  
 + \varepsilon_t  \circ \left( 
  { \bf x}_t +   {\bf u}_t  \right)
\eeq
where
\bea
\label{Omega_coeffs}
&& \Omega_0 = \left( 1 + r_f + \mbox{diag} \left(   {\bf W} \bar{{\bf z}}_t -   {\bf M} \bar{{\bf u}}_t  \right) \right) (  \bar{{\bf x}}_t +  \bar{{\bf u}}_t ) 
-   \bar{{\bf x}}_{t+1}  \nonumber \\
&& \Omega_x  = 1 + r_f + \mbox{diag} \left(  {\bf W} \bar{{\bf z}}_t -  {\bf M}  \bar{{\bf u}}_t  \right) \\
&& \Omega_u  = 1 + r_f + \mbox{diag} \left(   {\bf W} \bar{{\bf z}}_t -  {\bf M}  \bar{{\bf u}}_t  \right) 
-    ( \bar{{\bf x}}_t +  \bar{{\bf u}}_t ) \circ {\bf M}  
\nonumber \\
&& \Omega_z  =   ( \bar{{\bf x}}_t +  \bar{{\bf u}}_t ) \circ {\bf W}  \nonumber 
\eea
Here $  ( \bar{{\bf x}}_t +  \bar{{\bf u}}_t ) \circ {\bf M}  $ stands for an element-wise multiplication of a $ k $-th component of vector 
$  ( \bar{{\bf x}}_t +  \bar{{\bf u}}_t ) $ with a $k$-th row of matrix $ {\bf M} $, and a similar convention is used in the last relation.

Deviations can also be defined for the extended space  (\ref{y_t}). In this case, we expand around conditioning values  $ {\bar {\bf a}}_t, {\bar {\bf y}}_t $ 
in a similar way to Eq.(\ref{x_hat_u_hat}):
\beq
\label{y_hat_a_hat}
{\bf y}_t  = {\bar {\bf y}}_t  + \delta {\bf y}_t, \; \; 
{\bf a}_t    = {\bar {\bf a}}_t +  \delta {\bf a}_t
\eeq
so that linearization points in Eqs.(\ref{x_hat_u_hat}) and (\ref{y_hat_a_hat}) are related as follows: 
$ ( {\bar {\bf u}}_t,  {\bar {\bf x}}_t)  \equiv  ( {\bf 1}_{-1}^T {\bar {\bf a}}_t,  {\bf 1}_{0}^T {\bar {\bf y}}_t) $.

Stacking Eq.(\ref{delta_x_t}) and Eq.(\ref{OU_z}) written in terms of the increment $ \delta {\bf z}_t $ together, we can write a linearized equation for 
$ \delta {\bf y}_t $ as follows:
\beq
\label{x_t_1_hat}
 \delta {\bf y}_{t+1} = \Psi_0 + \Psi_y \delta {\bf y}_{t} +\Psi_a \delta {\bf a}_{t}  +   \varepsilon_t^{y}  \left( 
 \delta { \bf y}_t, \delta {\bf a}_t   \right)
\eeq 
where 
\bea
\label{hat_A_hat_B}
&& \Psi_0 = 
\left[ \begin{array}{l}
 \Omega_0 \\
  \left( {\bf I} - \Phi \right) \circ  \bar{{\bf z}}_t -  \bar{{\bf z}}_{t+1} 
 \end{array} \right] , \; 
\Psi_y = 
 \left[ \begin{array}{ll}
 \Omega_{x} & \Omega_{z} \\
 0 &  {\bf I} - \Phi 
 \end{array} \right]
 , \; 
 \Psi_a = 
 \left[ \begin{array}{c}
  \Omega_{u} {\bf 1}_{-1}^T \\
 0
 \end{array} \right]
\\
 && 
\varepsilon_t^{y}  \left(
 \delta { \bf y}_t, \delta {\bf a}_t   \right)
 = 
  \left[ \begin{array}{l}
 \varepsilon_{t}  \circ \left( {\bf x}_t + {\bf u}_t     
  \right)  \\
  \varepsilon_{t}^{z} 
 \end{array} \right] 
 = 
  \left[ \begin{array}{l}
 \varepsilon_{t}  \circ \left(  {\bf 1}_{0}^{T} {\bf y}_t +  {\bf 1}_{-1}^{T} {\bf a}_t \right)  \\
\varepsilon_{t}^{z}  
 \end{array} \right]  
 \nonumber 
 \eea 
Note that matrices $  \Psi_0,  \Psi_y, \,  \Psi_a $ implicitly depend on time via their dependence of $  \bar{{\bf y}}_t $ and $ \bar{{\bf a}}_t $.     
 Also note that Eq.(\ref{x_t_1_hat}) implies that
 \bea
 \label{mean_dx}
&&  \widehat{ \delta {\bf y}}_{t+1} \equiv \mathbb{E}_{t,a} \left[ \delta {\bf y}_{t+1} \right] = 
 \Psi_0 + \Psi_y \delta {\bf y}_{t} +\Psi_a \delta {\bf a}_{t} \nonumber \\
 &&  \Sigma_y \equiv \Cov \left[ \delta {\bf y}_{t+1}  \right] = 
 \left[ \begin{array}{ll}
 \Sigma_{xx}  &  0 \\
  0 &    \Sigma_z 
 \end{array} \right]  \\
&&  \Sigma_{xx}  = \Sigma_{r} \circ 
 \left[ \left(  {\bf 1}_{0}^{T} {\bf y}_t +  {\bf 1}_{-1}^{T} {\bf a}_t  \right) 
   \left(  {\bf 1}_{0}^{T} {\bf y}_t +  {\bf 1}_{-1}^{T} {\bf a}_t   \right)^T \right]  \nonumber
 \eea
We can also express the reward Eq.(\ref{R_t_y}) in terms of $  \delta {\bf y}_t  $ and $ \delta {\bf a}_t $:
 \beq
 \label{R_r_dx}
 \hat{R}_t({\bf y}_t, {\bf a}_t) 
 = 
  \delta  {\bf a}_t^T \hat{{\bf R}}_{aa}  \delta {\bf a}_t +  \delta {\bf y}_t^T \hat{{\bf R}}_{yy}  \delta {\bf y}_t  + 
   \delta {\bf a}_t^T  \hat{{\bf R}}_{ay}  \delta {\bf y}_t +  \delta  {\bf a}_t^T \hat{{\bf R}}_{a} + 
  \delta  {\bf y}_t^{T} \hat{{\bf R}}_{y} + r( \bar{{\bf y}}_t, \bar{{\bf a}}_t )  
\eeq
Here we defined
\bea
\label{R_hat}
 && \hat{{\bf R}}_{aa}  = {\bf R}_{aa}, \;  \hat{{\bf R}}_{yy}  = {\bf R}_{yy}, \; 
 \hat{{\bf R}}_{ay}  =  {\bf R}_{ay} ,  \nonumber \\
 && \hat{{\bf R}}_{a} =  {\bf R}_{a} + 2  {\bf R}_{aa} \bar{{\bf a}}_t +  {\bf R}_{ay}  \bar{{\bf y}}_t,  \; 
  \hat{{\bf R}}_{y} =   2  {\bf R}_{yy} \bar{{\bf y}}_t +  {\bf R}_{ay}^{T}  \bar{{\bf a}}_t  \\
 &&   r ( \bar{{\bf y}}_t, \bar{{\bf a}}_t ) = \bar{{\bf a}}_t^T  {\bf R}_{aa} \bar{{\bf a}}_t + \bar{{\bf y}}_t^T {\bf R}_{yy} \bar{{\bf y}}_t  + 
  \bar{{\bf a}}_t^T  {\bf R}_{ay} \bar{{\bf y}}_t +  \bar{{\bf a}}_t^T {\bf R}_{a} \nonumber 
  \eea
Recall that as the original parameters of the reward function coefficients $ {\bf R}_{aa} $ etc. were defined in terms of the original model parameters, the new 'hat' coefficients $  \hat{{\bf R}}_{aa} $ etc. are now functions of the original model parameters and conditioning variables $ \bar{{\bf y}}_t , \, \bar{{\bf a}}_t $.

 \subsection{Recursion for the G-function}
 \label{sect:G_recursion}
 
In this section, we consider a finite-horizon setting. In this case, a time dependence of coefficients will be implicit in equations to follow, and will be supplemented
by an additional upper script, e.g. $  {\bf F}_{yy}^{(t)} $, where needed for clarity.
 
For a finite-horizon setting with a planning horizon $ T $, as positions $ {\bf x}_T $ are fixed by (\ref{u_T}), we can use Eqs.(\ref{terminal_G_F}) 
 and (\ref{R_r_dx}) to get
\beq
 \label{F_T_delta_x_}
 F_T^{\pi}({\bf y}_T) =  \hat{R}_{T}( {\bar {\bf y}}_T  + \delta {\bf y}_T ,  {\bar {\bf a}}_T +  \delta {\bf a}_T  ) 
 \eeq    
 We use this to fix  $  {\bf F}_{yy}, \,  {\bf F}_{y}, \, F_0 (\bar{{\bf y}}_t)  $ 
 in Eq.(\ref{F_parametrization}) in terms of coefficient of reward function (\ref{R_r_dx}):
 \bea
 \label{D_T_H_T}
 && {\bf F}_{yy}^{(T)} = \hat{{\bf R}}_{yy}  = {\bf R}_{yy} \\
 &&  {\bf F}_{y}^{(T)} =   \hat{{\bf R}}_{ay}^{T} \delta {\bf a}_T +   \hat{{\bf R}}_{y} =  {\bf R}_{ay}^{T} \left( \bar{{\bf a}}_T +  \delta {\bf a}_T  \right) 
 + 2  {\bf R}_{yy} \bar{{\bf y}}_T    \nonumber \\
 && F_0 (\bar{{\bf y}}_T, \bar{{\bf a}}_T) =  \delta {\bf a}_T \hat{{\bf R}}_{aa}  \delta {\bf a}_T
 +  \delta {\bf a}_T^{T} \hat{{\bf R}}_{a} +    r ( \bar{{\bf y}}_T, \bar{{\bf a}}_T )  \nonumber 
 \eea

 
For values  $ t = T-1, \ldots, 0 $,  we use Eqs.(\ref{x_t_1_hat})  and (\ref{mean_dx}) to compute the conditional 
expectation of the next-period F-function as follows:
\beq
\label{F_next}
\mathbb{E}_{t, {\bf a}} \left[  F_{t+1}^{\pi}({\bf y}_{t+1})  \right] 
=
   F_0  (\bar{{\bf y}}_{t+1}, \bar{{\bf a}}_{t+1})
 + \widehat{ \delta {\bf y}}_{t+1}^T  {\bf F}_{y}^{(t+1)} + 
\widehat{ \delta {\bf y}}_{t+1}^T {\bf F}_{yy}^{(t+1)}  \widehat{  \delta {\bf y}}_{t+1}
+ \mbox{Tr} \left[ {\bf F}_{yy}^{(t+1)} \Sigma_y  \right]
\eeq
The last term can be expressed in a more convenient form using Eq.(\ref{F_yyt}):
\bea
\label{last_term}
\mbox{Tr} \left[ {\bf F}_{yy}^{(t+1)} \Sigma_y  \right] 
\hskip-0.5cm && = \mbox{Tr}\left[ \left( \left(  {\bf 1}_{0}^{T} {\bf y}_t +  {\bf 1}_{-1}^{T} {\bf a}_t 
 \right) \left(  {\bf 1}_{0}^{T} {\bf y}_t +  {\bf 1}_{-1}^{T} {\bf a}_t  \right)^T \right) 
\left( {\bf F}_{xx} \circ \Sigma_r \right) \right] + \mbox{Tr} \left[ {\bf F}_{zz} \Sigma_{z} \right] \nonumber \\
&& = \left(  {\bf 1}_{0}^{T} {\bf y}_t +  {\bf 1}_{-1}^{T} {\bf a}_t  \right)^T \left( {\bf F}_{xx} \circ \Sigma_r \right)
 \left(  {\bf 1}_{0}^{T} {\bf y}_t +  {\bf 1}_{-1}^{T} {\bf a}_t  \right) 
 + \mbox{Tr} \left[ {\bf F}_{zz} \Sigma_{z} \right] 
\eea
After some algebra, we put Eq.(\ref{F_next}) in a form similar to Eq.(\ref{R_r_dx}):
 \beq
\label{F_next_mat}
\mathbb{E}_{t, {\bf a}} \left[  F_{t+1}^{\pi}({\bf y}_{t+1})  \right]  = 
\delta  {\bf a}_t^T {\bf H}_{aa} \delta {\bf a}_t + \delta {\bf y}_t^T {\bf H}_{yy} \delta {\bf y}_t  + 
 \delta  {\bf a}_t^T  {\bf H}_{ay} \delta {\bf y}_t + \delta  {\bf a}_t^T {\bf H}_{a} + 
  \delta {\bf y}_t^{T} {\bf H}_{y} + \widehat{f} (\bar{{\bf y}}_t, \bar{{\bf a}}_t)  
\eeq 
where
\bea
\label{F_vect_aa}
&&  {\bf H}_{aa} = \Psi_a^{T} {\bf F}_{yy}  \Psi_a + {\bf 1}_{-1} \left( {\bf F}_{xx} \circ \Sigma_r \right)  {\bf 1}_{-1}^{T} \nonumber \\
&&  {\bf H}_{yy} = \Psi_y^{T} {\bf F}_{yy}  \Psi_y + {\bf 1}_{0} \left( {\bf F}_{xx} \circ \Sigma_r \right)  {\bf 1}_{0}^{T} \nonumber \\
&&  {\bf H}_{ay} = 2 \Psi_a^{T} {\bf F}_{yy}  \Psi_y +  2 \cdot {\bf 1}_{-1} \left( {\bf F}_{xx} \circ \Sigma_r \right)  {\bf 1}_{0}^{T}  \nonumber \\
&&  {\bf H}_{a} =   \Psi_a^{T} {\bf F}_{y} + 2 \Psi_a^{T} {\bf F}_{yy}  \Psi_0 +  2 \cdot {\bf 1}_{-1} 
\left( {\bf F}_{xx} \circ \Sigma_r \right)  \left( {\bf 1}_{0}^{T}  \bar{{\bf y}}_{t} + {\bf 1}_{-1}^{T}  \bar{{\bf a}}_{t} \right) \\
&&  {\bf H}_{y} =   \Psi_y^{T} {\bf F}_{y} + 2 \Psi_y^{T} {\bf F}_{yy}  \Psi_0 +  2 \cdot {\bf 1}_{0} 
\left( {\bf F}_{xx} \circ \Sigma_r \right)  \left( {\bf 1}_{0}^{T}  \bar{{\bf y}}_{t} + {\bf 1}_{-1}^{T}  \bar{{\bf a}}_{t} \right) \nonumber \\
&& \widehat{f} (\bar{{\bf y}}_t, \bar{{\bf a}}_t) =   F_0  (\bar{{\bf y}}_{t+1}, \bar{{\bf a}}_{t+1}) + 
\Psi_0^T {\bf F}_{y} + \Psi_0^T {\bf F}_{yy} \Psi_0 \nonumber  \\
&& + 
 \left( {\bf 1}_{0}^{T}  \bar{{\bf y}}_{t} + {\bf 1}_{-1}^{T}  \bar{{\bf a}}_{t} \right)^{T} \left( {\bf F}_{xx} \circ \Sigma_r \right) 
  \left( {\bf 1}_{0}^{T}  \bar{{\bf y}}_{t} + {\bf 1}_{-1}^{T}  \bar{{\bf a}}_{t} \right) + \mbox{Tr} \left[ {\bf F}_{zz} \Sigma_z \right] \nonumber 
\eea
These equations can be used for both the finite-horizon and infinite-horizon settings. For the former case, all parameters in the right-hand sides of 
Eqs.(\ref{F_vect_aa}) refer to the future time moment $ t + 1 $, so that Eqs.(\ref{F_vect_aa}) serve as a part of a backward recursion scheme to be completed below.
On the other hand, for an infinite-horizon case, they can be used as updates equations for time-independent parameters of the free energy function 
(\ref{F_parametrization}). 

 Next we take the Bellman equation for the G-function
 \beq
 \label{G_from_F_3}
 G_t^{\pi} ( {\bf y}_t , {\bf a}_t) 
 = \hat{R}_{t}(  {\bf y}_t , {\bf a}_t ) 
 +  \gamma  \mathbb{E}_{t, {\bf a}} \left[  F_{t+1}^{\pi} (  {\bf y}_{t+1} ) \right] 
 \eeq 
where we substitute Eqs.(\ref{G_dx_du}), (\ref{R_r_dx}) and (\ref{F_next_mat}). Equating coefficients in front of like powers of 
$  \delta {\bf x}_t  $ and $ \delta {\bf a}_t $ in the left-hand side and the right-hand side of the resulting equation, we get a set of recursive relations 
for matrix coefficients defining the G-function in  Eq.(\ref{G_dx_du}):
\bea
\label{recursive_G}
{\bf G}_{aa} 
\hskip-0.5cm && =  \hat{{\bf R}}_{aa}  +  {\bf H}_{aa}  , \; 
{\bf G}_{yy} =  \hat{{\bf R}}_{yy} +  {\bf H}_{yy}, \;
{\bf G}_{ay} =  \hat{{\bf R}}_{ay}  +  {\bf H}_{ay}
  \nonumber \\
{\bf G}_{a}  \hskip-0.5cm && =  \hat{{\bf R}}_{a}  +  {\bf H}_{a} , \;
{\bf G}_{y}   =  \hat{{\bf R}}_{y}  +  {\bf F}_{y}, \; 
g (\bar{{\bf y}}_t, \bar{{\bf a}}_t) =  r ( \bar{{\bf y}}_t, \bar{{\bf a}}_t ) +  \widehat{f} (\bar{{\bf y}}_t, \bar{{\bf a}}_t) 
\eea
In these equations, coefficients in the left-hand side and the right-hand side refer to the same time $ t $, therefore they can be used in the same way for both the finite- and infinite horizon cases.

\subsection{Backward recursion with observable rewards}
\label{sect:Backward_obs_rewards}

We first consider a complete backward recursion scheme for a finite-horizon case, that would apply if rewards were observed.
Below, we will modify this scheme to replace observed rewards by their estimated values. 
In both cases, Eqs.(\ref{recursive_G}) should be solved by backward recursion, starting at the planning horizon $ T $ with a terminal condition.  

For an arbitrary time step $ t < T $, we proceed as follows. 
First, we use Eqs.(\ref{recursive_G}) to obtain parameters of the Q-function at time $ t $ .
Note that parameters entering the right-hand of Eqs.(\ref{recursive_G})  are known at time $ t $, as they 
are computed using the values defined at time step $ t + 1 $.

Second, we use the computed Q-function as parametrized by Eq.(\ref{G_dx_du}) to compute the F-function at time $ t  $ according to Eq.(\ref{F_opt}).
To this end, we 
express the prior $ \pi_0 $ in Eq.(\ref{pi_0}) 
in terms of increments $ \delta{\bf a}_t $ with the mean  $ \widehat{\delta{\bf a}}_t = \hat{ {\bf a}}_t - \bar{{\bf a}}_t  $ (recall that we condition on the value of 
$ \bar{{\bf a}}_t $):
\beq
\label{pi_0_1}
\pi_0 ( \delta {\bf a}_t | {\bf y}_t ) = \frac{1}{ \sqrt{ (2 \pi)^{N}  \left| \Sigma_p \right| }} \exp\left( - \frac{1}{2} \left( \delta {\bf a}_t -  \widehat{\delta{\bf a}}_t\right)^{T} \Sigma_p^{-1}  
\left( {\bf a}_t - \widehat{\delta{\bf a}}_t \right) \right)
\eeq
where  
\beq
\label{prior_mean}
\widehat{\delta{\bf a}}_t = \hat{ {\bf a}}_t - \bar{{\bf a}}_t  =  \hat{ {\bf A}}_0 + \hat{ {\bf A}}_1 \left(\bar{{\bf y}}_t + \delta {\bf y}_t \right)
  - \bar{ {\bf a}}_t  
\eeq
Using this in Eq.(\ref{F_opt}) along with we Eqs.(\ref{recursive_G}) and replacing a discrete sum  by an integral\footnote{Recall that we used a discrete notation for convenience only, while working in fact in a continuous-action formulation.}, we obtain 
\bea
\label{F_opt_2}
F_t^{\pi}({\bf y}_t ) 
\hskip-0.5cm && =  \frac{1}{\beta} \log Z_t =  \frac{1}{\beta} \log \sum_{ \delta{\bf a}_t} \pi_0 
( \bar{{\bf a}}_t + \delta {\bf a}_t | {\bf y}_t) e^{ \beta G_t^{\pi} ( {\bf y}_t, {\bf a}_t)  }  \nonumber \\
&& = \frac{1}{\beta} \left[ - \frac{N_a}{2} \log (2 \pi) - \frac{1}{2} \log \left| \Sigma_p \right| + \beta \delta {\bf y}_t^T {\bf G}_{yy} \delta {\bf y}_t + 
\beta  \delta {\bf y}_t^T {\bf G}_{y} + \beta g (\bar{{\bf y}}_t, \bar{{\bf a}}_t) \right. \nonumber \\
&& \left. - \frac{1}{2}   \widehat{\delta{\bf a}}_t^T \Sigma_p^{-1} 
\widehat{\delta{\bf a}}_t + \log \int d \, {\bf a} e^{ - \frac{1}{2}  {\bf a}^T \left( \Sigma_p^{-1} - 2 \beta {\bf G}_{aa} \right)  {\bf a} 
+ {\bf a}^T \left(  \Sigma_p^{-1} \widehat{\delta{\bf a}}_t +  \beta{\bf G}_{ay}  \delta {\bf y}_t +  \beta {\bf G}_{a} \right) } \right]
\eea
To simplify formulae below, we introduce auxiliary quantities
\bea
\label{aux}
&& {\bf b}_t =  \bar{ {\bf a}}_t - \hat{ {\bf A}}_0  - \hat{ {\bf A}}_1 \bar{{\bf y}}_t   , \; 
 \tilde{\Sigma}_p = \Sigma_p^{-1} - 2 \beta {\bf G}_{aa},   \nonumber \\
 && \Gamma_{\beta} = \frac{1}{\beta} \left( {\bf I} - \left(\Sigma_p^{-1} \right)^T \tilde{\Sigma}_p^{-1} \right)  \Sigma_p^{-1}, \; 
 \Upsilon_{\beta} =   \tilde{\Sigma}_p^{-1} \Sigma_p^{-1} \nonumber \\
 && {\bf E}_{ay} = \Upsilon_{\beta} \hat{\bf A}_{1} + \frac{1}{2} \beta  \tilde{\Sigma}_p^{-1}  {\bf G}_{ay}, \; 
 {\bf D}_{ay} =  {\bf G}_{ay}^{T} \Upsilon_{\beta} - \hat{ {\bf A}}_1^{T} \Gamma_{\beta}   \\
 && {\bf E}_{a} = \hat{\bf A}_{1}^T \Upsilon_{\beta} {\bf G}_a + \beta {\bf G}_{ay}^T \tilde{\Sigma}_p^{-1} {\bf G}_a , \; 
 \mathcal{L}_{\beta} = \frac{1}{2 \beta} \left( \log  \left| \Sigma_p \right|  + \log \left| \tilde{\Sigma}_p \right|  \right) 
 \nonumber 
\eea
Note that $ \lim_{\beta \rightarrow 0}  \Gamma_{\beta}  = 0 $ and $ \lim_{\beta \rightarrow 0}  \Upsilon_{\beta}  = 1 $.  
Using Eqs.(\ref{aux}) for the Gaussian integral (\ref{F_opt_2}), we can express it as in the same form as in Eq.(\ref{F_parametrization}):
\beq
 \label{F_parametrization_2}
 F_t^{\pi}({\bf y}_t) = \delta  {\bf y}_t^T {\bf F}_{yy}  \delta  {\bf y}_t  
 +   \delta  {\bf y}_t^{T} {\bf F}_y +F_0 (\bar{{\bf y}}_t, \bar{{\bf a}}_t)
 \eeq 
where the coefficients are now {\it computed} as follows: 
\bea
\label{F_coeffs}
&&  {\bf F}_{yy} = {\bf G}_{yy} +  {\bf G}_{ay}^T  {\bf E}_{ay}
- \frac{1}{2}  \hat{ {\bf A}}_1^{T} \Gamma_{\beta}  \hat{ {\bf A}}_1 \nonumber \\
&&   {\bf F}_{y}  =  {\bf G}_{y}  - {\bf D}_{ay}  {\bf b}_t  
+ \hat{ {\bf A}}_1^{T} \Upsilon_{\beta} {\bf G}_a + \beta  {\bf G}_{ay}^T  \tilde{\Sigma}_p^{-1}  {\bf G}_a \\
&&  F_0 (\bar{{\bf y}}_t, \bar{{\bf a}}_t) = g (\bar{{\bf y}}_t, \bar{{\bf a}}_t)   
- \frac{1}{2} {\bf b}_t^T \Gamma_{\beta} {\bf b}_t -    {\bf G}_a^T \Upsilon_{\beta} {\bf b}_t + \frac{ \beta}{2}  {\bf G}_a^T \tilde{\Sigma}_p^{-1} {\bf G}_a
-  \mathcal{L}_{\beta}  \nonumber 
%
%
 \eea
%

\def\thesection{B}	
\setcounter{equation}{0}
\def\theequation{\thesection.\arabic{equation}}

\section*{Appendix B: IRL for a single investor case}
\label{sect:Appendix_B}


In this appendix, we consider the case of a single investor with observable actions as a special case of our model.
To recall, in this case, we build a probabilistic model of a specific trader, assuming that we have access to trader's trading record.
This model is given by the Gaussian policy of Eq.(\ref{pi_post_M}) where the mean and variance in Eq.(\ref{new_mean_var}) are computed using trader's trading data, interpreted as trader's observed actions $ {\bf a}_t $.  

A major simplification of the single investor inference in our model is that when actions are observed, we do not need an inner integral over $ {\bf a}_t $ 
in Eq.(\ref{F_Var_EM}).
The only integration that we need in this case is the outer integration over $ \bar{\bf a}_t $.  

For such setting with investor-specific actions and rewards, 
estimation of parameters of Eq.(\ref{R_r_dx}) amounts 
to the EM algorithm with the free energy for a set of $ N_b $ trajectories of length $ T $ of the following form (compare with Eq.(\ref{F_Var_EM}))
\beq
\label{F_Var_EM_single_agent}
 \mathcal{F}_s({\bf w}, \theta) 
 = \sum_{b=1}^{N_b} \sum_{t=0}^{T} \int d \bar{\bf a}_t  \, q_{\bar{a}}(  \bar{\bf a}_t | {\bf y}, {\bf w}  )
  \log  \frac{\pi_{\theta} \left( {\bf a}_t  | {\bf y}_t \right) 
  p_{\theta} \left( {\bf y}_{t+1},  | {\bf y}_t,  {\bf a}_t \right) }{ q_{\bar{a}}(  \bar{\bf a}_t | {\bf y}, {\bf w}  )} \nonumber \\  
\eeq 
where $ {\bf y}_t $ and $  {\bf a}_t  $ stands for observed values of investments, signals and trades in the investor portfolio, stored as a historical dataset, 
and conditional transition probability $  p_{\theta} \left( {\bf y}_{t+1} | {\bf y}_t,  {\bf a}_t   \right) $ is defined in Eq.(\ref{trans_prob_y}).
%


The complete variational EM IRL algorithm for a single investor is given by  Algorithm~\ref{Algorithm I|}.
 In step 1, the expectation of the next-time F-function is computed using Eq.(\ref{F_next_mat})  
 within a backward recursion that starts with a fixed terminal condition at time $ t = T $. 
In step 2, we compute the reward using Eq.(\ref{R_r_dx}). 
In step 3, an update of the Q-function is performed 
using Eq.(\ref{recursive_G}). The time-$t$ F-function is computed in step 4 using Eq.(\ref{F_coeffs}). 
Finally, in step 5, 
the optimal 
policy as a function of $ \theta $ is recomputed using Eq.(\ref{pi_post_M}).
Computing these quantities for all transitions in the mini-batch, we obtain the free energy (\ref{F_minib}) for the mini-batch. 
 This is used to produce an update of the current estimation of $ \theta $ using a 
learning rate  $ \alpha_{\theta} $. 
The new updated values of $ \theta $  are then used to update parameters  $ \hat{ {\bf A}}_1^{(k)}, \, \hat{ {\bf A}}_1^{(k)}, \,  \Sigma_p^{(k)} $ of the policy $ \pi_{\theta} $.
Then the algorithm proceeds to the next iteration.

\vskip0.5cm
\begin{algorithm}[H]
\label{Algorithm II}
    \SetAlgoLined
    \KwData{a sequence of states and signals}
    \KwResult{the reward function, optimal policy, and value function }
     Set  the learning rates $ \alpha_{\theta}, \, \alpha_{\omega} $, batch size $ N_b $, initial parameters $ \theta^{(0)}, \,  \omega^{(0)},  
     \hat{ {\bf A}}_0^{(0)}, \, \hat{ {\bf A}}_1^{(0)}, \,  \Sigma_p^{(0)} $  \\
    Set  $ k = 1 $ \\
    \While{not converged}{
    	Draw a new mini-batch of $ N_b $  $T$-step trajectories $ ( {\bf y}_t, \ldots, {\bf y}_{t+T} ) $  \\ 
    	{\it E-step}: \\
	 Compute the free energy $ \mathcal{F}_s(\omega, \theta^{(k-1)}) $  of the mini-batch using Eq.(\ref{F_Var_EM_single_agent}) \\
	Update recognition model parameters $ \omega^{(k)} =  (1-\alpha_{\omega})  \omega^{(k-1)} 
	+ \alpha_{\omega} \frac{\partial}{\partial \omega}  \mathcal{F}_s(\omega, \theta^{(k-1)}) $  \\
    	{\it M-step}:  Maximize $ \mathcal{F}_s(\omega^{(k)}, \theta) $ as a function of $ \theta $: \\
	\For{ \mbox{each transition} $ ( {\bf y}_t, {\bf y}_{t+1} ) $ for $ t = T-1, \ldots, 0 $ }{
	    	1. Compute the expected value at time $ t $ of the F-function at time $ t + 1 $. \\
		2. Compute the reward as a function of $ \theta $. \\
		3. Use steps 1 and 2 to update the Q-function at time $ t $ \\
		4. Compute the value of the F-function at time $ t $. \\
		5. Recompute the policy distribution  $ \pi_{\theta} ( {\bf a}_t| t, {\bf y}_t) $ as a function of $ \theta $ by updating its mean and variance.   \\     
        }
        Compute the free energy $ \mathcal{F}_s(\omega^{(k)}, \theta) $  of the mini-batch using Eq.(\ref{F_Var_EM_single_agent}) \\
        Update the parameter vector $ \theta^{(k)} = (1-\alpha_{\theta})  \theta^{(k-1)} + \alpha_{\theta} 
        \frac{\partial}{\partial \theta} \mathcal{F}_s(\omega^{(k)}, \theta)  $ \\
        Use the new value $ \theta^{(k)} $ to compute $  \hat{ {\bf A}}_1^{(k)}, \, \hat{ {\bf A}}_1^{(k)}, \,  \Sigma_p^{(k)} $ \\
        Increment $ k = k + 1 $
    }
\caption{IRL algorithm that learns the optimal policy, reward, and value function for a single investor.}
\end{algorithm}

\def\thesection{C}	
\setcounter{equation}{0}
\def\theequation{\thesection.\arabic{equation}}

\section*{Appendix C: Calibration results for the DJI portfolio}
\label{sect:Appendix_C}
Here we report results of Maximum Likelihood estimation of the market model (\ref{GMR}) for two sets of signals described in 
Sect.(\ref{sect:Experiments}). We show the results for  
the calibrated daily mean reversion parameter $ \kappa $ and variance $ \Sigma = \sigma^2 $ in Eq.(\ref{GMR}). Fitted weights of the signals are not shown to save space.
\begin{table}
\begin{center}
\begin{tabular}{|r||c|c|c|c|c|c|c|c|}
 \hline
& 2010 & 2011 & 2012 & 2013 & 2014 & 2015 & 2016 & 2017 \\
 \hline
AAPL &   0.7006 &   0.4707 &   0.3024 &   0.3621 &   0.2846 &   0.2403 &   0.2875 &   0.2036 \\ 
 \hline
AXP &   3.2127 &  -0.0447 &   2.5010 &   2.0031 &   1.7278 &   2.0213 &   2.6951 &   2.1649 \\ 
 \hline
BA &   3.3122 &   3.2815 &   2.9620 &   2.0125 &   1.7078 &   1.6523 &   1.9196 &   1.2189 \\ 
 \hline
CAT &   3.9048 &   2.6793 &   2.6194 &   2.8312 &   2.6242 &   3.4815 &   3.6305 &   2.4664 \\ 
 \hline
CSCO &   1.1863 &   1.6795 &   1.6632 &   1.3183 &   1.3022 &   1.1573 &   1.1861 &   0.9684 \\ 
 \hline
CVX &   1.0389 &   0.8057 &   0.7649 &   0.6913 &   0.7266 &   0.9405 &   0.9175 &   0.7601 \\ 
 \hline
DIS &   2.4381 &   2.3823 &   1.9216 &   1.4093 &   1.1324 &   0.8764 &   1.0116 &   0.9794 \\ 
 \hline
DWDP &   5.0047 &  -0.0785 &   4.2726 &   3.6760 &   2.7862 &   2.9058 &   2.8816 &   1.9451 \\ 
 \hline
GE &   0.9095 &   0.8770 &   0.7563 &   0.6556 &   0.6221 &   0.6057 &   0.5860 &   0.7540 \\ 
 \hline
GS &   1.9353 &   2.7332 &   2.9583 &   2.2502 &   2.0742 &   1.9744 &   2.3197 &   1.7672 \\ 
 \hline
HD &   3.0667 &   2.9361 &   1.9388 &   1.4898 &   1.3529 &   1.0774 &   1.0159 &   0.8540 \\ 
 \hline
IBM &   0.9639 &   0.7677 &   0.7085 &   0.7414 &   0.8652 &   1.0601 &   1.1543 &   1.0965 \\ 
 \hline
INTC &   1.3863 &   1.3800 &   1.2918 &   1.3920 &   1.0265 &   1.0671 &   1.0228 &   0.8517 \\ 
 \hline
JNJ &   0.9459 &   0.8814 &   0.8717 &   0.6484 &   0.5680 &   0.5876 &   0.5311 &   0.4577 \\ 
 \hline
JPM &   1.0068 &   1.1603 &   1.0930 &   0.8261 &   0.7321 &   0.6890 &   0.7049 &   0.4972 \\ 
 \hline
KO &   1.2571 &   1.0468 &   0.9509 &   0.9079 &   0.8886 &   0.8958 &   0.8481 &   0.8447 \\ 
 \hline
MCD &  -0.0194 &   1.8295 &   1.7101 &   1.6392 &   1.7204 &   1.7170 &   1.5119 &   1.3050 \\ 
 \hline
MMM &   2.7017 &   2.7410 &   2.6042 &   2.0222 &   1.7473 &   1.6580 &   1.6336 &   1.2609 \\ 
 \hline
MRK &   1.4462 &   1.5575 &   1.2520 &   1.1635 &   0.9826 &   1.0299 &   1.0075 &   0.9625 \\ 
 \hline
MSFT &   0.6883 &   0.7343 &   0.6487 &   0.5723 &   0.4595 &   0.4326 &   0.3776 &   0.2787 \\ 
 \hline
NKE &   5.4456 &   4.9397 &  -0.0221 &   3.6118 &   2.8779 &   2.0821 &   2.0890 &   2.1891 \\ 
 \hline
PFE &   1.1989 &   1.0767 &   0.9167 &   0.7890 &   0.8235 &   0.7841 &   0.8232 &   0.7923 \\ 
 \hline
PG &   0.9066 &   0.9232 &   0.8940 &   0.7440 &   0.7239 &   0.7564 &   0.7194 &   0.7010 \\ 
 \hline
TRV &   6.3443 &  -0.0210 &   6.4233 &   5.1396 &   5.0635 &   4.8245 &   4.9080 &   4.6598 \\ 
 \hline
UNH &   4.2802 &   3.2558 &   2.8095 &   2.3977 &   1.9617 &   1.4535 &   1.2777 &   0.8714 \\ 
 \hline
UTX &   2.4634 &   2.3068 &   2.2702 &   1.7594 &   1.5741 &   1.6942 &   1.9644 &   1.7134 \\ 
 \hline
V &   4.1773 &   3.7475 &   2.4586 &   1.6999 &   1.4692 &   1.1614 &   1.1130 &   0.8799 \\ 
 \hline
VZ &   1.8906 &   1.5762 &   1.3284 &   1.1525 &   1.1527 &   0.8486 &   0.7779 &   0.8242 \\ 
 \hline
WMT &   0.8198 &   0.8641 &   0.7082 &   0.6513 &   0.6365 &   0.6957 &   0.7640 &   0.6456 \\ 
 \hline
XOM &   0.5292 &   0.4262 &   0.4018 &   0.3992 &   0.3930 &   0.4779 &   0.4671 &   0.4656 \\ 
 \hline
\end{tabular}
\end{center}
\caption{Calibrated $ \kappa $ for a combination of a "perfect signal"  and a "noise signal"}
\label{Tab:kappa_perfect}
\end{table}

\newpage

\begin{table}
\begin{center}
\begin{tabular}{|r||c|c|c|c|c|c|c|c|}
 \hline
& 2010 & 2011 & 2012 & 2013 & 2014 & 2015 & 2016 & 2017 \\
 \hline
AAPL &   0.0001 &   0.0003 &   0.0001 &   0.0000 &   0.0000 &   0.0000 &   0.0000 &   0.0000 \\ 
 \hline
AXP &   0.0001 &   0.0016 &   0.0000 &   0.0000 &   0.0000 &   0.0000 &   0.0000 &   0.0000 \\ 
 \hline
BA &   0.0001 &   0.0003 &   0.0000 &   0.0000 &   0.0000 &   0.0000 &   0.0000 &   0.0001 \\ 
 \hline
CAT &   0.0001 &   0.0004 &   0.0001 &   0.0000 &   0.0000 &   0.0000 &   0.0000 &   0.0001 \\ 
 \hline
CSCO &   0.0001 &   0.0005 &   0.0001 &   0.0000 &   0.0000 &   0.0000 &   0.0000 &   0.0000 \\ 
 \hline
CVX &   0.0000 &   0.0002 &   0.0000 &   0.0000 &   0.0000 &   0.0001 &   0.0000 &   0.0000 \\ 
 \hline
DIS &   0.0000 &   0.0004 &   0.0001 &   0.0000 &   0.0000 &   0.0000 &   0.0000 &   0.0000 \\ 
 \hline
DWDP &   0.0001 &   0.0020 &   0.0001 &   0.0000 &   0.0000 &   0.0000 &   0.0000 &   0.0001 \\ 
 \hline
GE &   0.0001 &   0.0004 &   0.0000 &   0.0000 &   0.0000 &   0.0000 &   0.0000 &   0.0001 \\ 
 \hline
GS &   0.0001 &   0.0006 &   0.0001 &   0.0000 &   0.0000 &   0.0000 &   0.0000 &   0.0000 \\ 
 \hline
HD &   0.0001 &   0.0003 &   0.0001 &   0.0000 &   0.0000 &   0.0000 &   0.0000 &   0.0000 \\ 
 \hline
IBM &   0.0000 &   0.0002 &   0.0000 &   0.0000 &   0.0000 &   0.0000 &   0.0000 &   0.0000 \\ 
 \hline
INTC &   0.0001 &   0.0002 &   0.0001 &   0.0000 &   0.0000 &   0.0000 &   0.0000 &   0.0000 \\ 
 \hline
JNJ &   0.0000 &   0.0002 &   0.0000 &   0.0000 &   0.0000 &   0.0000 &   0.0000 &   0.0000 \\ 
 \hline
JPM &   0.0001 &   0.0005 &   0.0001 &   0.0000 &   0.0000 &   0.0000 &   0.0000 &   0.0000 \\ 
 \hline
KO &   0.0000 &   0.0001 &   0.0000 &   0.0000 &   0.0000 &   0.0000 &   0.0000 &   0.0000 \\ 
 \hline
MCD &   0.0002 &   0.0002 &   0.0000 &   0.0000 &   0.0000 &   0.0000 &   0.0000 &   0.0000 \\ 
 \hline
MMM &   0.0000 &   0.0003 &   0.0000 &   0.0000 &   0.0000 &   0.0000 &   0.0000 &   0.0000 \\ 
 \hline
MRK &   0.0000 &   0.0002 &   0.0000 &   0.0000 &   0.0000 &   0.0000 &   0.0000 &   0.0000 \\ 
 \hline
MSFT &   0.0001 &   0.0002 &   0.0000 &   0.0000 &   0.0000 &   0.0000 &   0.0000 &   0.0000 \\ 
 \hline
NKE &   0.0001 &   0.0002 &   0.0004 &   0.0000 &   0.0000 &   0.0000 &   0.0000 &   0.0000 \\ 
 \hline
PFE &   0.0001 &   0.0002 &   0.0000 &   0.0000 &   0.0000 &   0.0000 &   0.0000 &   0.0000 \\ 
 \hline
PG &   0.0000 &   0.0001 &   0.0000 &   0.0000 &   0.0000 &   0.0000 &   0.0000 &   0.0000 \\ 
 \hline
TRV &   0.0000 &   0.0015 &   0.0000 &   0.0000 &   0.0000 &   0.0000 &   0.0000 &   0.0000 \\ 
 \hline
UNH &   0.0001 &   0.0002 &   0.0000 &   0.0000 &   0.0000 &   0.0000 &   0.0000 &   0.0000 \\ 
 \hline
UTX &   0.0000 &   0.0003 &   0.0000 &   0.0000 &   0.0000 &   0.0000 &   0.0000 &   0.0000 \\ 
 \hline
V &   0.0001 &   0.0004 &   0.0001 &   0.0000 &   0.0000 &   0.0000 &   0.0000 &   0.0000 \\ 
 \hline
VZ &   0.0000 &   0.0001 &   0.0000 &   0.0000 &   0.0001 &   0.0000 &   0.0000 &   0.0000 \\ 
 \hline
WMT &   0.0000 &   0.0002 &   0.0001 &   0.0000 &   0.0000 &   0.0000 &   0.0000 &   0.0000 \\ 
 \hline
XOM &   0.0001 &   0.0002 &   0.0000 &   0.0000 &   0.0000 &   0.0000 &   0.0000 &   0.0000 \\ 
 \hline
\end{tabular}
\end{center}
\caption{Calibrated values of $ \sigma^2 $ for  a combination of a "perfect signal" and a "noise signal"}
\label{Tab:sigma_perfect}
\end{table}

\newpage

\begin{table}
\begin{center}
\begin{tabular}{|r||c|c|c|c|c|c|c|c|}
 \hline
& 2010 & 2011 & 2012 & 2013 & 2014 & 2015 & 2016 & 2017 \\
 \hline
AAPL &  -0.0122 &   0.0288 &  -0.0041 &   0.0047 &  -0.0038 &   0.0080 &   0.0045 &  -0.0038 \\ 
 \hline
AXP &   0.1524 &   0.1748 &   0.1317 &  -0.0354 &   0.0452 &   0.2109 &  -0.0387 &  -0.0299 \\ 
 \hline
BA &   0.0640 &   0.1488 &   0.2515 &  -0.0406 &   0.0542 &   0.0116 &   0.0550 &  -0.0307 \\ 
 \hline
CAT &  -0.0730 &   0.0430 &  -0.0278 &   0.0852 &  -0.0132 &   0.2088 &   0.1437 &  -0.0401 \\ 
 \hline
CSCO &   0.0309 &   0.1211 &  -0.0416 &   0.0326 &  -0.0234 &   0.0530 &   0.0277 &  -0.0177 \\ 
 \hline
CVX &  -0.0171 &   0.0376 &   0.0119 &   0.0276 &  -0.0092 &   0.0144 &   0.0884 &  -0.0070 \\ 
 \hline
DIS &   0.0971 &   0.0549 &   0.1424 &  -0.0284 &   0.0811 &  -0.0134 &   0.0196 &   0.0308 \\ 
 \hline
DWDP &   0.3140 &   0.2110 &   0.1000 &   0.1250 &   0.0446 &   0.1225 &   0.0668 &  -0.0277 \\ 
 \hline
GE &  -0.0152 &   0.0148 &   0.0287 &   0.0276 &   0.0194 &   0.0227 &   0.0141 &  -0.0182 \\ 
 \hline
GS &   0.0189 &   0.2639 &  -0.0738 &   0.0749 &   0.0348 &   0.0938 &  -0.0396 &   0.0526 \\ 
 \hline
HD &  -0.0557 &   0.0521 &   0.1519 &   0.0496 &   0.0260 &   0.1190 &   0.0225 &  -0.0193 \\ 
 \hline
IBM &   0.0938 &   0.0532 &   0.0128 &   0.0261 &  -0.0152 &   0.0801 &   0.0313 &  -0.0258 \\ 
 \hline
INTC &   0.0223 &   0.0315 &  -0.0071 &   0.0685 &  -0.0152 &   0.0055 &   0.0260 &  -0.0266 \\ 
 \hline
JNJ &   0.0081 &   0.0514 &   0.0191 &  -0.0163 &   0.0172 &   0.0665 &  -0.0052 &  -0.0090 \\ 
 \hline
JPM &   0.0326 &   0.1172 &  -0.0333 &   0.0286 &   0.0345 &   0.0470 &  -0.0071 &   0.0122 \\ 
 \hline
KO &   0.0519 &   0.0920 &   0.0188 &   0.0580 &   0.0246 &   0.0288 &   0.0788 &   0.0512 \\ 
 \hline
MCD &   0.1993 &   0.2217 &   0.0468 &   0.0669 &   0.0451 &   0.1489 &   0.0370 &  -0.0211 \\ 
 \hline
MMM &   0.1316 &   0.2049 &   0.1524 &  -0.0348 &  -0.0231 &   0.1088 &   0.0558 &  -0.0243 \\ 
 \hline
MRK &   0.0691 &   0.0122 &   0.0144 &   0.0530 &   0.0498 &   0.0572 &   0.1009 &  -0.0206 \\ 
 \hline
MSFT &  -0.0133 &   0.0349 &   0.0183 &   0.0273 &   0.0353 &   0.0084 &   0.0217 &  -0.0043 \\ 
 \hline
NKE &   0.2952 &   0.3174 &   0.2199 &  -0.0478 &   0.1105 &   0.1755 &   0.1088 &  -0.0258 \\ 
 \hline
PFE &   0.0197 &   0.0554 &   0.0293 &   0.0149 &   0.0264 &   0.0454 &   0.0208 &  -0.0141 \\ 
 \hline
PG &   0.0709 &   0.0314 &   0.0061 &   0.0340 &   0.0351 &   0.0116 &   0.0683 &   0.0246 \\ 
 \hline
TRV &   0.5598 &   0.3278 &   0.1963 &   0.3414 &  -0.1580 &   0.2416 &   0.1205 &   0.1552 \\ 
 \hline
UNH &   0.2908 &   0.1356 &   0.1251 &  -0.0413 &   0.0682 &   0.1874 &  -0.0204 &  -0.0139 \\ 
 \hline
UTX &   0.0452 &   0.0463 &   0.0799 &  -0.0270 &   0.0181 &  -0.0163 &   0.0469 &  -0.0223 \\ 
 \hline
V &   0.1820 &   0.4427 &   0.2446 &   0.1132 &  -0.0145 &   0.0954 &   0.1141 &  -0.0130 \\ 
 \hline
VZ &  -0.0785 &   0.1175 &   0.0159 &   0.0113 &   0.0170 &   0.0722 &  -0.0274 &  -0.0328 \\ 
 \hline
WMT &   0.0607 &   0.0184 &  -0.0087 &   0.0095 &  -0.0074 &   0.0456 &   0.1101 &  -0.0103 \\ 
 \hline
XOM &   0.0144 &   0.0103 &   0.0163 &   0.0040 &   0.0115 &   0.0209 &   0.0380 &   0.0068 \\ 
 \hline
\end{tabular}
\end{center}
\caption{Calibrated values of $ \kappa $ for exponential moving averages signals ($\gamma =  0.9 $ and $ 0.96 $)}
\label{Tab:kappa_exp}
\end{table}

\newpage

\begin{table}
\begin{center}
\begin{tabular}{|r||c|c|c|c|c|c|c|c|}
 \hline
& 2010 & 2011 & 2012 & 2013 & 2014 & 2015 & 2016 & 2017 \\
 \hline
AAPL &   0.0073 &   0.0088 &   0.0060 &   0.0053 &   0.0045 &   0.0065 &   0.0053 &   0.0041 \\ 
 \hline
AXP &   0.0082 &   0.0097 &   0.0047 &   0.0039 &   0.0041 &   0.0056 &   0.0055 &   0.0033 \\ 
 \hline
BA &   0.0080 &   0.0098 &   0.0044 &   0.0044 &   0.0044 &   0.0057 &   0.0054 &   0.0040 \\ 
 \hline
CAT &   0.0082 &   0.0112 &   0.0055 &   0.0039 &   0.0045 &   0.0062 &   0.0058 &   0.0045 \\ 
 \hline
CSCO &   0.0080 &   0.0108 &   0.0055 &   0.0049 &   0.0040 &   0.0059 &   0.0050 &   0.0038 \\ 
 \hline
CVX &   0.0061 &   0.0092 &   0.0042 &   0.0032 &   0.0042 &   0.0065 &   0.0052 &   0.0036 \\ 
 \hline
DIS &   0.0067 &   0.0101 &   0.0043 &   0.0039 &   0.0041 &   0.0059 &   0.0043 &   0.0039 \\ 
 \hline
DWDP &   0.0091 &   0.0123 &   0.0055 &   0.0047 &   0.0051 &   0.0070 &   0.0047 &   0.0126 \\ 
 \hline
GE &   0.0074 &   0.0098 &   0.0044 &   0.0038 &   0.0036 &   0.0059 &   0.0044 &   0.0045 \\ 
 \hline
GS &   0.0080 &   0.0114 &   0.0058 &   0.0044 &   0.0041 &   0.0057 &   0.0058 &   0.0044 \\ 
 \hline
HD &   0.0067 &   0.0089 &   0.0044 &   0.0039 &   0.0040 &   0.0053 &   0.0045 &   0.0034 \\ 
 \hline
IBM &   0.0057 &   0.0086 &   0.0040 &   0.0040 &   0.0041 &   0.0055 &   0.0047 &   0.0039 \\ 
 \hline
INTC &   0.0070 &   0.0091 &   0.0048 &   0.0043 &   0.0048 &   0.0061 &   0.0052 &   0.0041 \\ 
 \hline
JNJ &   0.0045 &   0.0084 &   0.0028 &   0.0031 &   0.0036 &   0.0047 &   0.0036 &   0.0031 \\ 
 \hline
JPM &   0.0080 &   0.0117 &   0.0058 &   0.0041 &   0.0041 &   0.0057 &   0.0055 &   0.0039 \\ 
 \hline
KO &   0.0051 &   0.0067 &   0.0035 &   0.0035 &   0.0036 &   0.0043 &   0.0038 &   0.0027 \\ 
 \hline
MCD &   0.0050 &   0.0066 &   0.0036 &   0.0030 &   0.0033 &   0.0052 &   0.0040 &   0.0034 \\ 
 \hline
MMM &   0.0061 &   0.0092 &   0.0037 &   0.0033 &   0.0037 &   0.0051 &   0.0038 &   0.0033 \\ 
 \hline
MRK &   0.0061 &   0.0081 &   0.0039 &   0.0037 &   0.0043 &   0.0056 &   0.0049 &   0.0038 \\ 
 \hline
MSFT &   0.0064 &   0.0083 &   0.0047 &   0.0048 &   0.0042 &   0.0067 &   0.0051 &   0.0037 \\ 
 \hline
NKE &   0.0065 &   0.0098 &   0.0051 &   0.0043 &   0.0046 &   0.0058 &   0.0051 &   0.0047 \\ 
 \hline
PFE &   0.0062 &   0.0084 &   0.0035 &   0.0039 &   0.0039 &   0.0053 &   0.0046 &   0.0030 \\ 
 \hline
PG &   0.0046 &   0.0062 &   0.0035 &   0.0036 &   0.0031 &   0.0046 &   0.0038 &   0.0032 \\ 
 \hline
TRV &   0.0058 &   0.0091 &   0.0041 &   0.0035 &   0.0034 &   0.0049 &   0.0044 &   0.0034 \\ 
 \hline
UNH &   0.0070 &   0.0099 &   0.0048 &   0.0042 &   0.0042 &   0.0062 &   0.0046 &   0.0036 \\ 
 \hline
UTX &   0.0061 &   0.0092 &   0.0045 &   0.0036 &   0.0037 &   0.0053 &   0.0046 &   0.0034 \\ 
 \hline
V &   0.0082 &   0.0101 &   0.0047 &   0.0043 &   0.0047 &   0.0056 &   0.0049 &   0.0033 \\ 
 \hline
VZ &   0.0054 &   0.0073 &   0.0037 &   0.0038 &   0.0078 &   0.0046 &   0.0041 &   0.0041 \\ 
 \hline
WMT &   0.0048 &   0.0067 &   0.0040 &   0.0031 &   0.0034 &   0.0055 &   0.0046 &   0.0042 \\ 
 \hline
XOM &   0.0059 &   0.0088 &   0.0038 &   0.0032 &   0.0039 &   0.0058 &   0.0046 &   0.0031 \\ 
 \hline
\end{tabular}
\end{center}
\caption{Calibrated values of $ \sigma^2 $ for exponential moving averages signals ($\gamma =  0.9 $ and $ 0.96 $)}
\label{Tab:sigma_exp}
\end{table}


\end{document}